\documentclass[a4paper, 11pt]{article}
\pdfoutput=1 
 
\usepackage{jheppub} 
                     
\usepackage[T1]{fontenc} 
                     
\usepackage{amsmath}  
\usepackage{amssymb}   
\usepackage{latexsym}
\usepackage{graphicx}
\usepackage{subfig} 
\usepackage{empheq}
\usepackage{upgreek} 
 
\usepackage{braket}
\usepackage{amssymb}
\usepackage{amsmath}
\usepackage{amsthm}
\usepackage{mathrsfs}
\usepackage{skak}
\usepackage{framed}
\usepackage{hyperref}
\usepackage{slashed}
\usepackage{array}

\hypersetup{colorlinks=true}
\hypersetup{linkcolor=blue}
\hypersetup{citecolor=blue}
\hypersetup{urlcolor=blue}

\numberwithin{equation}{section}

\usepackage
{datetime}
\usepackage{fancyhdr}
\usepackage{tikz, pgf}
\usetikzlibrary{shapes.misc}
\usetikzlibrary{shapes}
\usetikzlibrary{fit}
\usetikzlibrary{arrows}

\usetikzlibrary{decorations.pathmorphing}	
\usetikzlibrary{decorations.markings}
 \usetikzlibrary{patterns}

\tikzset{
    sugra/.style={decorate, decoration={snake}, draw=black},
    scalarphi/.style={dashed,draw=black, postaction={decorate},
        },
    scalarchi/.style={draw=brown}, 
    hwbou/.style={draw=blue, postaction={decorate}, ultra thick
        },
    vector/.style={draw=blue,decorate, decoration={snake}, draw},
	provector/.style={decorate, decoration={snake,amplitude=2.5pt}, draw},
	antivector/.style={decorate, decoration={snake,amplitude=-2.5pt}, draw},
   	 fermion/.style={draw=cyan, postaction={decorate},
        decoration={markings,mark=at position .55 with {\arrow[draw=black]{>}}}},
    fermionbar/.style={draw=cyan, postaction={decorate},
        decoration={markings,mark=at position .55 with {\arrow[draw=black]{<}}}},
    fermionnoarrow/.style={draw=black},
    gluon/.style={decorate, draw=red,
        decoration={coil, amplitude=4pt, segment length=5pt}},
    scalar/.style={dashed,draw=black, postaction={decorate},
        decoration={markings,mark=at position .55 with {\arrow[draw=black]{>}}}},
    scalarbar/.style={dashed,draw=black, postaction={decorate},
        decoration={markings,mark=at position .55 with {\arrow[draw=black]{<}}}},
    electron/.style={draw=black, postaction={decorate},
        decoration={markings,mark=at position .55 with {\arrow[draw=black]{>}}}},
    scalarnoarrow/.style={dashed, draw=black},
    electron/.style={draw=black, postaction={decorate},
        decoration={markings, mark=at position .55 with {\arrow[draw=black]{>}}}},
	bigvector/.style={decorate, decoration={snake, amplitude=4pt}, draw},
    photon/.style={draw=violet, decorate, decoration={snake}, draw},
    higgs/.style={dashed, draw=black, postaction={decorate},
        },	
        goldstone/.style={draw=brown, postaction={decorate},
        },    
          ghost/.style={dashed, draw=magenta, postaction={decorate},
        decoration={markings, mark=at position .55 with {\arrow[draw=black]{>}}}
        },  
          antighost/.style={dashed, draw=magenta, postaction={decorate},
        decoration={markings, mark=at position .55 with {\arrow[draw=black]{<}}}
        }, 
            scalartwo/.style={dashed,draw=brown, postaction={decorate},
        decoration={markings,mark=at position .55 with {\arrow[draw=black]{>}}}},
    scalarbartwo/.style={dashed,draw=brown, postaction={decorate},
        decoration={markings,mark=at position .55 with {\arrow[draw=black]{<}}}}, 
    fermiontwo/.style={draw=purple, postaction={decorate},
        decoration={markings,mark=at position .55 with {\arrow[draw=black]{>}}}},
    fermionbartwo/.style={draw=purple, postaction={decorate},
        decoration={markings,mark=at position .55 with {\arrow[draw=black]{<}}}},    
        mphoton/.style={decorate, decoration={snake}, draw=violet},
        realscalar/.style={draw=black}, 
        fakerealscalar/.style={draw=white}, 
        realscalarone/.style={ draw=black},
    	realscalartwo/.style={draw=brown},    	    pseudoscalar/.style={draw=brown},
        mgluon/.style={decorate, draw=blue,
        	decoration={coil,amplitude=4pt, segment length=5pt}},
         weylfermion/.style={draw=orange, postaction={decorate},
        decoration={markings,mark=at position .55 with {\arrow[draw=black]{>}}}},
         weylfermionbar/.style={draw=orange, postaction={decorate},
        decoration={markings,mark=at position .55 with {\arrow[draw=black]{<}}}}, 
    majorana/.style={draw=cyan, postaction={decorate},
        decoration={markings,mark=at position .55 with {\arrow[draw=black]{><}}}},
    majoranabar/.style={draw=cyan, postaction={decorate},
        decoration={markings,mark=at position .55 with {\arrow[draw=black]{><}}}},    
   	wboson/.style={draw=blue,decorate, decoration={snake,amplitude=4pt}, draw},  
    zboson/.style={draw=violet, decorate, decoration={snake}, draw},   
    lepton/.style={draw=black, postaction={decorate},
        decoration={markings, mark=at position .55 with {\arrow[draw=black]{>}}}},
    leptonbar/.style={draw=black, postaction={decorate},
        decoration={markings, mark=at position .55 with {\arrow[draw=black]{<}}}}, 
    clepton/.style={draw=cyan, postaction={decorate},
        decoration={markings, mark=at position .55 with {\arrow[draw= black]{>}}}},
    cleptonbar/.style={draw=cyan, postaction={decorate},
        decoration={markings, mark=at position .55 with {\arrow[draw=black]{<}}}},   
   nlepton/.style={draw=orange, postaction={decorate},
        decoration={markings, mark=at position .55 with {\arrow[draw=black]{>}}}},
    nleptonbar/.style={draw=orange, postaction={decorate},
        decoration={markings, mark=at position .55 with {\arrow[draw=black]{<}}}},              
        graviton/.style={draw=blue, decorate, decoration={snake, amplitude=4pt}, draw},  
        bgraviton/.style={draw=blue, decorate, decoration={snake, amplitude=4pt}, draw},  
        gravitino/.style={draw=red, postaction={decorate}, 
        decoration={snake,  markings, mark=at position .55 with {\arrow[draw=black]{><}}}},
    	gravitinobar/.style={draw=red, postaction={decorate},
        decoration={snake, markings, mark=at position .55 with {\arrow[draw=black]{><}}} },  
    phir/.style={draw=blue, postaction={decorate},},
   phil/.style={dashed,draw=blue,},
     phiav/.style={draw=cyan, postaction={decorate},},
   phidif/.style={dashed,draw=cyan,},  
    chir/.style={draw=red, postaction={decorate},},
   chil/.style={dashed,draw=red,},  
}

\tikzstyle{block} = [draw, rectangle, 
    minimum height=3em, minimum width=6em]

\def\iimg{ {\bf i}} 

\def\ucharge{\mathcal{U}} 

\def\tsfactor{{\widetilde {\mathcal{S}} }}
\def\tmfactor{\widetilde {\mathcal {M}}}
\def\tcfactor{\widetilde{\mathcal{ C}}} 
\def\sfactor{\mathcal{S}}
\def\mfactor{\mathcal{M}}
\def\cfactor{\mathcal{C}} 
\def\scharge{\mathcal{Q}}
\def\zfactor{\mathcal{Z}}

\newcommand{\scri}{{\mathscr I}}

\hbadness=\maxdimen
\vbadness=\maxdimen

\allowdisplaybreaks[1]

\title{\boldmath Leading soft theorem for multiple gravitini}

\vspace{10mm}

\author[ \symrook,\,    \symknight ]{Diksha Jain }
\author[\symknight]{, Arnab Rudra }
\author[]{\\ }
\vspace*{4.0ex} 

\vspace{10mm}

\affiliation[\symrook]{SISSA International School for Advanced Studies and INFN Trieste,\\ Via Bonomea 265, Trieste 34136 Italy. \\ }

\vspace*{4.0ex} 

\affiliation[\symknight]{International Centre for Theoretical Physics\\ 
Strada Costiera 11, Trieste 34151 Italy. \\ }

\vspace*{2.0ex}

\emailAdd{djain@sissa.it}  
\emailAdd{ rudra.arnab@gmail.com } 

\abstract{We compute leading soft theorem for multiple gravitini (and gravitons) in an arbitrary theory of supergravity with an arbitrary number of finite energy particles by extending Sen's approach \cite{Sen:2017xjn, Sen:2017nim}  to fermionic symmetry. Our results are independent of the mass and the spin of the external particles. Our results are valid for any compactification of type II and Heterotic superstring theory. Our results are valid at all orders in perturbation theory for  four and higher spacetime dimensions.}

\begin{document}
\maketitle
\raggedbottom

\newpage

\section{Introduction}
\label{sec:jrintro}
In a scattering event, a particle whose momentum in the center of mass frame is much lower than other particles is called a soft particle and the relation between the $S$-matrix with and without the soft particles is known as the soft theorem. The soft theorems capture certain universal features of the theories. Study of soft theorems is an old subject \cite{GellMann:1954kc, Low:1954kd, Low:1958sn,  Weinberg:1964ew, Weinberg:1965nx, Gross:1968in, Jackiw:1968zza, Burnett:1967km, Bell:1969yw, Saito:1970yq, DelDuca:1990gz}. However, in the last few years the interest on soft theorems has been renewed because of its connection to asymptotic symmetry \cite{Strominger:2013lka, Strominger:2013jfa, He:2014laa, Strominger:2014pwa, Kapec:2015vwa, Strominger:2017zoo}. It has been argued that the flat space $S$-matrix in four dimensions should possess the symmetry of an asymptotically flat space. This symmetry is spontaneously broken and the graviton is the Goldstone boson of this spontaneously broken symmetry. Similarly, soft photon theorem can also be understood as a consequence of the large gauge transformations. These studies established the relation between seemingly different phenomena - Asymptotic symmetry, soft theorems, and memory effect. In subsequent papers the study of asymptotic symmetry was extended to higher than four dimensions \cite{Kapec:2014zla, Kapec:2015vwa, Pate:2017fgt} but our understanding of asymptotic symmetries in an arbitrary dimension is far from being complete. The spacetime dimension independent treatment mostly relies on Feynman diagrammatic techniques. In this approach, one starts from a specific Lagrangian and then computes only a  subclass of the Feynman diagrams which contribute to the (sub-)leading soft theorem(s). Soft photon and soft graviton theorem were computed in  \cite{White:2011yy,  Casali:2014xpa, Schwab:2014xua, Bern:2014oka,  He:2014bga, Larkoski:2014hta, White:2014qia, Cachazo:2014dia, Afkhami-Jeddi:2014fia, Schwab:2014fia, DiVecchia:2015oba, Bianchi:2014gla, Broedel:2014fsa, Vera:2014tda, Zlotnikov:2014sva, Du:2014eca, Cachazo:2015ksa, Kalousios:2014uva, Bern:2014vva, Bonocore:2014wua, Schwab:2014sla, Klose:2015xoa, Lipstein:2015rxa, Volovich:2015yoa, Rao:2016tgx, DiVecchia:2015bfa, Bianchi:2015yta, Guerrieri:2015eea, Huang:2015sla, Alston:2015gea, Bianchi:2015lnw, DiVecchia:2015srk, Bianchi:2016tju, DiVecchia:2016amo, He:2016vfi, Cachazo:2016njl, Saha:2016kjr, DiVecchia:2016szw, DiVecchia:2017uqn, Cheung:2016drk, Luna:2016idw, Elvang:2016qvq, Saha:2017yqi, Chakrabarti:2017ltl, Hamada:2018vrw, Li:2018gnc}. The new impetus in this direction came from Sen's work \cite{Sen:2017xjn, Sen:2017nim}. This method relies on covariantization of 1PI effective action with respect to the soft field. So the result does not depend on any particular Lagrangian or on asymptotic symmetry. This powerful method was used to compute sub-sub-leading soft graviton theorem \cite{Laddha:2017ygw} and also to compute multiple (sub-)leading soft graviton theorem \cite{Chakrabarti:2017ltl}. It has been noted that the soft-photon theorem is universal at leading order \cite{Weinberg:1964ew, Weinberg:1965nx} and the soft-graviton theorem is universal not only in the leading order but also in the sub-leading order \cite{Cachazo:2014fwa}. In a recent paper \cite{AtulBhatkar:2018kfi}, the soft theorems has been investigated when two different types of massless particles are present. 

In four and higher dimensions the theories of massless particles are severely constrained by Poincare invariance and Unitarity. Massless particles with spin $>2$ cannot couple minimally; they can couple only through the field strength. So the only particles which possess gauge invariance and can have minimal coupling are spin $1,3/2$ and $2$.  We already have a complete understanding of soft photon and soft graviton theorem. However, our knowledge of the soft gluon and the soft gravitino theorems is limited. These computations involve subtlety in the sense that the leading soft factors do not commute and their commutator is also leading order in soft momenta. At the level complexity, the soft gravitino theorem is more subtle than photon or graviton but significantly less subtle than that of the gluon. This is because even though the commutator of two soft factors is non-vanishing, the commutator of three soft factors vanishes in the case of gravitino but it does not vanish for gluon. However, for  specific type of theories, soft gluon theorem can be conveniently computed using CHY formalism \cite{Cachazo:2013hca, Cachazo:2013iea, Cachazo:2014xea, Cachazo:2013gna}. This advantage is not currently available for soft gravitino/photino. In this paper, we wish to derive the leading order soft theorem for gravitino in a general quantum field theory with local Supersymmetry in an arbitrary number of dimensions. Soft gravitino operator is a fermionic soft operator. Though a lot is known about bosonic soft theorems, the available literature for fermionic soft theorems is significantly little. Single soft photino theorem was computed in \cite{Dumitrescu:2015fej}.

Amplitudes with one and two soft gravitini for four-dimensional supergravity theories were computed for a particular model in \cite{Grisaru:1976vm, Grisaru:1977kk, Grisaru:1977px}.  The result for single soft gravitino in $D=4$ can also be obtained from asymptotic symmetry \cite{Lysov:2015jrs, Avery:2015iix}.  We generalize the result to the case with an arbitrary number of soft gravitini. In our work, we follow Sen's covariantization approach \cite{Sen:2017xjn, Sen:2017nim, Laddha:2017ygw}. The advantage of this method is that it is valid for arbitrary theories, to all orders in perturbation theory and in arbitrary dimensions, as long as there is no infrared divergence.  In this paper, we mostly follow the notation and conventions of \cite{Laddha:2017ygw}. We have summarized our notation and convention in section \ref{sec:gravitinonotation}. We find that for multiple gravitini the leading order result is universal.  
 
 An important aspect of quantum theories with massless particles is IR divergence. In $D=4$ loop diagrams suffer from IR divergences. In QED, there is a procedure to write IR finite S matrix element \cite{Kulish:1970ut}. This procedure has also been understood from the perspective of asymptotic symmetries \cite{Kapec:2017tkm, Gabai:2016kuf, Choi:2017ylo}. There has been some recent progress for quantum gravity \cite{Ware:2013zja, Choi:2017bna}. In section \ref{sec:irdivergence} we discuss the IR divergence of supergravity theories. Our result is valid for any theory of supergravity in $D\geq   4$  \footnote{In $D\leq 3$ there is no graviton and gravitino. }.  
 
Background independence of String field theory implies that String field theory in the presence of a soft field is obtained simply by deforming the world-sheet CFT by a marginal super-conformal operator which corresponds to that field.  Recently Sen has proved background independence in superstring field theory \cite{Sen:2017szq}. So our analysis is also valid for any supersymmetric compactification of superstring theory.


\subsection{Main result}
\label{subsec:jrmainresult}
Our main result is equation \eqref{multiplegravitinosoft} where we have written the soft factor for arbitrary number of external soft gravitini. Consider an amplitude $\Gamma_{M+N}(\{p_i\},\{k_u\})$ for $M$ soft gravitini and $N$ hard particles. It is  related to the amplitude of $N$ hard particles $\Gamma_N(\{p_i\})$ in the following way  
\begin{eqnarray}\label{jrmainresult}
     \Gamma_{M+N}(\{p_i\},\{k_{u_i}\}) =   \left[\prod_{i=1}^M \sfactor_{u_i} + \sum_{A = 1}^{\lfloor M/2\rfloor }\prod_{i =1}^{A}\mfactor_{u_iv_i}\prod_{j=1}^{M -2A}\sfactor_{r_j} \right]\Gamma_N(\{p_i\}) +\mathcal{O}(1/k^{M-1})
\end{eqnarray}
Various terms in this expression are explained below  
\begin{enumerate}
    \item  $p_i$ are the momenta of the hard particles, $k_u$ are momenta of the soft particles    
    \item $\sfactor_u$ is the soft factor for single soft gravitino. It is given by 
\begin{eqnarray}
        \sfactor_u &=&  \kappa  \sum_{i=1}^N  \left(  \frac{ \epsilon^{(u)\, \alpha}_{\mu}p_i^{\mu}}{p_i\cdot k_u} \scharge_\alpha \right)
\end{eqnarray}
Here $\kappa$ is the gravitational coupling constant. $\epsilon^{(u)\, \alpha}_{\mu}$ is the polarization of the $u^{\textrm{th}}$ gravitino; it has a Lorentz vector index \& a majorana spinor index and it is grassmann odd. The gravitino polarization (in the harmonic gauge) satisfies the transversality condition and gamma traceless condition. 
\begin{eqnarray}
    (k_u)^\mu \epsilon^{(u)\, \alpha}_{\mu}=0 
    \qquad,\qquad 
\gamma^{\mu}_{ \alpha \beta}\,     \epsilon^{(u)\, \beta}_{\mu} =0 
\end{eqnarray}
$\scharge_\alpha$ are the supersymmetry charges/generators.  The  single soft gravitino factor in four dimensional theories was also computed in \cite{Grisaru:1977kk, Lysov:2015jrs, Avery:2015iix}. Since $\sfactor_u$ is a product of two grassmann odd quantities, it is grassmann even. Two single soft factors do not commute with each other. 
\begin{eqnarray}
 \sfactor_u\, \sfactor_v\ne \sfactor_v\, \sfactor_u   
\end{eqnarray}

    \item Whenever there is more than one gravitino, they can combine pairwise to give a soft graviton which in-turn couples to the hard particles.  $\mfactor_{uv}$ encodes these type of contributions. The explicit expression for $\mfactor_{uv}$ is given by 
\begin{eqnarray}
\mfactor_{uv}&=&  \kappa^2  
\sum_{i=1}^N \frac{1}{2} \frac{\epsilon^{(u)}_\mu \slashed{p_i}\epsilon^{(v)}_\nu}{p_i \cdot (k_u+k_v)}\left[ \frac{p_i^\mu p_i^\nu}{p_i \cdot k_v } + \frac{1}{2} \frac{\eta^{\mu \nu}p_i \cdot(k_u-k_v)}{k_u\cdot k_v} + \frac{(k_v^\mu p_i^\nu - k_u^\nu p_i^\mu) }{k_u \cdot k_v}\right]
\end{eqnarray}
$\mfactor_{uv}$ is neither symmetric nor anti-symmetric in its (particle-)indices
\begin{eqnarray}
\mfactor_{uv}&\ne & \pm \mfactor_{vu}
\end{eqnarray}

    \item Since the single soft factors for gravitino do not commute, the final expression for the arbitrary number of soft gravitini depends on the choice of ordering of external soft gravitini. In section \ref{subsec:rearrangementarbitrarygravitino}, we demonstrate that any order can be obtained from any other ordering. However, our expression is not manifestly symmetric in various soft gravitini. 

    \item The first term is the product of single-soft gravitino factors. The single-soft factors appear in a particular order and the explicit form of second piece changes depending on the ordering of soft factors because two soft factors do not commute.
    \item In the second term, $\lfloor M/2\rfloor  $ denotes the greatest integer which is less than or equal to $M/2$. $A$ counts the number of pairs of gravitini giving a soft graviton. The subscripts $\{r_j, u_i,v_i\}$ can take values from $1,...,M$ and $v_i > u_i$ and $r_j$'s are also ordered with the largest $r_j$ appearing on the right. 
  
    \item The supersymmetry algebra may contain central charges. In this case, the gravitino super-multiplet contains graviphoton. In the presence of central charge,  there are additional contributions to $\mfactor_{uv}$ due to graviphoton couplings. In the presence of central charge the expression of $\mfactor_{uv}$ is modified as follows
\begin{eqnarray}
    \mfactor_{uv}\longrightarrow\tmfactor_{uv}=\mfactor_{uv}+
\frac{\kappa^2}{2}\sum_{i=1}^N       e_i\,\frac{   \epsilon^{(u)}_\mu \zfactor\epsilon^{(v)}_\nu}{p_i\cdot (k_u+k_v)} \left[\frac{p_i^\mu p_i^\nu}{p_i \cdot  k_v } + \frac{1}{2} \frac{\eta^{\mu \nu}p_i \cdot(k_u-k_v)}{k_u\cdot k_v} + \frac{(k_v^\mu p_i^\nu - k_u^\nu p_i^\mu) }{k_u \cdot k_v}\right]
\nonumber\\
\end{eqnarray}    
$e_i$ is the charge of the $i^{\textrm{th}}$  external state under symmetry generated by graviphoton. $\zfactor$ is an element of the Clifford algebra such that ${\zfactor\,_{\alpha}}^\beta$ commutes with all other element of the Clifford algebra and \footnote{ Spinor indices are raised and lowered using charge-conjugation matrix}
\begin{eqnarray}
 \zfactor_{\alpha \beta}=\zfactor_{\beta \alpha}   
\end{eqnarray}

\end{enumerate}
We checked the gauge invariance of \eqref{jrmainresult}.  

In the presence of soft graviton, we have to multiply the above expression by soft factors of the graviton. For $M_1$ soft gravitini and $M_2$ soft gravitons equation \eqref{jrmainresult} is modified as given below 
\begin{eqnarray}
     \Gamma_{N+M_1+M_2}(\{p_i\},\{k_r\}) =    \left[\prod_{j=1}^{M_2} \tsfactor_{u_j}\right] \left[\prod_{i=1}^{M_1} \sfactor_{u_i} + \sum_{A = 1}^{\lfloor M_1/2\rfloor }\prod_{i =1}^{A}\mfactor_{u_iv_i}\prod_{j=1}^{M_1 -2A}\sfactor_{r_j} \right]\Gamma_N(\{p_i\})
\end{eqnarray} 
$ \tsfactor_{u}$ is the leading soft factor for graviton. It given by 
\begin{eqnarray}
     \tsfactor_{u}&=& \kappa  \sum_{i=1}^N \left(\frac{\zeta^{(u)}_{\mu \nu}p_i^{\mu} p_i^{\nu} }{p_i\cdot k_u} \right)
\label{softgravitonfactor1} 
\end{eqnarray}
here $\zeta_{\mu \nu}$ is the polarization of soft graviton. 

The rest of the paper is organized as follows. In section \ref{sec:setup}, we derive the vertices from the 1PI effective action. Then we start with the simplest case of single soft gravitino in section \ref{sec:singlesoftgravitino}.  We show the gauge invariance of the expression. Then we compute the expression for the two soft gravitini in section \ref{sec:doublesoftgravitino}. The coupling of the gravitini to the graviton is essential to show the gauge invariance of the expression for two soft gravitini. Then we write down the expression for multiple soft gravitini in section \ref{sec:arbitrarygravitino}. We do not present any derivation of this result. We check the gauge invariance of this expression. Our conjectured result is based on the computation in section   \ref{sec:singlesoftgravitino}, section   \ref{sec:doublesoftgravitino} and appendix \ref{sec:threesoft}. In section \ref{sec:gravitinocentralterm}, we derive the contribution to the soft theorem, when the supersymmetry algebra contains central charges.  In section \ref{sec:irdivergence} we discuss infrared divergence in supergravity and we show that the soft gravitino theorem is not affected by the IR divergence. Then we present our brief conclusion and potential future directions.

\section{Set-up} 
\label{sec:setup}
We are interested in deriving the leading order soft theorem for gravitino in an arbitrary theory of supergravity.  We follow the approach of Sen \cite{Sen:2017xjn, Sen:2017nim}. As in \cite{Sen:2017xjn, Sen:2017nim} we treat finite energy gravitino and soft gravitino differently. One can always do that at the level of tree amplitudes. So, we consider 1PI effective action and replace all the derivatives by covariant derivative with respect to the soft gravitino.

Our starting point is a globally super-symmetric 1PI effective action which is invariant under some number of Majorana supersymmetries\footnote{From Coleman-Mandula theorem and HLS theorem, the maximum number of super-charge is  $32$.}. So the usual (dimension-dependent) restriction for the existence of a globally super-symmetric action applies. We promote the global supersymmetry to a local one by replacing all the derivatives with covariant derivatives. At the leading order, only the minimal coupling of gravitino with matter fields contribute. We do not assume anything about the multiplet in which matter fields are sitting. Our analysis is valid for the matter in any supersymmetry multiplet.

Let $\Phi_m$ be any quantum field which transforms under some reducible representation of the Poincare group, supersymmetry, and the internal symmetry group(s). The transform of the fields under the global supersymmetry  is given by  
\begin{eqnarray}
    \Phi_m \longrightarrow \Phi_m+ \iimg {(\theta^\alpha\, \scharge_\alpha)_{m}}^n\, \Phi_n 
\end{eqnarray}
$\scharge_\alpha$ are supersymmetry generators. They satisfy the following algebra
\begin{eqnarray}
    \Big\{\scharge_\alpha,\scharge_\beta\Big\}= -\frac{1}{2}\gamma^\mu_{\alpha \beta}P_\mu 
\label{drsusyalgebra1}    
\end{eqnarray}
Here $P_\mu$ is the momentum generator. The indices $\alpha,\beta$ are the collection of all possible spinor indices, not the indices for the minimal spinor (of that dimension). So, in a theory with more than one supersymmetry, $\scharge_\alpha$ are the collection of all the super-charges. Gamma matrices are in Majorana representation and symmetric in the spinor indices. 

Now we will  evaluate the vertex that describes the coupling of a soft gravitino to any hard particle. We start from the quadratic term of the 1PI effective action \cite{Sen:2017nim}   
\begin{eqnarray}
S = \frac{1}{2}\int \frac{d^dp_1}{(2\pi)^D}\frac{d^dp_2}{(2\pi)^D}     \Phi_m(p_1)\mathcal{K}^{mn}(p_2)\Phi_n (p_2) (2\pi)^D \delta^{(D)}(p_1+p_2)  
\label{sgravsetup2}
\end{eqnarray}
The kinetic term is invariant under global supersymmetry  transformation. This implies 
\begin{eqnarray}
\mathcal{K}^{m_1m_3}
{({\scharge_{\alpha}} )_{m_3}}^{m_2}    
+ 
\mathcal{K}^{m_3m_2}
{({\scharge_{\alpha}})_{m_3}}^{m_1}    
&=& 0 
\label{sgravsetup3}
\end{eqnarray}
Later we will need the propagator. Let us assume it has the following form:
\begin{equation}
   \Upxi (q)(q^2+M^2)^{-1}
\label{sgravsetup4}
\end{equation}
where $\Upxi(q)$ is defined as 
\begin{equation}
    \Upxi(q) = \iimg (q^2+M^2)\mathcal{K}^{-1}(q)
\label{sgravsetup5}
\end{equation}
and $M$ is some arbitrary mass parameter \footnote{We have already used $M$ for number of soft-particles. Since the mass-parameter does not appear extensively, we also $M$ for mass-parameter.} . From \eqref{sgravsetup3} we get,
\begin{eqnarray}
\Upxi^{m_1m_3}
{({\scharge_{\alpha}} )_{m_3}}^{m_2}    
+
\Upxi^{m_3m_2}
{({\scharge_{\alpha}})_{m_3}}^{m_1}    
&=& 0 
\label{sgravsetup6}
\end{eqnarray}
We write down two more relations which will be useful later 
\begin{eqnarray}
   \mathcal{K}^{m_1 m_2}(-p)\,  \Upxi_{m_2m_3}(-p)  &=& \iimg (p^2 +M^2) \, {\delta^{m_1}}_{m_3}
  \nonumber\\ 
    \frac{\partial\, \mathcal{K}^{m_1 m_2}(-p)}{\partial p_\mu}\Upxi_{m_2m_3}(-p) &=& - \mathcal{K}^{m_1 m_2}(-p)\frac{\partial\, \Upxi_{m_2m_3} (-p)}{\partial p_\mu} + 2 \iimg  p^\mu \, {\delta^{m_1}}_{m_3}
   \label{sgravsetup6.1} 
\end{eqnarray}

\subsection{Covariant derivative}
In super-gravity theories, the super-covariant derivative \cite{Ferrara:1976kg} is given by
\begin{eqnarray}
\mathcal{D}_a=     {E_a}^\mu \left(\partial_\mu  -\iimg\,  { \Psi_\mu}^\alpha\scharge_\alpha - \iimg\,  \frac{1}{2}{\omega_\mu}^{cd}\mathcal{J}_{cd}
\right)
\label{sucovariantderi1}
\end{eqnarray}   
 The local-supersymmetry transformation of the vielbein $e^{a}_\mu $ and the gravitino $ \Psi_{\mu\alpha}$ are given by 
\begin{subequations}
\begin{eqnarray}
    \delta e^{a}_\mu &=&\frac{1}{2}  \theta \gamma^a \Psi_\mu 
\label{sucovariantderi2a}
\\
\delta \Psi_{\mu\alpha}  &=& \mathcal{D}_\mu\, \theta_\alpha  =\partial_\mu \,\theta_\alpha+\frac{1}{4}\omega_{\mu ab}\gamma^{ab}\theta_\alpha     
\label{sucovariantderi2b}
\end{eqnarray}    
\end{subequations}
Here $\theta$ is the local supersymmetry parameter. We consider the covariant derivative with respect to the soft fields only.  So we consider a small fluctuation with soft momenta \cite{Sen:2017nim}
\begin{subequations}
\begin{eqnarray}
{E_a}^\mu&=& {\delta_a}^\mu-\kappa\,  {\zeta_a}^\mu e^{\iimg k\cdot x}     
\label{sucovariantderi5a}
\\
{  \Psi_\mu}^\alpha
&=&\kappa\,  { \epsilon_\mu}^\alpha e^{\iimg k\cdot x}
\label{sucovariantderi5b}
\end{eqnarray}
\end{subequations}
Here $\kappa $ is the gravitational coupling constant. At linear order in fluctuation  of the soft fields we get the following expression for the super-covariant derivative
\begin{eqnarray}
\mathcal{D}_a =     \partial_a - \kappa\,  {\zeta_a}^\mu \partial_\mu - \iimg \kappa\,  \epsilon_a^\alpha \scharge_\alpha-  \iimg \kappa\,  \frac{1}{2}{\omega_a}^{cd}({\zeta_a}^\mu)\, \mathcal{J}_{cd}
\label{sucovariantderi6}
\end{eqnarray}
 
\subsection{Vertex of one soft gravitino to matter}
\label{subsec:mattervertex}
The coupling of one soft gravitino to matter fields at linear order can be found by covariantizing the derivative in \eqref{sgravsetup2}. Due to the interaction with gravitino, the momenta of hard particle changes by $\delta q = - \kappa\,  \epsilon_{\mu}^{\alpha}\scharge_\alpha$. So the quadratic part of the 1PI effective action  changes as follows \eqref{sgravsetup2} \cite{ Sen:2017nim} :
\begin{itemize}
    \item $\delta^{(D)}(p_1+p_2)$ gets replaced by $\delta^{(D)}(p_1+p_2 +k)$ where $k$ is the momenta of soft gravitino.
    \item The change in kinetic operator $\mathcal{K}^{mn}$ due to shift in momenta has to be substituted.
\end{itemize}
So we get 
\begin{eqnarray}
S^{(L)} 
&=& \frac{1}{2}\int \frac{d^dp_1}{(2\pi)^D}\frac{d^dp_2}{(2\pi)^D}     \Phi_m(p_1)\left[- \frac{\partial\mathcal{K}(p_2)}{\partial p_{2\mu}}\kappa\epsilon_{\mu}^{\alpha}\scharge_\alpha\right]^{mn}\Phi_n (p_2) (2\pi)^D \delta^{(D)}(p_1+p_2 +k)
\qquad
\label{sgravsetup21} 
\end{eqnarray}
So the vertex is given by 
\begin{eqnarray}
- \left[\iimg\,  \kappa \, \frac{\partial\mathcal{K}(p_i)}{\partial\,  p_{i\mu}}\epsilon_{\mu}^{\alpha}\scharge_\alpha\right]^{mn} 
\label{sgravsetup22}
\end{eqnarray}  
\subsection{External particles}
\label{subsec:externalparticle}
Since we compute only $S$-matrix elements,  all the external particles satisfy on-shell and transversality condition.     The external particle of polarization $\varepsilon_{i,m}$ and momenta $p_i$ satisfies the conditions:
\begin{subequations}
\begin{eqnarray}
\varepsilon_{i,m}\mathcal{K}^{mn}(q) &=& 0 
\label{sgravsetup51}
\\
p_i^2 + M_i^2 &=& 0
\label{sgravsetup52}
\end{eqnarray}     
\end{subequations}    

\subsection{Coupling of two soft gravitini to a soft graviton }
\label{subsec:sugrathreepoint}
When we have more than one soft gravitino, we need to consider the minimal coupling of gravitino with graviton. To derive this vertex, we followed  \cite{Sen:2017nim}. The graviton coupling to any matter field can be written as :
\begin{eqnarray}
S =& &\frac{1}{2}\int\frac{d^D k_1}{(2\pi)^D} \frac{d^D k_2}{(2\pi)^D}(2\pi)^D \delta^{(D)}(k_1+k_2+p)     
\nonumber\\
&&\Phi_m (k_1) \left[ -\zeta_{\mu \nu}k_2^\nu \frac{\partial }{\partial k_{2\mu }}\mathcal{K}^{mn}(k_2)+\frac{1}{2}(p_b\zeta_{a\mu }-p_a\zeta_{b\mu } )\frac{\partial }{\partial k_{2\mu }}\mathcal{K}^{mp }(k_2){(\mathcal{J}^{ab})_p}^{n}\right]\Phi_n (k_2)
\nonumber\\
\label{sengravitino1}
\end{eqnarray}
where $\zeta_{\mu\nu}$ is the graviton polarization. 

The kinetic term for the gravitino, in the harmonic gauge, is given by 
\begin{eqnarray}
\mathcal{K}^{\mu\alpha;\nu\beta}(p)    = (p_\rho\gamma^\rho)^{\alpha\beta}\eta^{\mu\nu}
\label{sengravitino2}
\end{eqnarray}
The angular momentum generator is 
\begin{eqnarray}
{(\mathcal{J}^{ab})_{\mu,\alpha}}^{\nu,\beta}=    {(\mathcal{J}^{ab}_{_V})_{\mu}}^{\nu}\, {\delta_{\alpha}}^{\beta}+{(\mathcal{J}_{_S}^{ab})_{\alpha}}^{\beta}{\delta_\mu}^\nu
\label{sengravitino3} 
\end{eqnarray}
where $\mathcal{J}^{ab}_{_V}$ and $\mathcal{J}^{ab}_{_S}$ are angular momentum generator in vector and spinor representations respectively. 
\begin{subequations}
\begin{eqnarray}
    {(\mathcal{J}^{ab}_{_V})_{\mu}}^{\nu}&=&{\delta^{a}}_{\mu}\eta^{b\nu}-{\delta^{b}}_{\mu}\eta^{a\nu}
\label{sengravitino4a}
\\
{(\mathcal{J}_{_S}^{ab})_{\alpha}}^{\beta}&=&-\frac{1}{2}{(\gamma^{ab})_{\alpha}}^{\beta}
\qquad\qquad\qquad
\gamma^{ab}\equiv \frac{1}{2}(\gamma^a\gamma^b-\gamma^b\gamma^a)
\label{sengravitino4b}
\end{eqnarray}
\end{subequations}
Our gamma matrix convention is given in \eqref{gammaconvention1}. 
Our convention is that all the particles are incoming; the gravitino has momentum $k_1$ and $k_2$ and the graviton has momenta $p$. The momentum conservation implies 
\begin{eqnarray}
    p+k_1+k_2=0 
\label{sengravitino6.1}
\end{eqnarray}
So the vertex $(\mathcal{V}^{\mu\nu;\mu_1\mu_2} )^{\alpha \beta}$  is given by 
\begin{eqnarray}
&&    -\iimg \kappa \Big[k_2^{\mu_2} (\gamma^{\mu_1})^{\alpha\beta}\eta^{\mu\nu}
+\frac{1}{4}(p_d\,\delta_{c}^{\mu_2}-p_c\, \delta_{d}^{\mu_2 } )(\gamma^{\mu_1 }\gamma^{cd})^{\alpha\beta} \eta^{\mu\nu}
+(p^\mu\eta^{\nu\mu_2 }-p^\nu\eta^{\mu\mu_2 } )(\gamma^{\mu_1 } )^{\alpha \beta} \Big]
\label{sengravitino7}
\end{eqnarray}

\subsection{Note on Feynman diagrams}
We use a red double-arrowed line for soft-gravitino, a blue wavy line to denote soft gravitons, a violet wavy line for graviphoton, Cyan double arrowed\footnote{We use a double arrowed line for Majorana particles because they are their own anti-particle; they only have $\mathbb{Z}_2$ charge. } line for hard fermionic particles (including hard gravitini) and black line to denote hard bosonic particles. 

\begin{figure}[h]
\begin{center}
\begin{tikzpicture}[line width=1.5 pt, scale=1]

\begin{scope}[shift={(6,0)}]

\draw[realscalar] (-1,0)--(1,0); 
\node at (3.5,0) {Hard bosonic particle}; 

\draw[majorana] (-1,-.75)--(1,-.75); 
\node at (3.5,-.75) {Hard fermionic particle}; 
 
\draw[gravitino] (-1,-1.5)--(1,-1.5); 
\node at (3.5,-1.5) {Soft gravitino}; 

\draw[graviton] (-1,-2.25)--(1,-2.25); 
\node at (3.5,-2.25) {Soft graviton}; 

\draw[photon] (-1,-3)--(1,-3); 
\node at (3.5,-3) {Soft gravi-photon}; 
 
\end{scope}

\end{tikzpicture}
\end{center}
\caption{Conventions for Feynman diagrams} 
\label{fig:feyndiagconv}
\end{figure}
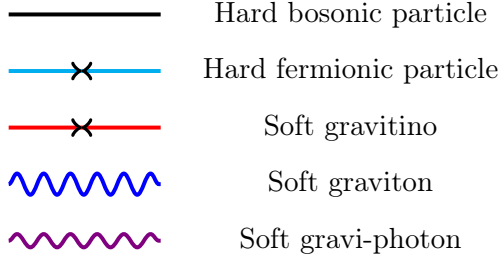

\newpage
\section{Single soft gravitino} 
\label{sec:singlesoftgravitino}
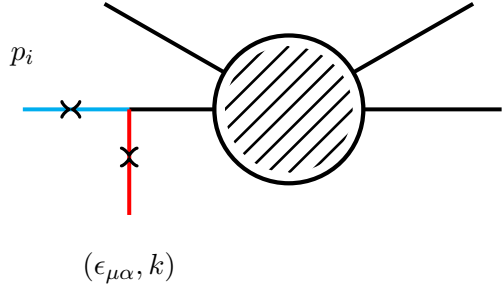
\begin{figure}[h]
\begin{center}
\begin{tikzpicture}[line width=1.5 pt, scale=1.4]

\begin{scope}[shift={(6,0)}]

\draw[realscalar] (.7,0)--(2,0);

\draw[realscalar][rotate=30] (.7,0)--(2,0);

\draw[realscalar][rotate=180] (.7,0)--(1.5,0);

\draw[majorana][rotate=180] (2.5,0)--(1.5,0);

\draw[realscalar][rotate=150] (.7,0)--(2,0);

\draw[gravitino][rotate=180] (1.5,0)--(1.5,1);

\node at (-1.5,-1.5) {$(\epsilon_{\mu \alpha}, k)$};

\node at (-2.5,.5) {$p_i$}; 

\begin{scope}[shift={(0,0)}, scale=2]
    \draw [ultra thick] (0,0) circle (.35);
    \clip (0,0) circle (.3cm);
    \foreach \x in {-.9,-.8,...,.3}
    \draw[line width=1 pt] (\x,-.3) -- (\x+.6,.3);
\end{scope}
 
\end{scope}

\end{tikzpicture}
\end{center}
\caption{Feynman diagram for single soft gravitino} 
\label{fig:onegravitino}
\end{figure}

In this section, we compute the leading order contribution to soft gravitino theorem due to one soft gravitino.  This result for $D=4$ was first derived in \cite{Grisaru:1977kk} and was reproduced from the analysis of asymptotic symmetries in \cite{Avery:2015iix, Lysov:2015jrs}. The only diagram that contributes to this process is depicted in figure \ref{fig:onegravitino}.

The expression for the propagator is given in equation \eqref{sgravsetup4}.  In this diagram, the propagator carries momenta $p_i + k$ and $M_i$ is the mass of the $i$-th particle. Let us denote the corresponding propagator by $\Upxi_{m_i n_i}(p_i+k)$.  
The contribution to figure \ref{fig:onegravitino} is given by:
\begin{eqnarray} 
\Gamma_{N+1}^{m_1...m_N}(\{p_i\},k) &=&\left[ \iimg \kappa\sum_{i=1}^N \left(\frac{\partial\mathcal{K}(-p_i)}{\partial p_{i\mu} }\epsilon_{\mu}^{\alpha}\, \scharge_\alpha \right)^{m_in_i}\frac{\Upxi_{n_i \tilde n_i  }}{(p_i + k)^2 + M_i^2} \right] \Gamma_{N}^{m_1...m_{i-1}\tilde n_i m_{i+1}...m_N}(\{p_i\})
\nonumber\\
&=&\left[ \iimg \kappa\sum_{i=1}^N \left(\frac{\partial\mathcal{K}(-p_i)}{\partial p_{i\mu} }\epsilon_{\mu}^{\alpha}\, \scharge_\alpha \right)^{m_in_i}\frac{\Upxi_{n_i \tilde n_i  }}{(2\, p_i \cdot k)} \right] \Gamma_{N}^{m_1...m_{i-1}\tilde n_i m_{i+1}...m_N}(\{p_i\})
\label{singlesoftgrav1}
\end{eqnarray}
where in the second step, we have used the on-shell condition \eqref{sgravsetup52} for external hard particle and the fact that gravitino is soft. Now we will use \eqref{sgravsetup6} and \eqref{sgravsetup6.1} to simplify the expression
\begin{eqnarray}
\left(\frac{\partial\mathcal{K}(-p_i)}{\partial p_{i\mu}} \epsilon_{\mu}^{\alpha}\, \scharge_\alpha \right)^{m_in_i}\Upxi_{n_i \tilde n_i  } 
&=&
\epsilon_{\mu}^{\alpha}\left(\frac{\partial\mathcal{K}(-p_i)}{\partial p_{i\mu}}\scharge_\alpha\,  \Upxi \right)^{m_i}{}_{ \tilde n_i  } 
=
- \epsilon_{\mu}^{\alpha}\left(\frac{\partial \mathcal{K}(-p_i)}{\partial p_{i\mu}}  \Upxi\,   \scharge_\alpha\right)^{m_i}{}_{ \tilde n_i  } 
\nonumber\\
&=&
- \epsilon_{\mu}^{\alpha}\left(-\mathcal{K}(-p_i)\frac{\partial\, \Upxi }{\partial p_{i\mu}}\,    \scharge_\alpha+ 2\, \iimg\,  p_i^\mu\,  \scharge_\alpha\right)^{m_i}{}_{ \tilde n_i  } 
\label{singlesoftgrav2}
\end{eqnarray}
From first step to second step we have used \eqref{sgravsetup6} and from second step to third step we have used \eqref{sgravsetup6.1}. Now the first term drops out because of the on-shell condition \eqref{sgravsetup51}. Hence we obtain \cite{Grisaru:1977kk} 
\begin{eqnarray}
\Gamma_{N+1}^{m_1...m_N}(\{p_i\},k) &=&  \left[ \kappa  \sum_{i=1}^N  \left(  \frac{p_i^{\mu}\,\epsilon_{\mu}^{\alpha}}{p_i\cdot k} \scharge_\alpha \right)^{m_i}{}_{ \tilde n_i  }\, \right]  \Gamma_{N}^{m_1...m_{i-1}\tilde n_i m_{i+1}...m_N}(\{p_i\}) 
\label{singlesoftgrav3} 
\end{eqnarray}
\paragraph{Soft operator}
We define the soft operator $\sfactor_u$ \cite{Grisaru:1977kk} as
\begin{eqnarray}
    \sfactor_u =   \kappa  \sum_{i=1}^N  \left(  \frac{p_i^{\mu} \epsilon^{(u)\, \alpha}_{\mu}}{p_i\cdot k_u} \scharge_\alpha \right)
\label{singlesoftgrav11} 
\end{eqnarray}
where $u$ labels the soft gravitino. So the above result can be re-written as: 
\begin{eqnarray}
\Gamma_{N+1}^{m_1...m_N}(\{p_i\},k) =  \Big[   \,\, \sfactor^{\, m_i}{}_{ \tilde n_i  }\, \Big]  \Gamma_{N}^{m_1...m_{i-1}\tilde n_i m_{i+1}...m_N}(\{p_i\})
\label{singlesoftgrav4} 
\end{eqnarray}
\subsection{Gauge invariance}
\label{subsec:singlesoftgaugeinvariance}
As a consistency check, we check the gauge invariance of equation \eqref{singlesoftgrav3}. We put pure gauge polarization for the gravitino 
\begin{eqnarray}
   \epsilon_{\alpha \mu }=k_\mu\,  \theta_\alpha 
\end{eqnarray}
Here $\theta_\alpha$ is a Majorana spinor. For pure gauge gravitino the amplitude should vanish. From \eqref{singlesoftgrav3}, we obtain 
\begin{eqnarray} 
 \theta^\alpha\,  \sum_{i=1}^N\left(   \scharge_\alpha \right)^{m_i}{}_{ \tilde n_i  } \Gamma_{N}^{m_1...m_{i-1}\tilde n_i m_{i+1}...m_N}(p_i)=0
\label{singlesoftgrav41} 
\end{eqnarray}
This is the ward-identity for the global super-symmetry.

\section{Two soft gravitini}  
\label{sec:doublesoftgravitino}

\begin{figure}[h]
\begin{center}
\begin{tikzpicture}[line width=1.5 pt, scale=.7]

\begin{scope}[shift={(0,0)}, scale=2]

\draw[realscalar] (.7,0)--(2,0);

\draw[realscalar][rotate=30] (.7,0)--(2,0);

\draw[realscalar][rotate=180] (1.5,0)--(2.1,0);

\draw[majoranabar][rotate=180] (.7,0)--(1.5,0);

\draw[realscalar][rotate=0] (1.5,0)--(2.1,0);

\draw[majorana][rotate=0] (.7,0)--(1.5,0);

\draw[realscalar][rotate=150] (.7,0)--(2,0);

\draw[gravitino][rotate=180] (1.5,0)--(1.5,1);

\draw[gravitinobar][rotate=0] (1.5,0)--(1.5,-1);

\node at (-1.5,-1.5) {$(\epsilon^{(1)}_{\mu \alpha}, k_1)$};
\node at (1.5,-1.5) {$(\epsilon^{(2)}_{\nu \beta}, k_2)$};

\node at (-2,.35) {$p_i$}; 
\node at (2,.35) {$p_j$}; 


\begin{scope}[shift={(0,0)}, scale=2]
    \draw [ultra thick] (0,0) circle (.35);
    \clip (0,0) circle (.3cm);
    \foreach \x in {-.9,-.8,...,.3}
    \draw[line width=1 pt] (\x,-.3) -- (\x+.6,.3);
\end{scope}
 
\end{scope}

\end{tikzpicture}
\end{center}
\caption{Feynman diagram for double soft gravitini - I}
\label{fig:twogravitino0}
\end{figure}
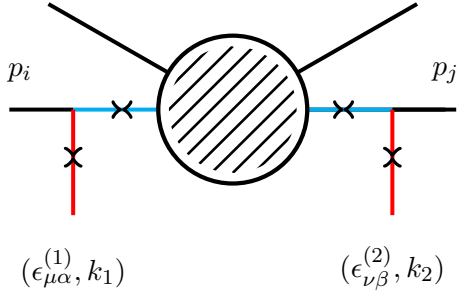

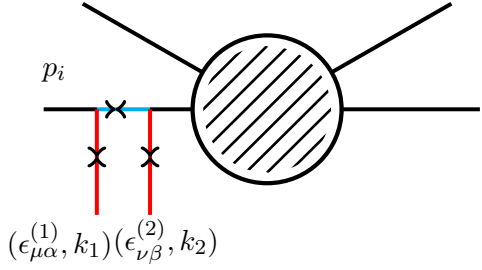
\begin{figure}[h] 
\begin{center}
\begin{tikzpicture}[line width=1.5 pt, scale=.7]

\begin{scope}[shift={(0,0)}, scale=2]

\draw[realscalar] (.7,0)--(2,0);

\draw[realscalar][rotate=30] (.7,0)--(2,0);

\draw[realscalar][rotate=180] (.7,0)--(1.1,0);

\draw[majorana][rotate=180] (1.6,0)--(1.1,0);

\draw[realscalar][rotate=180] (1.6,0)--(2.1,0);

\draw[realscalar][rotate=150] (.7,0)--(2,0);

\draw[gravitino][rotate=180] (1.1,0)--(1.1,1);

\draw[gravitinobar][rotate=180] (1.6,0)--(1.6,1);
\node at (-1.95,-1.25) {$(\epsilon^{(1)}_{\mu \alpha}, k_1)$};
\node at (-0.95,-1.25) {$(\epsilon^{(2)}_{\nu \beta}, k_2)$};

\node at (-2,.35) {$p_i$};


\begin{scope}[shift={(0,0)}, scale=2]
    \draw [ultra thick] (0,0) circle (.35);
    \clip (0,0) circle (.3cm);
    \foreach \x in {-.9,-.8,...,.3}
    \draw[line width=1 pt] (\x,-.3) -- (\x+.6,.3);
\end{scope}
 
\end{scope}

\end{tikzpicture}
\end{center}
\caption{Feynman diagram for double soft gravitini - II}
\label{fig:twogravitinoI}
\end{figure}

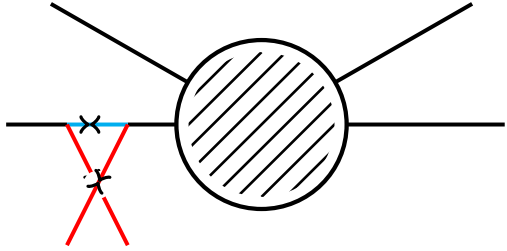
\begin{figure}[h]
\begin{center}
\begin{tikzpicture}[line width=1.5 pt, scale=.8]

\begin{scope}[shift={(0,-6)}, scale=2]

\draw[realscalar] (.7,0)--(2,0);

\draw[realscalar][rotate=30] (.7,0)--(2,0);

\draw[realscalar][rotate=180] (.7,0)--(1.1,0);

\draw[majorana][rotate=180] (1.6,0)--(1.1,0);

\draw[realscalar][rotate=180] (1.6,0)--(2.1,0);

\draw[realscalar][rotate=150] (.7,0)--(2,0);

\draw[gravitinobar][rotate=180] (1.6,0)--(1.1,1);

\filldraw [white] (-1.35,-.5) circle (.1);

\draw[gravitino][rotate=180] (1.1,0)--(1.6,1);


\begin{scope}[shift={(0,0)}, scale=2]
    \draw [ultra thick] (0,0) circle (.35);
    \clip (0,0) circle (.3cm);
    \foreach \x in {-.9,-.8,...,.3}
    \draw[line width=1 pt] (\x,-.3) -- (\x+.6,.3);
\end{scope}
 
\end{scope}

\end{tikzpicture}
\end{center}
\caption{Feynman diagram for double soft gravitini - III}
\label{fig:twogravitinoII}
\end{figure}

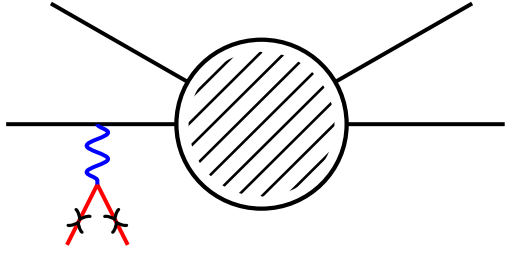
\begin{figure}[h]
\begin{center}
\begin{tikzpicture}[line width=1.5 pt, scale=.8]

\begin{scope}[shift={(0,-12)}, scale=2]

\draw[realscalar] (.7,0)--(2,0);

\draw[realscalar][rotate=30] (.7,0)--(2,0);

\draw[realscalar][rotate=180] (.7,0)--(2.1,0);

\draw[realscalar][rotate=150] (.7,0)--(2,0);

\draw[gravitinobar][rotate=180] (1.6,1)--(1.35,.5);


\draw[gravitino][rotate=180] (1.1,1)--(1.35,.5);

\draw[graviton][rotate=180] (1.35,0)--(1.35,.5);


\begin{scope}[shift={(0,0)}, scale=2]
    \draw [ultra thick] (0,0) circle (.35);
    \clip (0,0) circle (.3cm);
    \foreach \x in {-.9,-.8,...,.3}
    \draw[line width=1 pt] (\x,-.3) -- (\x+.6,.3);
\end{scope}
 
\end{scope}

\end{tikzpicture}
\end{center}
\caption{Feynman diagram for double soft gravitini - IV}
\label{fig:twogravitinoIII}
\end{figure}

Now we will consider the amplitude with $N$ hard particles and $2$ soft gravitini. There are essentially four different types of Feynman diagrams which contribute to this process 
\begin{enumerate}
    \item The class of diagrams where the two soft gravitini are attached to different external legs (for example, figure \ref{fig:twogravitino0}). These diagrams are easy to evaluate. The computation for these type of diagrams is essentially the same as single soft gravitino. The contribution from figure \ref{fig:twogravitino0} is given by 
\begin{eqnarray}
      \kappa^2 \sum_{i=1}^N   \frac{ \epsilon ^{(1);\, \alpha}_{\mu}p_i^{\mu}}{p_i\cdot k_1} \scharge_\alpha   \sum_{j=1;j\ne i}^N \frac{ \epsilon^{(2);\, \beta}_{\nu}p_i^{\nu}}{p_j \cdot  k_2} \scharge_\beta \,  \Gamma(\{p_i\})
\label{doublesoftgravi1}
\end{eqnarray}

    \item The class of diagrams where both of the soft gravitini are attached to the same external leg. There are mainly three types of such diagrams - (figure \ref{fig:twogravitinoI}, figure \ref{fig:twogravitinoII}, figure \ref{fig:twogravitinoIII}). Figure \ref{fig:twogravitinoI}, Figure \ref{fig:twogravitinoII} denote the diagrams where the soft gravitino directly attaches the same hard-particles. These two diagrams differ only in order of attaching to the hard particle. Figure \ref{fig:twogravitinoIII} captures the process when the soft gravitini combine to give a soft graviton and then the soft graviton attaches to the hard particles. 
    
\end{enumerate}
Now will evaluate these diagrams 

\paragraph{Evaluation of figure \ref{fig:twogravitinoI} } The contribution from the Feynman diagram in figure \ref{fig:twogravitinoI} is given by 
\begin{eqnarray}
\Gamma_{N+2}^{(1)} = \kappa^2 \sum_{i=1}^N \frac{\partial\mathcal{K}^{mp}(-p_i)}{\partial p_{i\mu}}\frac{[\epsilon^{(1);\alpha}_{\mu}  \scharge_\alpha \Upxi(-p_i - k_1)]_{pq}}{(2p_i\cdot k_1)} \frac{\partial\mathcal{K}^{qr}(-p_i-k_1)}{\partial p_{i\nu} }\frac{[\epsilon^{(2); \beta}_{\nu} \scharge_\beta\,  \Upxi(-p_i - k_1-k_2)]_{rs}}{(2p_i\cdot (k_1+k_2))}  \Gamma_{N}(\{p_i\})
\nonumber\\
\label{doublesoftgravi11}
\end{eqnarray}
Using \eqref{sgravsetup6} and \eqref{sgravsetup6.1} we can simplify this expression and we get 
\begin{eqnarray} 
  \kappa^2 \sum_{i=1}^N  \frac{ \epsilon ^{(1);\alpha}_{\mu}p_i^{\mu}}{p_i\cdot k_1} \frac{ \epsilon^{(2);\beta}_{\nu}p_i^{\nu}}{p_i\cdot (k_1+k_2)} \scharge_\alpha\, \scharge_\beta\, \Gamma_{N}(\{p_i\})
\label{doublesoftgravi12}
\end{eqnarray}

\paragraph{Evaluation of figure \ref{fig:twogravitinoII} } The contribution due to figure \ref{fig:twogravitinoII}  can obtained from equation \eqref{doublesoftgravi11} by interchanging $1\longleftrightarrow 2$
\begin{eqnarray}
\Gamma_{N+2}^{(2)}(\{p_i\},k_1,k_2) &=& \kappa^2 \sum_{i=1}^N  \frac{ \epsilon ^{(2); \beta}_{\nu}p_i^{\nu}}{p_i\cdot k_2} \frac{ \epsilon^{(1); \alpha}_{\mu}p_i^{\mu}}{p_i\cdot (k_1+k_2)} \scharge_\beta \scharge_\alpha\,  \Gamma_{N}(\{p_i\})\nonumber\\
&=& \kappa^2 \sum_{i=1}^N  \frac{\epsilon^{(1); \alpha}_{\mu}p_i^{\mu}}{p_i\cdot k_2} \frac{  \epsilon ^{(2); \beta}_{\nu}p_i^{\nu}}{p_i\cdot (k_1+k_2)} \left[\scharge_\alpha\scharge_\beta + \frac{1}{2} (\slashed{p}_i)_{\alpha\beta}\right] \Gamma_{N}(\{p_i\})
\label{doublesoftgravi21}
\end{eqnarray}

\paragraph{Evaluation of figure \ref{fig:twogravitinoIII} } Now we would like to evaluate the figure \ref{fig:twogravitinoIII}. This diagram denotes the process when two soft gravitini interact first to produce a soft graviton which then attaches to any of the external legs.  The contributions from these kinds of processes are given by  
\begin{eqnarray}
  \Gamma_{N+2}^{(3)} (\{p_i\},k_1,k_2)
    &=& - \left[  \epsilon^{(1)\,  \alpha}_{\mu}(\mathcal{V}^{\mu\nu;\mu_1\mu_2} )_{\alpha \beta}\,   \epsilon^{(2)\, \beta}_{\nu}  \right]
    \left[\left(\frac{\iimg }{2}\right)\frac{\eta_{\mu_1\nu_1 }\eta_{\mu_2\nu_2 }+ \eta_{\mu_1\nu_2 }\eta_{\mu_2\nu_1 }- \frac{2}{D-2}\eta_{\mu_1\mu_2 }\eta_{\nu_1\nu_2 }}{2(k_1\cdot k_2)}\right]  
\nonumber\\
&&
     \left[-\iimg \kappa\,  {p_i^{\nu_1}}\frac{\partial\mathcal{K}}{\partial p_{i\nu_2}} \frac{\Upxi}{2 p_i\cdot (k_1+k_2)}\right] \Gamma_{N}(\{p_i\})
\end{eqnarray} 
where the first square bracket denotes gravitino-gravitino-graviton vertex, the second one is the graviton propagator and the third one is the matter- soft graviton -matter vertex.
Using the explicit expression for $(\mathcal{V}^{\mu\nu;\mu_1\mu_2} )_{\alpha \beta}$ from \eqref{sengravitino7} and simplifying the above expression, we get
\begin{eqnarray}  
  \Gamma_{N+2}^{(3)} (\{p_i\},k_1,k_2)
    &=&  \frac{ \kappa^2}{2} \sum_{i=1}^N \epsilon^{(1)\alpha}_{\mu}  (  \slashed{p}_{i})_{\alpha \beta}\left[- \eta^{\mu\nu} p_i\cdot k_2 - \frac{1}{2}\eta^{\mu \nu} (k_1+k_2)_d\,   p_{ie}\gamma^{de}+ (k_2^\mu p_i^\nu -k_1^\nu p_i^\mu) \right] \epsilon^{(2)\beta}_{\nu}
\nonumber\\    
 &&   \left[\frac{1}{( p_i\cdot (k_2 + k_1))(k_1\cdot k_2)}\right]\Gamma_{N}(\{p_i\})
\label{doublesoftgravi34}
\end{eqnarray}
After simplifying the second term and using gamma-traceless condition for gravitino, we get
\begin{eqnarray}  
 \Gamma_{N+2}^{(3)}
    &=&    \kappa^2 \left[ \sum_{i=1}^N\cfactor_{12}(p_i)
 \frac{1}{( p_i\cdot (k_2 + k_1))}\right]\Gamma_{N}(\{p_i\})
\label{doublesoftgravi36}
\end{eqnarray} 
here we have introduced  $\cfactor_{12}$ where  $\cfactor_{uv}(p_i)$ is defined as follows
\begin{eqnarray}
    \cfactor_{uv}(p_i)= \frac{1}{2}  \epsilon^{(u)}_\mu \slashed{p_i}\epsilon^{(v)}_\nu \left[ \frac{1}{2} \frac{\eta^{\mu \nu}p_i \cdot(k_u-k_v)}{k_u\cdot k_v} + \frac{(k_v^\mu p_i^\nu - k_u^\nu p_i^\mu) }{k_u \cdot k_v}\right]    
\label{doublesoftgravi37}
\end{eqnarray} 
From the property of the gamma matrices it follows that $\cfactor_{uv}(p_i)$ is symmetric in its particle indices 
\begin{eqnarray}
\epsilon^{(u)}_\mu \slashed{p_i}\epsilon^{(v)}_\nu=-\epsilon^{(v)}_\mu \slashed{p_i}\epsilon^{(u)}_\nu
\implies
    \cfactor_{uv}(p_i)=\cfactor_{vu}(p_i)    
\label{doublesoftgravi38}
\end{eqnarray}

\paragraph{Total contribution}
Now we add the contributions from  \eqref{doublesoftgravi1}, \eqref{doublesoftgravi12}, \eqref{doublesoftgravi21} and \eqref{doublesoftgravi36} to get the full answer for two soft gravitini. The total contribution can be written as 
\begin{eqnarray}
\Gamma_{N+2}(\{p_i\},k_1,k_2) = \Big[ \sfactor_1\, \sfactor_2 +\mfactor_{12}\Big] \Gamma_N(\{p_i\})
\label{doublesoftgravi61}
\end{eqnarray}
we have already defined $\sfactor_u$ in \eqref{singlesoftgrav11}. $\mfactor_{uv}$ is defined as follows
\begin{eqnarray}
\mfactor_{uv}&=& \kappa^2
\sum_{i=1}^N \frac{1}{2} \frac{\epsilon^{(u)}_\mu \slashed{p_i}\epsilon^{(v)}_\nu}{p_i \cdot (k_u+k_v)}\left[ \frac{p_i^\mu p_i^\nu}{p_i \cdot k_v } + \frac{1}{2} \frac{\eta^{\mu \nu}p_i \cdot(k_u-k_v)}{k_u\cdot k_v} + \frac{(k_v^\mu p_i^\nu - k_u^\nu p_i^\mu) }{k_u \cdot k_v}\right]     
\label{doublesoftgravi62}
\end{eqnarray}

\paragraph{Some properties of $\sfactor_u$ and $\mfactor_{uv}$ }
\begin{itemize}
\item Two soft operators do not commute 
\begin{eqnarray}
\left[\sfactor_u,\sfactor_v    \right]    = -\frac{\kappa^2}{2} \sum_{i=1}^N  \left[ \left(\epsilon_{\mu}^{(u);\alpha}\slashed{p}_{i; \alpha \beta}  \epsilon_{\nu}^{(v);\beta}\right)  \frac{p_i^\mu  }{p_i\cdot k_u}\frac{p_i^\nu  }{p_i\cdot k_v}   \right] 
\label{singlesoftgrav12}  
\end{eqnarray}
\item While writing the result for two soft gravitini, we could have chosen the other ordering of soft factors but both results should match i.e.
\begin{eqnarray}
\sfactor_u\, \sfactor_v + \mfactor_{uv}    = \sfactor_v\, \sfactor_u+ \mfactor_{vu}    \label{doublesoftgravi65}
\end{eqnarray}
Above equation can be explicitly verified by noting that:
\begin{eqnarray}
    \mfactor_{vu}-\mfactor_{uv} &=& \kappa^2\sum_{i=1}^N \frac{\epsilon^{(u)} \cdot p_i} {p_i \cdot k_u} \frac{\epsilon^{(v)} \cdot p_i} {p_i \cdot k_v}\left(-\frac{1}{2}\slashed{p_i}\right)
\label{doublesoftgravi66}
\end{eqnarray}
We already computed $\sfactor_u\, \sfactor_v-\sfactor_v\, \sfactor_u$ in \eqref{singlesoftgrav12}.  Hence \eqref{doublesoftgravi65} is satisfied.

\item Three soft operators satisfy Jacobi identity. 
\begin{eqnarray}
    [\sfactor_u,[\sfactor_v,\sfactor_w]]+    [\sfactor_v,[\sfactor_w,\sfactor_u]]+    [\sfactor_w,[\sfactor_u,\sfactor_v]]=0 
\label{softfactorcommutator1}
\end{eqnarray}
In this particular case, each term in the above equation is individually zero. 
\begin{eqnarray}
\left[ \sfactor_u , 
\left[\sfactor_v,\sfactor_w    \right]\right]    = 0
\label{softfactorcommutator2}
\end{eqnarray}
This is not true for soft gluon operator(s). Though \eqref{softfactorcommutator1} is true for soft gluon operator, \eqref{softfactorcommutator2} does not hold for soft gluon operator.  This fact makes the computation of the soft factors for multiple soft gluon even more cumbersome. 
\item Some more properties of $\mfactor_{uv}$ are listed below 
\begin{subequations} 
\begin{eqnarray}
\mfactor_{uv}&\ne&\pm  \mfactor_{vu}    
\label{doublesoftgravi63a}
\\
\mfactor_{u_1v_1}\, \mfactor_{u_2v_2}&=&\mfactor_{u_2v_2}\,     \mfactor_{u_1v_1}
\label{doublesoftgravi63b}
\\
\sfactor_w\, \mfactor_{uv}
&=&\mfactor_{uv}\, \sfactor_w
\label{doublesoftgravi63c}
\end{eqnarray}
\end{subequations}
\end{itemize}

\subsection{Gauge invariance }
\label{subsec:doublesoftgaugeinvariance}
As a consistency check, we check the gauge invariance of the result obtained in \eqref{doublesoftgravi61}. The right-hand side should vanish when one puts any of the gravitini as a pure gauge. Here we will put $\epsilon^{(2)}$ as a pure gauge and check if RHS vanishes or not.
\begin{eqnarray}
    \epsilon^{(2)\alpha}_{\mu} = k_{2\mu}\, \theta_{2}^{\alpha}  
\label{doublesoftgravi72}
\end{eqnarray}
So for pure gauge, the first term in \eqref{doublesoftgravi61} vanishes because $\scharge_\beta$ directly hits $\Gamma_N(\{p_i\})$ and gives zero due to supersymmetry ward-identity \eqref{singlesoftgrav41}. The second piece gives:
\begin{eqnarray}
      \mfactor_{12} (\epsilon_1^{\mu\alpha}, k_2^\mu \theta^\alpha_2)&=&  \kappa^2 \sum_{i=1}^N\frac{1}{2} \frac{\epsilon^{(1)}_\mu \slashed{p_i}\theta^{(2)} }{p_i \cdot (k_1+k_2)}\left[ \frac{p_i^\mu p_i\cdot k_2}{p_i \cdot  k_2 } + \frac{1}{2} \frac{k_2^\mu p_i \cdot(k_1-k_2)}{k_1\cdot k_2} + \frac{(k_2^\mu (k_2\cdot p_i)  - k_2\cdot k_1  p_i^\mu) }{k_1 \cdot k_2}\right]  
\nonumber\\
&=&  \kappa^2 \sum_{i=1}^N\frac{1}{2} \epsilon^{(1)}_\mu \slashed{p_i}\theta^{(2)} \left[ \frac{1}{2} \frac{k_2^\mu}{k_1\cdot k_2}  \right]
=0
\label{doublesoftgravi73}
\end{eqnarray} 
where in the last step we have used momentum conservation $\sum_{i=1}^N p_i = 0$. \\
One should be able to show the gauge invariance when $\epsilon^{(1)}$ is pure gauge. But in this case, first term in \eqref{doublesoftgravi61} does not give ward-identity directly and also $\mfactor_{12}$ term does not vanish. But one can check that the sum is gauge invariant. Alternative we can use  \eqref{doublesoftgravi65} to express the amplitude in the other ordering of soft factors 
\begin{eqnarray}
\Gamma_{N+2}(\{p_i\},k_1,k_2) = \Big[ \sfactor_2\,  \sfactor_1 +\mfactor_{21}\Big] \Gamma_N(\{p_i\})
\end{eqnarray}
In this representation, it is obvious that the RHS vanishes for pure-gauge $\epsilon^{(1)}$. In general,
\begin{subequations}
\begin{eqnarray}
      \mfactor_{uv} ( \epsilon_u^{\mu\, \alpha}, k_v^\mu \theta^\alpha_v ) &=& 0
\label{doublesoftgravi173a}
      \\
      \mfactor_{uv} ( k_u^\mu\,  \theta^\alpha_u, \epsilon_v^{\mu\alpha})&\ne& 0
\label{doublesoftgravi173b}
\end{eqnarray}
\end{subequations}
To express the result for an arbitrary number of soft gravitini, we always choose an ordering amongst the external gravitini. The gauge invariance would be manifest when one puts the last gravitino as pure gauge. Using relations of the form \eqref{doublesoftgravi65}, one can check the gauge invariance for pure gauge configuration of any soft particle.

At this point, we would like to emphasize that the combined contribution from figure \ref{fig:twogravitinoI} and \ref{fig:twogravitinoII} is not gauge-invariant. Only after adding the contribution from figure \ref{fig:twogravitinoIII} the answer becomes gauge invariant. A different way to state the same result is that the existence of massless spin $3/2$ particles which interact with other fields at low momenta requires an interacting massless spin $2$ particle at low energy. This point was first elucidated in \cite{Grisaru:1977kk}.  

\paragraph{Symmetrized form the amplitude}
The expression for the soft factor in \eqref{doublesoftgravi61} is not manifestly symmetric on the gravitini. That form was useful to prove gauge invariance. Now we use \eqref{softfactorcommutator1} and \eqref{softfactorcommutator2} to  write the answer in a form which is manifestly symmetric on the gravitini    
\begin{eqnarray}
&&\Gamma_{N+2}(\{p_i\},k_1,k_2)
\nonumber\\
&=& \frac{1}{2}\Big[ \sfactor_1\,  \sfactor_2+\sfactor_2\,  \sfactor_1 +\mfactor_{12}+\mfactor_{21}\Big] \Gamma_N(\{p_i\})
\\
&=&
\left[\frac{1}{2}( \sfactor_1\,  \sfactor_2+\sfactor_2\,  \sfactor_1) + \kappa^2 \sum_{i=1}^N   \frac{1}{p_i \cdot (k_1+k_2)}\left[ \cfactor_{12}(p_i) 
+ \frac{1}{4}  (p_i\cdot  \epsilon^{(1)}) \slashed{p_i}(\epsilon^{(2)}\cdot  p_i)
\frac{p_i\cdot (k_1-k_2) }{(p_i \cdot k_2)(p_i \cdot k_1) }
\right]
\right] \Gamma_N(\{p_i\})
\nonumber
\end{eqnarray}
Apart from the last term, other terms are clearly symmetric under the exchange $1\longleftrightarrow 2$.

\subsection{Simultaneous and consecutive soft limit }

When there are more than one soft particles, there are various ways in which one can take the soft limit. \footnote{We are thankful to the unknown referee for pointing out this issue.}  Consider the amplitude with $N$ hard particles with momenta $\{p_i\}$ and two soft particles with momenta $k_1$ and $k_2$ ($    \Gamma_{N+2}(\{p_i\},k_1,k_2)$). 
Then the consecutive soft limit is defined as the limit in which the momenta are taken to be soft one after another. So for two soft particles, this can be done in two different ways 
\begin{eqnarray}
    \lim_{k_1\rightarrow 0 }\lim_{k_2\rightarrow 0 }\Gamma_{N+2}(\{p_i\},k_1,k_2)
\qquad,\qquad
    \lim_{k_2\rightarrow 0 }\lim_{k_1\rightarrow 0 }\Gamma_{N+2}(\{p_i\},k_1,k_2)
\end{eqnarray}
Alternatively, one can take simultaneous limit where one takes both $k_1$ and $k_2$ to zero keeping $k_1/k_2$ fixed 
\begin{eqnarray}
    \lim_{k_1,\,  k_2\rightarrow 0 }\Gamma_{N+2}(\{p_i\},k_1,k_2)
\end{eqnarray}
In this paper we have focused on the simultaneous limit.  If the single soft factors mutually commute (i. e. if the generators of the gauge symmetry commute) then the simultaneous limit is the same as the consecutive limit. For example, in the case of photon these two limits give the same answer. However, if the symmetry generators do not commute then these two limits differ. In our case, the supersymmetry generators do not commute.
For example, if we take the consecutive limit by taking $k_1$ to be soft first then the Feynman diagram in figure \ref{fig:twogravitinoII} does not contribute because the soft particle (with momentum $k_1$) in figure \ref{fig:twogravitinoII} is being emitted from an internal line. Hence the total contribution in this case is different from the case when we take simultaneous soft limit.

\section{Arbitrary number of soft gravitini}
\label{sec:arbitrarygravitino}
Now we consider the amplitude with an arbitrary number of soft gravitini.  In this case, the following type of diagrams can contribute:
\begin{itemize}
    \item When some soft gravitini attach on one external leg and some on another external leg(s), but none of them form pairs to give soft graviton as shown in figure \ref{fig:multiplegravitinoI}.

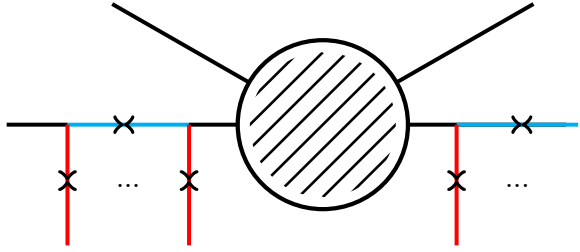
\begin{figure}[h]
\begin{center}
\begin{tikzpicture}[line width=1.5 pt, scale=.8]

\begin{scope}[shift={(0,0)}, scale=2]

\draw[realscalar] (.7,0)--(2,0);

\draw[realscalar][rotate=30] (.7,0)--(2,0);

\draw[realscalar][rotate=180] (.7,0)--(1.1,0);

\draw[majorana][rotate=180] (2.1,0)--(1.1,0);
\draw[majorana][rotate= 0] (2.1,0)--(1.1,0);
\draw[realscalar][rotate=180] (2.1,0)--(2.6,0);
 
\draw[realscalar][rotate=150] (.7,0)--(2,0);

\draw[gravitino][rotate=180] (1.1,0)--(1.1,1);
\draw[gravitino][rotate=0] (1.1,0)--(1.1,-1);
\node at (-1.6,-.5) {$...$};
\node at (1.6,-.5) {$...$};
\draw[gravitino][rotate=180] (2.1,0)--(2.1,1);


\begin{scope}[shift={(0,0)}, scale=2]
    \draw [ultra thick] (0,0) circle (.35);
    \clip (0,0) circle (.3cm);
    \foreach \x in {-.9,-.8,...,.3}
    \draw[line width=1 pt] (\x,-.3) -- (\x+.6,.3);
\end{scope}
 
\end{scope}

\end{tikzpicture} 
\end{center}
\caption{Feynman diagram for multiple soft gravitini - I}
\label{fig:multiplegravitinoI}
\end{figure} 
\item When some soft gravitini attach on one external leg and some on another external leg(s) and some form pairs to give soft graviton as shown in figure \ref{fig:multiplegravitinoII}.

\begin{figure}[h]
\begin{center}
\begin{tikzpicture}[line width=1.5 pt, scale=.8]

\begin{scope}[shift={(0,-12)}, scale=2]

\draw[realscalar] (.7,0)--(2,0);

\draw[realscalar][rotate=30] (.7,0)--(2,0);

\draw[realscalar][rotate=180] (.7,0)--(2.1,0);
\draw[majorana][rotate= 0] (2.1,0)--(1.1,0);

\draw[realscalar][rotate=150] (.7,0)--(2,0);

\draw[gravitinobar][rotate=180] (1.5,1)--(1.25,.5);


\draw[gravitino][rotate=180] (1.1,1)--(1.25,.5);

\draw[graviton][rotate=180] (1.25,0)--(1.25,.5);
\draw[gravitino][rotate=0] (1.1,0)--(1.1,-1);
\draw[gravitino][rotate=180] (1.9,0)--(1.9, 1);
\draw[majorana][rotate= 180] (2.5,0)--(1.9,0);
\node at (-1.6,-.5) {$...$};
\node at (1.6,-.5) {$...$};


\begin{scope}[shift={(0,0)}, scale=2]
    \draw [ultra thick] (0,0) circle (.35);
    \clip (0,0) circle (.3cm);
    \foreach \x in {-.9,-.8,...,.3}
    \draw[line width=1 pt] (\x,-.3) -- (\x+.6,.3);
\end{scope}
 
\end{scope}

\end{tikzpicture}
\end{center}
\caption{Feynman diagram for multiple soft gravitini - II}
\label{fig:multiplegravitinoII}
\end{figure}
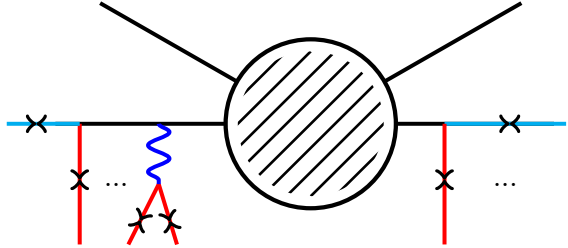
    \item All gravitini attach on the same external leg, but none of them form pairs to give soft graviton as shown in figure \ref{fig:multiplegravitinoIII}.
\begin{figure}[h]
\begin{center}
\begin{tikzpicture}[line width=1.5 pt, scale=.8]

\begin{scope}[shift={(0,0)}, scale=2]

\draw[realscalar] (.7,0)--(2,0);

\draw[realscalar][rotate=30] (.7,0)--(2,0);

\draw[realscalar][rotate=180] (.7,0)--(1.1,0);

\draw[majorana][rotate=180] (2.1,0)--(1.1,0);

\draw[realscalar][rotate=180] (2.1,0)--(2.6,0);
 
\draw[realscalar][rotate=150] (.7,0)--(2,0);

\draw[gravitino][rotate=180] (1.1,0)--(1.1,1);

\node at (-1.6,-.5) {$...$};

\draw[gravitino][rotate=180] (2.1,0)--(2.1,1);


\begin{scope}[shift={(0,0)}, scale=2]
    \draw [ultra thick] (0,0) circle (.35);
    \clip (0,0) circle (.3cm);
    \foreach \x in {-.9,-.8,...,.3}
    \draw[line width=1 pt] (\x,-.3) -- (\x+.6,.3);
\end{scope}
 
\end{scope}

\end{tikzpicture} 
\end{center}
\caption{Feynman diagram for multiple soft gravitini - III}
\label{fig:multiplegravitinoIII}
\end{figure}
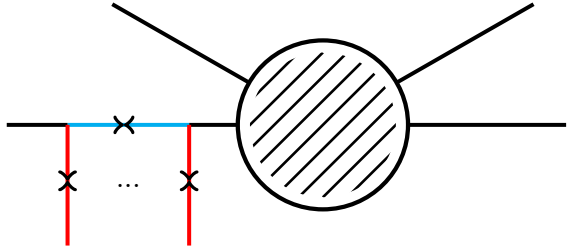      

    \item Some gravitini form pairs and give a soft graviton while some attach directly to external leg as shown in figure \ref{fig:multiplegravitinoIV}.
\begin{figure}[h]
\begin{center}
\begin{tikzpicture}[line width=1.5 pt, scale=.8]

\begin{scope}[shift={(0,0)}, scale=2]

\draw[realscalar] (.7,0)--(2,0);

\draw[realscalar][rotate=30] (.7,0)--(2,0);

\draw[realscalar][rotate=180] (.7,0)--(1.1,0);

\draw[realscalar][rotate=180] ( 1.9,0)--(1.1,0);

\draw[majorana][rotate=180] (1.9,0)--(2.6,0);
\draw[majorana][rotate=180] (2.6,0)--(3,0);

\draw[gravitino][rotate=180] (2.6,.5)--(2.25,1);

\draw[graviton][rotate=180] (2.6,0)--(2.6,.5);

\draw[gravitino][rotate=180] (2.6,.5)--(2.95, 1); 
\draw[realscalar][rotate=150] (.7,0)--(2,0);
\node at (-2.35,-.5) {$...$};

\draw[gravitino][rotate=180] (1.9,0)--(1.9,1);
\node at (-1.65,-.5) {$....$};
\draw[gravitino][rotate=180] (1.0,1)--(1.25,.5);
\draw[graviton][rotate=180] (1.25,0)--(1.25,.5);
\draw[gravitino][rotate=180] (1.5,1)--(1.25,.5);
\node at (-0.9,-.5) {$....$};
\begin{scope}[shift={(0,0)}, scale=2]
    \draw [ultra thick] (0,0) circle (.35);
    \clip (0,0) circle (.3cm);
    \foreach \x in {-.9,-.8,...,.3}
    \draw[line width=1 pt] (\x,-.3) -- (\x+.6,.3);
\end{scope}
 
\end{scope}

\end{tikzpicture} 
\end{center}
\caption{Feynman diagram for multiple soft gravitini - IV}
\label{fig:multiplegravitinoIV}
\end{figure}
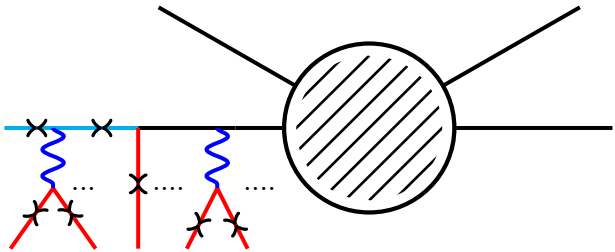  
\end{itemize}
We performed explicit computation for three soft gravitini, which is shown in appendix \ref{sec:threesoft}. By looking at the pattern followed in two and three soft gravitini case, we propose the following expression for $M$-soft gravitini.
\begin{eqnarray}
\label{multiplegravitinosoft}
     \Gamma_{N+M}(\{p_i\},\{k_{u_i}\})&& =    \left[\prod_{i=1}^M \sfactor_{u_i} + \sum_{A = 1}^{\lfloor M/2\rfloor }\prod_{i =1}^{A}\mfactor_{u_iv_i}\prod_{j=1}^{M -2A}\sfactor_{r_j} \right]\Gamma_{N}(\{p_i\}) 
\end{eqnarray}
where $\lfloor M/2\rfloor  $ denotes the greatest integer which is less than or equal to $M/2$. Now we will explain various terms of the above expression. 

\begin{enumerate}
    \item The first term is very similar to the leading soft factor for multiple soft photons or multiple soft gravitons. The other terms are there because of the fact that soft gravitino factors do not commute. We always write the first factor in a particular order, for example, $\sfactor_{u_1},....,\sfactor_{u_M}$ $u_1< u_2...<u_M$ and then the particular form of the second term depends on this choice of ordering for the first term. This way to write in particular ordering also turns out to be convenient to check gauge invariance.

    \item In the second term, $A$ counts the number of pairs of gravitini giving soft gravitons. For each pair, we have a factor of $\cfactor_{uv}$ coming from gravitino-graviton-gravitino vertex which combines with a factor due to the use of anti-commutation relation to bring the first term in particular order, to give $\mfactor_{uv}$. The subscripts $\{r_j, u_i,v_i\}$ can take values from $1,...,M$ and $v_i > u_i$ and $r_j$'s are also ordered with the largest $r_j$ appearing on the right. 
    
\end{enumerate}
 The disadvantage of the expression \eqref{multiplegravitinosoft} is that it depends on the ordering of the external soft gravitini. The expression is not manifestly invariant under alternation of the ordering. Now, we will show that the expression is actually invariant under rearrangement. We can go to any particular ordering starting from any other ordering.  Our strategy is as follows: 
\begin{enumerate}
    \item We first show that any two consecutive entries can be interchanged. 

    \item By repeating this operation (of interchanging any two consecutive entries) many times, we can obtain any ordering starting from any other ordering\footnote{Theorem 2.1 in this \href{http://www.math.uconn.edu/~kconrad/blurbs/grouptheory/genset.pdf}{note} gives a proof of the above statement.}.
\end{enumerate}

\subsection{Re-arrangement}
\label{subsec:rearrangementarbitrarygravitino}

Here we show that any two consecutive terms of equation \eqref{multiplegravitinosoft} can be interchanged. Consider the $i^{\textrm{th}}$ and $(i+1)^{\textrm{th}}$ particle. We write the expression \eqref{multiplegravitinosoft}
\begin{eqnarray}
  \Gamma_{N+M}(\{p_i\},\{k_u\})&=&  \Bigg[ \sfactor_{u_1}...\sfactor_{u_i}\sfactor_{u_{i+1}}... \sfactor_{u_M}+ \mfactor_{u_1u_2}\, \sfactor_{u_3}...\sfactor_{u_i}\sfactor_{u_{i+1}}... \sfactor_{u_M}
\nonumber    \\
  && \mfactor_{u_2u_3}\, \sfactor_{u_1}...\sfactor_{u_i}\sfactor_{u_{i+1}}...\sfactor_{u_{M-1}}\sfactor_{u_M}+....+\mfactor_{u_1u_i}\sfactor_{u_2}...\sfactor_{u_{i+1}}...\sfactor_{u_{M-1}}\sfactor_{u_M}
\nonumber    \\
  && \mfactor_{u_2u_i}\, \sfactor_{u_1}...\sfactor_{u_{i+1}}...\sfactor_{u_{M-1}}\sfactor_{u_M}+...+\mfactor_{u_iu_{i+1}}\sfactor_{u_1}...\sfactor_{u_{M-1}}\sfactor_{u_M}+
\nonumber    \\
  && \mfactor_{u_1u_2}\mfactor_{u_3u_4}...\sfactor_{u_i}\sfactor_{u_{i+1}}...\sfactor_{u_{M-1}}\sfactor_{u_M}+...+\mfactor_{u_1u_2}...\mfactor_{u_iu_{i+1}}...\mfactor_{u_{M-1}u_M}\Bigg] \Gamma_{N}(\{p_i\})
\nonumber\\  
\label{multiplegravitinosoftrearrangement1}
\end{eqnarray}
Here the $i^{\textrm{th}}$ and $(i+1)^{\textrm{th}}$ particle can appear only in three different ways 
\begin{itemize}
    \item {\bf Possibility I:} Both the $i^{\textrm{th}}$ and $(i+1)^{\textrm{th}}$ gravitini appear as $\sfactor$ factor 
 \begin{eqnarray}
  \Big[  A\, \sfactor_{u_i}\sfactor_{u_{i+1}}\, B \Big]\Gamma_{N}(\{p_i\}) 
\label{multiplegravitinosoftrearrangement2}
 \end{eqnarray}  
 where $A$ and $B$ involves all the other $M-2$ gravitini. The other gravitions appear as ordered multiplications of $\sfactor_u$ and $\mfactor_{vw}$'s in all possible ways.
 
    \item {\bf Possibility II:}  Both the $i^{\textrm{th}}$ and $(i+1)^{\textrm{th}}$ gravitino appear in $\mfactor_{uv}$ together 
\begin{eqnarray}
\Big[ \widetilde A\,  \mfactor_{u_i u_{i+1}}\, \widetilde B \Big] \Gamma_{N}(\{p_i\})
\label{multiplegravitinosoftrearrangement3}
\end{eqnarray} 
Here $\tilde A$ and $\tilde B$ involves all the other $M-2$ gravitini. Again the other gravitions appear as ordered multiplications of $\sfactor_u$ and $\mfactor_{vw}$'s in all possible ways. This would imply 
\begin{eqnarray}
    A=\widetilde A \qquad,\qquad B=\widetilde  B
\end{eqnarray}
So same $A$ and $B$ appear in \eqref{multiplegravitinosoftrearrangement2} and in \eqref{multiplegravitinosoftrearrangement3}. 
Adding \eqref{multiplegravitinosoftrearrangement2} and \eqref{multiplegravitinosoftrearrangement3} we get 
\begin{eqnarray}
\Big[    A(\sfactor_{u_{i}}\, \sfactor_{u_{i+1}} + \mfactor_{u_i u_{i+1}})B     \Big]\Gamma_{N}(\{p_i\})
\label{multiplegravitinosoftrearrangement3.1}
\end{eqnarray}

    \item {\bf Possibility III:} At least one of them appears as $\mfactor$ and if both of them appear in $\mfactor_{uv}$,  they do not appear together. The possibility of both of them to appear together in  $\mfactor_{uv}$  has already been taken into account in possibility II. 
\begin{eqnarray}
 \sum_{j=1, j\ne i,i+1}^N\Big[ \mfactor_{u_{j}u_i}C_{i+1}(\epsilon_{ u_{i+1}})  + \mfactor_{u_{j}u_{i+1}}C_{i} (\epsilon_{u_i})  \Big] \Gamma_{N}(\{p_i\})
\label{multiplegravitinosoftrearrangement4}
\end{eqnarray}    
Here $C_{i+1}( \epsilon_{u_{i+1}})$ is the all possible arrangements of all the gravitini except $u_{j}$and $u_i$ and similarly $C_{i}( \epsilon_{u_{i}})$ is the all possible arrangements of all the gravitini except $u_{j}$ and $u_{i+1}$.

\end{itemize}
Now if we started with an ordering in which $u_{i+1}$ appeared before $u_{i}$ then  we can repeat the same analysis. Equation \eqref{multiplegravitinosoftrearrangement4} is same in both cases, but in \eqref{multiplegravitinosoftrearrangement1} and in \eqref{multiplegravitinosoftrearrangement2} will $i$ and $i+1$ will be interchanged (i.e. $i\longleftrightarrow i+1$). Hence instead of \eqref{multiplegravitinosoftrearrangement3.1} we would get 
\begin{eqnarray}
\Big[A\left(\sfactor_{u_{i+1}}\sfactor_{u_{i}} + \mfactor_{u_{i+1}u_i }\right)B \Big]\Gamma_{N}(\{p_i\})
\label{multiplegravitinosoftrearrangement4.1}
\end{eqnarray}
But now we can use  \eqref{doublesoftgravi65} to see that \eqref{multiplegravitinosoftrearrangement3.1} and \eqref{multiplegravitinosoftrearrangement4.1} a essentially same.  
Hence the final answer is same irrespective of ordering of the soft factors. 

\subsection{Gauge invariance}
We have proved the expression for multiple soft gravitini can be rearranged to any particular ordering. Using this, we can bring any gravitino to be the rightmost. So we will show the gauge invariance of the expression only when the rightmost gravitino is pure gauge. 

The right-most gravitino can appear only in two ways 
\begin{enumerate}
    \item It  can appear in $\sfactor_u$. Since it is the right-most gravitino, it will directly hit the amplitude of the hard-particle and hence zero by \eqref{singlesoftgrav41}. 
    
    \item Or it can appear in $\mfactor_{uv}$. Again it will always appear as the 2nd index. But this vanishes because of  \eqref{doublesoftgravi173a}. 
\end{enumerate}

\section{Two soft gravitini in the presence of central charge} 
\label{sec:gravitinocentralterm}
In case of extended supersymmetries \footnote{We are thankful to Matteo Bertolini,  Atish Dabholkar, Kyriakos Papadodimas, Cumrun Vafa for discussion on this point.}, one can have central charges in the supersymmetry algebra. The supersymmetry algebra in \eqref{drsusyalgebra1} modifies to 
\begin{eqnarray}
    \Big\{\scharge_\alpha,\scharge_\beta\Big\}= -\frac{1}{2}\gamma^\mu_{\alpha \beta}P_\mu -\frac{1}{2} \zfactor_{\alpha \beta }\,    \ucharge 
\label{doublesoftgravi101}
\end{eqnarray}
$\ucharge $ is (are) the generator(s) of $U(1)$ symmetry(-ies) generated by the central charge(s). As explained below equation \eqref{drsusyalgebra1}, $\alpha, \beta$ are some (ir-)reducible spinor indices. In this language the existence of central charge is equivalent to the condition that there exists an element(s) ${Z_{\alpha}}^\beta$ in the Clifford algebra such that, $\zfactor_{\alpha \beta }$ satisfies 
\begin{eqnarray}
 \zfactor_{\alpha \beta } = \zfactor_{\beta \alpha } 
\label{doublesoftgravi102}
\end{eqnarray}
In general, there can be higher form central charges. For example, in $D=11$, the supersymmetry algebra is of the form 
\begin{eqnarray}
     \Big\{\scharge_\alpha,\scharge_\beta\Big\}= -\frac{1}{2}\gamma^\mu_{\alpha \beta}P_\mu + \gamma^{\mu_1\mu_2\mu_3}_{\alpha \beta }A_{\mu_1\mu_2\mu_3}   
\end{eqnarray}
But for our purpose, we ignore any higher form central charges. This is because the higher form central charges can only minimally couple to extended objects (of appropriate dimensions),  whereas we consider the scattering of point-like states only.  

In this case the commutator of two soft operators in \eqref{singlesoftgrav12} is modified as follows 
\begin{eqnarray}
\left[\sfactor_u,\sfactor_v    \right]    = -\frac{\kappa^2}{2} \sum_{i=1}^N  \left[ \epsilon_{\mu}^{(u);\alpha}(\slashed{p}_{i; \alpha \beta} + e_i\,  \zfactor_{\alpha\beta}) \epsilon_{\nu}^{(v);\beta}  \frac{p_i^\mu  }{p_i\cdot k_u}\frac{p_i^\nu  }{p_i\cdot k_v}   \right]  
\label{doublesoftgravi103}
\end{eqnarray}
In presence of the central term the computation in section \ref{sec:doublesoftgravitino} will be modified. In presence of particles charged under the central charge, the combined contribution from figure \ref{fig:twogravitino0}, {\bf \ref{fig:twogravitinoI}}, \ref{fig:twogravitinoII} and \ref{fig:twogravitinoIII} is not gauge invariant. We need a new interaction to make it gauge invariant. In fact, it is possible to extend the argument in \cite{Grisaru:1977kk} to argue that gauge invariance in presence of central charge implies the existence of massless photon which interacts at low energy.

\paragraph{Graviphoton and new interaction  \cite{Zachos:1978iw} }

When we gauge the global supersymmetry with central charge to get supergravity, we get a $U(1)^N$ gauge symmetry generated by spin $1$ bosons (graviphoton) present in the graviton multiplet.  These graviphotons couple to the gravitino and to any matter which carries the central charge. The coupling of the graviphoton to gravitino is completely fixed by supersymmetry and is related to that of graviton. The gravitino-gravitino-graviphoton three point function $(\widetilde{\mathcal{V}}^{\mu\nu;\mu_1 } )_{\alpha \beta}$ is given by 
\begin{eqnarray}
&&    -\iimg \kappa  \Big[k_2^{\mu_1}  ( \zfactor)^{\alpha\beta}\eta^{\mu\nu}\Big]  
-  \frac{\iimg \kappa}{2}    
\Big[[(k_1+k_2)_c\, \delta^{\mu_1}_{d} ]( \zfactor\gamma^{cd})^{\alpha\beta} \eta^{\mu\nu}\Big]  
+\iimg \kappa   \Big[(k_2^\mu\, \eta^{\mu_1 \nu}-k_1^\nu\, \eta^{\mu_1 \mu} )( \zfactor )^{\alpha\beta} \Big]  
\label{doublesoftgravi112}
\end{eqnarray}

\begin{figure}[h]
\begin{center}
\begin{tikzpicture}[line width=1.5 pt, scale=.8]

\begin{scope}[shift={(0,-12)}, scale=2]

\draw[realscalar] (.7,0)--(2,0);

\draw[realscalar][rotate=30] (.7,0)--(2,0);

\draw[realscalar][rotate=180] (.7,0)--(2.1,0);

\draw[realscalar][rotate=150] (.7,0)--(2,0);

\draw[gravitinobar][rotate=180] (1.6,1)--(1.35,.5);


\draw[gravitino][rotate=180] (1.1,1)--(1.35,.5);

\draw[photon][rotate=180] (1.35,0)--(1.35,.5);

\begin{scope}[shift={(0,0)}, scale=2]
    \draw [ultra thick] (0,0) circle (.35);
    \clip (0,0) circle (.3cm);
    \foreach \x in {-.9,-.8,...,.3}
    \draw[line width=1 pt] (\x,-.3) -- (\x+.6,.3);
\end{scope}
 
\end{scope}

\end{tikzpicture}
\end{center}
\caption{Feynman diagram for double soft gravitini - V}
\label{fig:twogravitinogaugeI}
\end{figure}
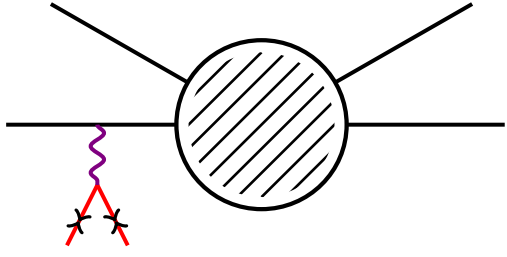
Whenever we have more than one soft gravitini, the vertex in \eqref{doublesoftgravi112} contributes. In particular, consider the case of two soft gravitini. We already evaluated it in section \ref{sec:doublesoftgravitino}. In presence of the central charge(s) we have a new contribution from the diagram \ref{fig:twogravitinogaugeI} 

\paragraph{Evaluation of figure \ref{fig:twogravitinogaugeI} } Now we would like to evaluate the figure \ref{fig:twogravitinogaugeI}. The evaluation of this diagram very similar to the evaluation of the figure \ref{fig:twogravitinoIII} . It is given by
 \begin{eqnarray}
  \Gamma_{N+2}^{(4)} (\{p_i\},k_1,k_2)
    &=&   \left[   \epsilon^{(1) \alpha}_{\mu}(\widetilde{\mathcal{V}}^{\mu\nu;\mu_1 } )_{\alpha \beta} \, \epsilon^{(2)\beta}_{\nu}  \right]\left[\frac{\iimg \eta_{\mu_1\mu_2}}{2k_1\cdot k_2}\right]
 \left[   -2\kappa \,  e_i \,   {p_i^{\mu_2}  }\right]\left[\frac{1}{2p_i\cdot(k_1+k_2)} \right]   \Gamma_N(\{p_i\})
 \nonumber\\
 \label{doublesoftgravi111}
 \end{eqnarray} 
Here the first square bracket denotes gravitino-gravitino-graviphoton vertex, the second one is the graviphoton propagator and the third one is the matter- soft graviphoton -matter vertex and the last one is the internal propagator. Now we substitute the explicit expression for $(\widetilde{\mathcal{V}}^{ab;\mu } )_{\alpha \beta}$ from equation \eqref{doublesoftgravi112} and we obtain
\begin{eqnarray}  
  \Gamma_{N+2}^{(4)} (\{p_i\},k_1,k_2)
    &=&  \frac{(-\iimg \kappa)^2}{2} \sum_{i=1}^Ne_i  \,  \epsilon^{(1)}_{\mu \alpha} (  \zfactor)^{\alpha \beta}\left[\eta^{\mu\nu} p_i\cdot k_2 + \frac{1}{2}\eta^{\mu \nu} (k_1+k_2)_d\,   p_{ie}\gamma^{de}- (k_2^\mu p_i^\nu -k_1^\nu p_i^\mu) \right] \epsilon^{(2)}_{\nu\beta}
\nonumber\\    
 &&   \left[\frac{1}{( p_i\cdot (k_2 + k_1))(k_1\cdot k_2)}\right]\Gamma_N(\{p_i\})
\label{doublesoftgravi113}
\end{eqnarray}
We simplify the above expression to get  
\begin{eqnarray}  
  \Gamma_{N+2}^{(4)} (\{p_i\},k_1,k_2)
    &=&  \frac{\kappa^2}{2} \sum_{i=1}^Ne_i\,   \epsilon^{(1)}_{\mu \alpha} (  \zfactor )^{\alpha \beta}\left[- \eta^{\mu\nu} p_i\cdot k_2 + \frac{1}{2}\eta^{\mu \nu} p_{i}\cdot  (k_1+k_2)   + (k_2^\mu p_i^\nu -k_1^\nu p_i^\mu) \right] \epsilon^{(2)}_{\nu\beta}
\nonumber\\    
 &&   \left[\frac{1}{( p_i\cdot (k_2 + k_1))(k_1\cdot k_2)}\right]\Gamma_{N}(\{p_i\})
\label{doublesoftgravi115}
\end{eqnarray} 
In this case the definition of $\cfactor_{ uv}(p_i)$ in \eqref{doublesoftgravi37} will be modified as follows 
\begin{eqnarray}
\tcfactor_{uv}(p_i)=\cfactor_{uv}(p_i) + \frac{\kappa^2}{2} e_i\,   \epsilon^{(u)}_\mu \zfactor\epsilon^{(v)}_\nu \left[ \frac{1}{2} \frac{\eta^{\mu \nu}p_i \cdot(k_u-k_v)}{k_u\cdot k_v} + \frac{(k_v^\mu p_i^\nu - k_u^\nu p_i^\mu) }{k_u \cdot k_v}\right]    
\label{doublesoftgravi116}
\end{eqnarray} 
In equation \eqref{doublesoftgravi38} we show that $\cfactor_{uv}$ is symmetric in its particle induces. The same property holds for $\tcfactor_{uv}$
\begin{eqnarray}
    \tcfactor_{uv}(p_i)    =\tcfactor_{vu}(p_i)    
\label{doublesoftgravi117}
\end{eqnarray}
We add the contribution from \eqref{doublesoftgravi116} to \eqref{doublesoftgravi61} to get the final answer. It is given by  
\begin{eqnarray}
\Gamma_{N+2}(\{p_i\},k_1,k_2)=\Big[ \sfactor_1\, \sfactor_2 +\tmfactor_{12}\Big] \Gamma_{N}(\{p_i\})
\label{doublesoftgravi161}
\end{eqnarray}
Here we have introduced $\tmfactor_{uv}$. It is defined as 
\begin{eqnarray}
\tmfactor_{uv}&=& \mfactor_{uv} +
\frac{\kappa^2}{2}\sum_{i=1}^N       e_i\, 
\frac{\epsilon^{(u)}_\mu \zfactor\epsilon^{(v)}_\nu}{p_i \cdot (k_u+k_v)} \left[\frac{p_i^\mu p_i^\nu}{p_i \cdot  k_v } + \frac{1}{2} \frac{\eta^{\mu \nu}p_i \cdot(k_u-k_v)}{k_u\cdot k_v} + \frac{(k_v^\mu p_i^\nu - k_u^\nu p_i^\mu) }{k_u \cdot k_v}\right]    
\label{doublesoftgravi162} 
\end{eqnarray}
Note that the relations in equations \eqref{doublesoftgravi63a}, \eqref{doublesoftgravi63b}, \eqref{doublesoftgravi63c} remain the same if we replace $\mfactor_{uv}$ with $\tmfactor_{uv}$. In this particular case 
\begin{eqnarray}
\sfactor_u\, \sfactor_v-\sfactor_v\, \sfactor_u=-\tmfactor_{uv}    +\tmfactor_{vu}    
\label{doublesoftgravi163}
\end{eqnarray}
We check this explicitly. We have already the LHS, i.e. $\sfactor_u\, \sfactor_v-\sfactor_v\, \sfactor_u$ in \eqref{doublesoftgravi103}. Now we compute the RHS
\begin{eqnarray}
    \tmfactor_{vu}-\tmfactor_{uv} &=&     \mfactor_{vu}-\mfactor_{uv} +\frac{\kappa^2}{2} \sum_{i=1}^N e_i\,  \frac{\epsilon^{(v)}_\mu \zfactor\epsilon^{(u)}_\nu}{p_i \cdot (k_u+k_v)}\left[ \frac{p_i^\mu p_i^\nu}{p_i \cdot  k_u } + \frac{1}{2} \frac{\eta^{\mu \nu}p_i \cdot(k_v-k_u)}{k_u\cdot k_v} + \frac{(k_u^\mu p_i^\nu - k_v^\nu p_i^\mu) }{k_u \cdot k_v}\right]\nonumber\\
    &&- \frac{\kappa^2}{2}\sum_{i=1}^N e_i\,  \frac{\epsilon^{(u)}_\mu \zfactor\epsilon^{(v)}_\nu}{p_i \cdot (k_u+k_v)}\left[ \frac{p_i^\mu p_i^\nu}{p_i \cdot  k_v } + \frac{1}{2} \frac{\eta^{\mu \nu}p_i \cdot(k_u-k_v)}{k_u\cdot k_v} + \frac{(k_v^\mu p_i^\nu - k_u^\nu p_i^\mu) }{k_u \cdot k_v}\right]\nonumber\\
    &=& -  \frac{\kappa^2}{2} \sum_{i=1}^N \frac{\epsilon^{(u)} \cdot p_i} {p_i \cdot k_u} \frac{\epsilon^{(v)} \cdot p_i} {p_i \cdot k_v}\left(\slashed{p_i}+e_i\, \zfactor \right)
\label{doublesoftgravi166}
\end{eqnarray}
Hence \eqref{doublesoftgravi163} is satisfied.
  
\subsection{Gauge invariance }
As explained in the section \ref{subsec:doublesoftgaugeinvariance}, it is easier to prove gauge invariance if we put pure gauge polarization for the gravitino adjacent to $\Gamma_N$. So we consider pure gauge polarization for the second gravitino
\begin{eqnarray}
    \epsilon_2^{\mu\alpha}= k_2^\mu \theta^\alpha_2  
\label{doublesoftgravi172}
\end{eqnarray}
For pure gauge 
\begin{eqnarray}
      \tmfactor_{uv} 
      &=&  \frac{\kappa^2}{2} \sum_{i=1}^N  \frac{1}{p_i \cdot (k_u+k_v)} \left[   \epsilon^{(u)}_\mu \slashed{p}_i \theta^{(v)} + e_i\,  \epsilon^{(u)}_\mu \zfactor\theta^{(v)} \right]
\nonumber\\      
&&      \Bigg[ \frac{p_i^\mu p_i\cdot k_v}{p_i \cdot  k_v } + \frac{1}{2} \frac{k_v^\mu p_i \cdot(k_u-k_v)}{k_u\cdot k_v} + \frac{(k_v^\mu k_v\cdot p_i  - k_v\cdot k_u  p_i^\mu) }{k_u \cdot k_v}\Bigg]  
\nonumber\\
&=&  \frac{\kappa^2}{2} \sum_{i=1}^N\frac{1}{p_i \cdot (k_u+k_v)} \left[  \epsilon^{(u)}_\mu \slashed{p}_i \theta^{(v)}  + e_i\,  \epsilon^{(u)}_\mu \zfactor\theta^{(v)} \right]  \left[ \frac{1}{2} \frac{k_v^\mu}{k_u\cdot k_v}  \right]
=0  
\label{doublesoftgravi173}
\end{eqnarray} 
where in the last step we have used momentum conservation and (central-)charge conservation
\begin{eqnarray}
\sum_{i=1}^N p_i = 0
\qquad,\qquad
\sum_{i=1}^N e_i = 0    
\end{eqnarray}

\subsection{Presence of soft graviton}

Following \cite{Sen:2017xjn, Sen:2017nim, Laddha:2017ygw} it is easy to include soft graviton into this computation. The vertex for the leading soft graviton ($\zeta_{\mu \nu}P^\mu P^\nu$) commutes with the vertex for soft gravitino and also commutes with the vertex for any other soft graviton. So, in the presence of $M_1$ soft gravitini and $M_2$ soft gravitons equation \eqref{multiplegravitinosoft} is modified as follows
\begin{eqnarray}
     \Gamma_{N+M_1+M_2}(\{p_i\},\{k_r\}) =    \left[\prod_{j=1}^{M_2} \tsfactor_{u_j}\right] \left[\prod_{i=1}^{M_1} \sfactor_{u_i} + \sum_{A = 1}^{\lfloor M_1/2\rfloor }\prod_{i =1}^{A}\mfactor_{u_iv_i}\prod_{j=1}^{M_1 -2A}\sfactor_{r_j} \right]\Gamma_N(\{p_i\})
\end{eqnarray}
$ \tsfactor_{u}$ is the leading soft factor for graviton, given in equation \eqref{softgravitonfactor1}.  

 
We know that the leading and sub-leading soft factors for multiple gravitons are universal. In this paper, we derived the leading order expression for multiple soft gravitini and we found that it is also universal. These three soft theorems are inter-related by supersymmetry. One way to argue this is to observe that all these three soft theorems follow from covariantizing the action with respect to the soft field.  In supergravity, the structure of the covariant derivative is uniquely fixed by supersymmetry.

\section{Infrared divergence and soft gravitino theorem}
\label{sec:irdivergence}

Now we briefly discuss infrared divergences in supergravity theories \footnote{ We are thankful to Ashoke Sen for discussion on this section and correcting one mistake in an earlier version of the draft. We are thankful to the unknown referee for suggesting various improvements to this section.}. We are using 1PI effective action for our computation but this approach fails when 1PI vertices are IR divergent. The presence of the massless particles in the loops can potentially give rise to these divergences, hence in the supergravity theories, graviton, gravi-photon and gravitino\footnote{If there is a massless matter multiplet then in principle it can also contribute to infrared divergence} can contribute. There are no IR divergences in the 1PI effective action for $D \geq 5$. We will show below that the virtual gravitino does not give rise to any IR divergence in any dimensions. In $D=4$, 1PI vertices suffer from IR divergences only due to graviton and graviphoton. However, a more careful analysis shows that the leading soft gravitino factor is not altered by IR divergences.

First, we discuss the case of $D\geq 5$. Then we discuss the case of $D=4$ which needs more careful analysis.  We show that IR divergence does not alter leading soft gravitino theorem.

\subsection{Infrared divergences in $D \geq 5$}

\label{subsec:irdivergencevirtualgravitino}
We want to check if the approach based on 1PI effective action remains valid in $D \geq 5$ even after taking the soft limit for the external gravitino.

\begin{figure}[h]
\begin{center}
\begin{tikzpicture}[line width=1.5 pt, scale=1.4]

\begin{scope}[shift={(0,0)}]

\draw[realscalar] (.7,0)--(2,0);

\draw[realscalar][rotate=30] (.7,0)--(2,0);

\draw[realscalar][rotate=180] (1.8,0)--(2.7,0);
\draw[majorana][rotate=180] (.7,0)--(1.8,0);

 \draw[majorana][rotate=180] (3.5,0)--(2.7,0);

\draw[majorana][rotate=120] (.7,0)--(1.4,0);
\draw[realscalar][rotate=120] (1.4,0)--(2.1,0);

\draw[gravitino][rotate=180] (1.8,0)--(1.8,1);

\draw[gravitino] (-2.7,0)--(-0.7,1.2);


 \node at (-1.6,1) {$\ell $}; 
  \node at (-2.4,-.25) {$p_i +\ell $}; 
\node at (-3.3,-.25) {$p_i $}; 
\node at (-1.5,-.5) {$k $}; 
\node at (-1.3,.25) {$p_i +k +\ell $};
\node at (-.6,1.8) {$p_j $}; 
\node at (0,1) {$p_j-\ell  $};

\begin{scope}[shift={(0,0)}, scale=2]
    \draw [ultra thick] (0,0) circle (.35);
    \clip (0,0) circle (.3cm);
    \foreach \x in {-.9,-.8,...,.3}
    \draw[line width=1 pt] (\x,-.3) -- (\x+.6,.3);
\end{scope}
 \node at (0,-1.5) {(a)}; 
\end{scope}

\begin{scope}[shift={(6.5,0)}]

\draw[realscalar] (.7,0)--(2,0);

\draw[realscalar][rotate=30] (.7,0)--(2,0);

\draw[majorana][rotate=180] (1.8,0)--(2.7,0);
\draw[realscalar][rotate=180] (.7,0)--(1.8,0);

\draw[majorana][rotate=180] (3.5,0)--(2.7,0);

\draw[realscalar][rotate=120] (.7,0)--(1.4,0);
\draw[realscalar][rotate=120] (1.4,0)--(2.1,0);

\draw[gravitino][rotate=180] (1.8,0)--(1.8,1);

\draw[graviton] (-2.7,0)--(-0.7,1.2);


 \node at (-1.6,1) {$\ell $}; 
  \node at (-2.4,-.25) {$p_i +\ell $}; 
\node at (-3.3,-.25) {$p_i $}; 
\node at (-1.5,-.5) {$k $}; 
\node at (-1.3,.25) {$p_i +k +\ell $};
\node at (-.6,1.8) {$p_j $}; 
\node at (0,1) {$p_j-\ell  $};

\begin{scope}[shift={(0,0)}, scale=2]
    \draw [ultra thick] (0,0) circle (.35);
    \clip (0,0) circle (.3cm);
    \foreach \x in {-.9,-.8,...,.3}
    \draw[line width=1 pt] (\x,-.3) -- (\x+.6,.3);
\end{scope}
 \node at (0,-1.5) {(b)};  
 
\end{scope}

\end{tikzpicture}
\end{center}
\caption{Infrared divergence in supergravity I} 
\label{fig:gravitinoirdivergenceI}
\end{figure}
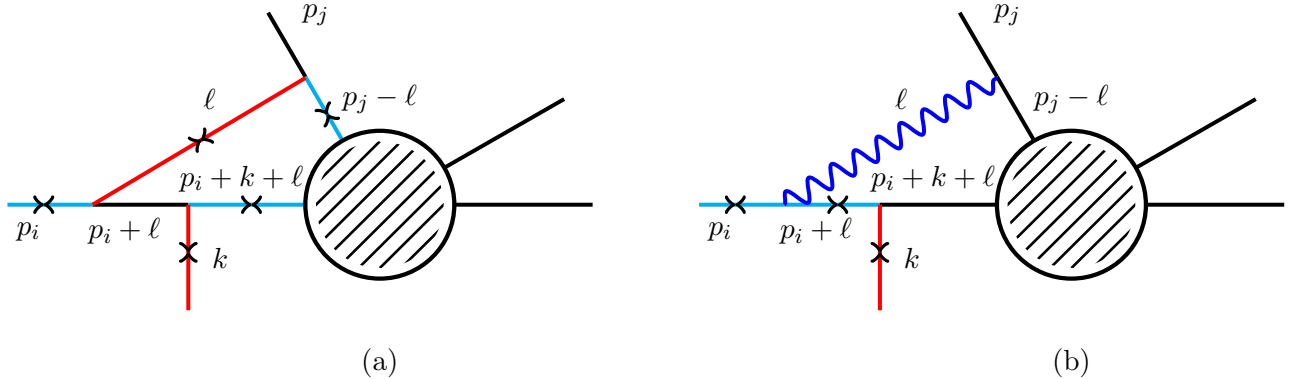

Consider the Feynman diagram in figure \ref{fig:gravitinoirdivergenceI}(a), if the external momenta are finite, then by naive power-counting, we can see that the amplitude does not have IR divergence for $D \geq 4$. We have three powers of $\ell$ in the denominator, one from each of the propagators with momenta $p_i + \ell$, $p_j -\ell$ and $\ell$. The last propagator gives one power of $\ell$ because it is a fermionic particle. In $D$ dimensions, we have $D$ powers of $\ell$ in numerator due to the loop integral and hence the amplitude goes like $\ell^{D-3}$ for small loop momentum $\ell$. So the diagram is free of IR divergence in $D \geq 4$. Hence any virtual gravitino does not give rise to IR divergences. But when the momenta $k \rightarrow 0$, then the propagator carrying momentum $p_i + k+ \ell$ gives another power of $\ell$ and makes the result logarithmically divergent in $D = 4$  but there is no additional divergence in $D \geq 5$. So our results are still valid for $D \geq 5$.

 Next we consider the Feynman diagram in figure  \ref{fig:gravitinoirdivergenceI}(b). In this case, the internal massless particle is graviton (it can also be photon/graviphoton). From power-counting we see that there are four powers of $\ell$ in the denominator, one from each of the propagators with momenta $p_i + \ell$, $p_j -\ell$ and two powers of $\ell$ coming from the graviton propagator.  Now in $k \rightarrow 0$ limit, the propagator carrying momentum $p_i + k+ \ell$ gives another power of $\ell$ and the diagram is logarithmic divergent in $D=5$. But the leading order answer is $O(k^{-1})$ and hence it still holds for $D\geq 5$.





\begin{figure}[h]
\begin{center}
\begin{tikzpicture}[line width=1.5 pt, scale=1.4]

\begin{scope}[shift={(0,0)}]

\draw[realscalar] (.7,0)--(2,0);

\draw[realscalar][rotate=30] (.7,0)--(2,0);

\draw[realscalar][rotate=180] (1.8,0)--(2.7,0);
\draw[realscalar][rotate=180] (.7,0)--(1.8,0);

\draw[majorana][rotate=180] (3.5,0)--(2.7,0);

\draw[realscalar][rotate=120] (.7,0)--(1.4,0);
\draw[realscalar][rotate=120] (1.4,0)--(2.1,0);

 \draw[gravitino][rotate=180] (2.7,0)--(2.7,1);

\draw[graviton] (-1.8,0)--(-0.7,1.2);


 \node at (-1.6,1) {$\ell $}; 
  \node at (-2.4,.25) {$p_i +k $}; 
\node at (-3.3,-.25) {$p_i $}; 
\node at (-2.5,-.5) {$k $}; 
 \node at (-1.3,-.25) {$p_i +k +\ell $};
\node at (-.6,1.8) {$p_j $}; 
\node at (0,1) {$p_j-\ell  $};

\begin{scope}[shift={(0,0)}, scale=2]
    \draw [ultra thick] (0,0) circle (.35);
    \clip (0,0) circle (.3cm);
    \foreach \x in {-.9,-.8,...,.3}
    \draw[line width=1 pt] (\x,-.3) -- (\x+.6,.3);
\end{scope}
 
\node at (0,-1.25) {(a)};  
\end{scope}

\begin{scope}[shift={(6.5,0)}]

\draw[realscalar] (.7,0)--(2,0);

\draw[realscalar][rotate=30] (.7,0)--(2,0);

\draw[majorana][rotate=180] (1.8,0)--(2.7,0);
\draw[realscalar][rotate=180] (.7,0)--(1.8,0);

\draw[majorana][rotate=180] (3.5,0)--(2.7,0);

\draw[realscalar][rotate=120] (.7,0)--(1.4,0);
\draw[realscalar][rotate=120] (1.4,0)--(2.1,0);

\draw[gravitino][rotate=180] (1.8,0)--(1.8,1);

\draw[graviton] (-2.7,0)--(-0.7,1.2);


 \node at (-1.6,1) {$\ell $}; 
  \node at (-2.4,-.25) {$p_i +\ell $}; 
\node at (-3.3,-.25) {$p_i $}; 
\node at (-1.5,-.5) {$k $}; 
\node at (-1.3,.25) {$p_i +k +\ell $};
\node at (-.6,1.8) {$p_j $}; 
\node at (0,1) {$p_j-\ell  $};

\begin{scope}[shift={(0,0)}, scale=2]
    \draw [ultra thick] (0,0) circle (.35);
    \clip (0,0) circle (.3cm);
    \foreach \x in {-.9,-.8,...,.3}
    \draw[line width=1 pt] (\x,-.3) -- (\x+.6,.3);
\end{scope}
 \node at (0,-1.25) {(b)};  
 
\end{scope}

\begin{scope}[shift={(3,-2.5)}]

\draw[realscalar] (.7,0)--(2,0);

\draw[realscalar][rotate=30] (.7,0)--(2,0);

\draw[realscalar][rotate=180] (.7,0)--(1.5,0);

\draw[majorana][rotate=180] (2.5,0)--(1.5,0);

\draw[realscalar][rotate=120] (.7,0)--(1.4,0);
\draw[realscalar][rotate=120] (1.4,0)--(2.1,0);

\draw[gravitino][rotate=180] (1.5,0)--(1.5,1);

\draw[graviton] (1.7,0)--(1.15,.7);


\node at (-2.6,-.25) {$p_i $}; 
\node at (-1.8,-.5) {$k $}; 
\node at (-.6,1.8) {$p_j $}; 

\begin{scope}[shift={(0,0)}, scale=2]
    \draw [ultra thick] (0,0) circle (.35);
    \clip (0,0) circle (.3cm);
    \foreach \x in {-.9,-.8,...,.3}
    \draw[line width=1 pt] (\x,-.3) -- (\x+.6,.3);
\end{scope}
 \node at (0,-1.25) {(c)};  
 
\end{scope}

\end{tikzpicture}
\end{center}
\caption{Infrared divergence in supergravity II} 
\label{fig:gravitinoirdivergenceII}
\end{figure}
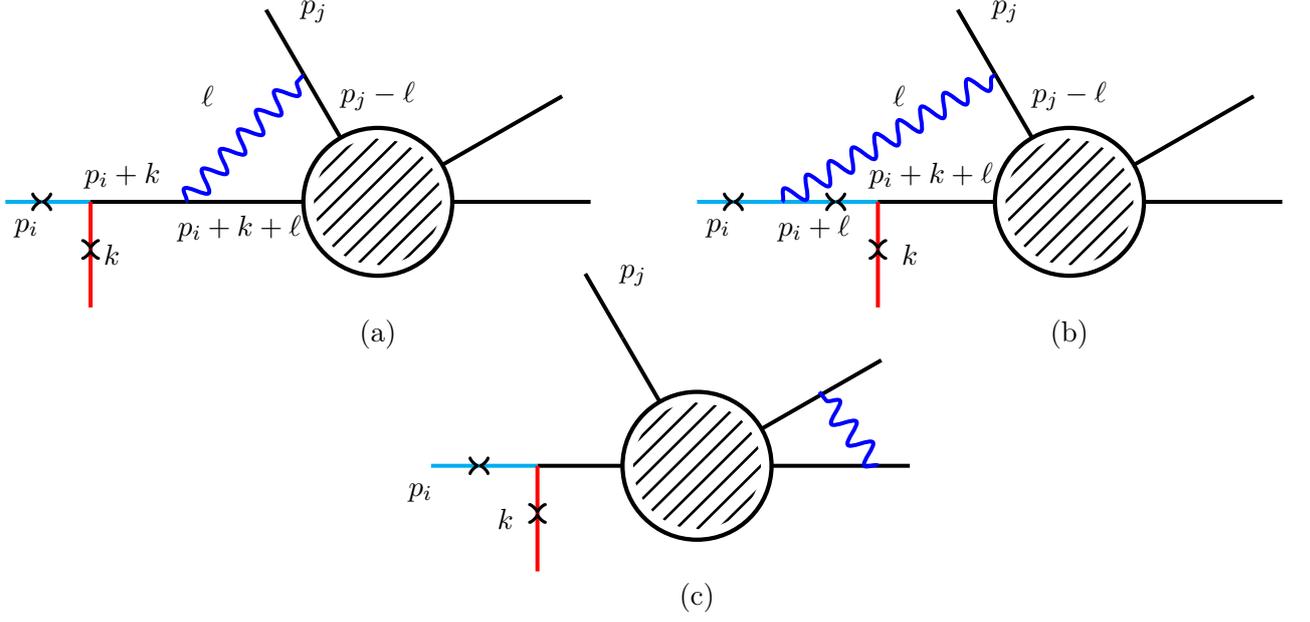
 










 


\subsection{Infrared divergences in $D = 4$ }
\label{subsec:irdivergencevirtualgraviton}
In $D=4$, the 1PI effective action suffers from IR divergences due to the presence of graviton and photon in the loop (We already argued that there is no IR divergence in the 1PI vertex due to the presence of gravitino in the loop). So we cannot use it to compute the $S$-matrix. But one can use the tree level action to derive soft theorems order by order in perturbation theory. So in four dimensions, we use the tree level action instead of 1PI action in equation \eqref{sgravsetup2}.

Now the question is whether loop corrections can alter the results of leading soft theorems. In case of soft graviton (and photon \footnote{with massive matter}), it has been shown that even though the amplitudes with and without soft particles suffer from IR divergences but at leading order, when one sum over diagrams, the divergence factorizes out and cancels from both sides \cite{Weinberg:1965nx, Bern:2014oka, Bern:2014vva}. In this section, we show that the same result holds for soft gravitino. We will show that the IR divergence due to graviton and graviphoton is same for amplitudes with and without soft gravitino. 

\subsubsection{Single real soft gravitino in presence of virtual graviton }  

First we consider the loop correction to the soft gravitino factor in $D=4$ in presence of a graviton running in the loop. We denote the contributions from these diagrams as $\Gamma^{(i;j,k)}_{N+1}(k,\{p_i\}) $; here the superscripts $j$ and $k$ denote the legs to which the virtual graviton attaches and $i$ denote the one to which the soft gravitino attaches. The total contribution is given by 
\begin{eqnarray}\label{totgamma}
\Gamma_{N+1}(k,\{p_i\}) = \sum_{i=1}^N\sum_{j=1}^N \sum_{k=1;k\ne j}^N
\, \Gamma^{(i;j,k)}_{N+1}(k,\{p_i\}) 
\end{eqnarray}
First we  evaluate $\Gamma^{(i;i,j)}_{N+1}(k,\{p_i\}) $. It is given by
\begin{eqnarray}
\Gamma^{(i;i,j)}_{N+1}(k,\{p_i\}) = \Big[\widetilde A_1(p_i,p_j;k)+\widetilde A_2(p_i,p_j;k)\Big]\, \, 
\Gamma_N(\{p_i\})
\label{softgravitinoirdiv1}
\end{eqnarray}
$\widetilde A_1(p_i,p_j;k)$ and $\widetilde A_2(p_i,p_j;k)$ are contributions from diagram $(a)$ and $(b)$ respectively in fig \ref{fig:gravitinoirdivergenceII}.  In small $k$ and small $\ell$ limit, these contributions are given by 
\begin{eqnarray}
\widetilde A_1(p_i,p_j;k) 
&=& \kappa^3 \beta_{ij}\,  (p_i \cdot \epsilon^\alpha\scharge_\alpha) \int \frac{d^4\ell}{(2\pi)^4}  \frac{1}{\ell^2} \frac{1}{p_i\cdot \ell}\frac{1}{p_j\cdot \ell }\frac{1}{p_i\cdot (k+\ell) }
\label{softgravitinoirdiv1.1}
\\
\widetilde A_2(p_i,p_j;k) 
&=&\kappa^3 \beta_{ij}\,  (p_i \cdot \epsilon^\alpha\scharge_\alpha) \int \frac{d^4\ell}{(2\pi)^4}\frac{1 }{\ell^2} \,    \frac{1}{p_i\cdot k}\frac{1}{p_j\cdot \ell }\frac{1}{p_i\cdot (k+\ell)}
\label{softgravitinoirdiv1.2}
\end{eqnarray}
where $\beta_{ij}$ is given by 
\begin{eqnarray}
\beta_{ij} = \left(\frac{\iimg}{2}\right) \, \Big[2(p_i \cdot p_j)^2 - p_i^2p_j^2 \Big]
\label{softgravitinoirdiv3}
\end{eqnarray}
Adding the contributions from \eqref{softgravitinoirdiv1.1} and \eqref{softgravitinoirdiv1.2}, we get
\begin{eqnarray}
\Gamma^{(i;i,j)}_{N+1}(k,\{p_i\})&=& \kappa^2\beta_{ij} \int \frac{d^4\ell}{(2\pi)^4}      \frac{1 }{\ell^2} \,    \frac{\kappa\,  p_i \cdot \epsilon^\alpha\scharge_\alpha}{p_i\cdot k}\frac{1}{p_j\cdot \ell }\frac{1}{p_i\cdot \ell} \, 
\Gamma_N(\{p_i\} +\mathcal{O}(k^0)
\nonumber\\
&=& A(p_i,p_j)\frac{\kappa\,  p_i \cdot \epsilon^\alpha\scharge_\alpha}{p_i\cdot k}\, \Gamma_N(\{p_i\}) +\mathcal{O}(k^0)
\label{softgravitinoirdiv7}
\end{eqnarray}
where  $A(p_i,p_j) $ is the IR divergence that appears in diagram without soft gravitino which is depicted in fig. \ref{fig:gravitinoirdivergenceIII}. It is given by:
\begin{eqnarray}
A(p_i,p_j) 
= \kappa^2 \beta_{ij} \int \frac{d^4\ell}{(2\pi)^4} \frac{1}{\ell^2} \frac{1}{p_i\cdot \ell}\frac{1}{p_j\cdot \ell }
\label{softgravitinoirdiv8}
\end{eqnarray}
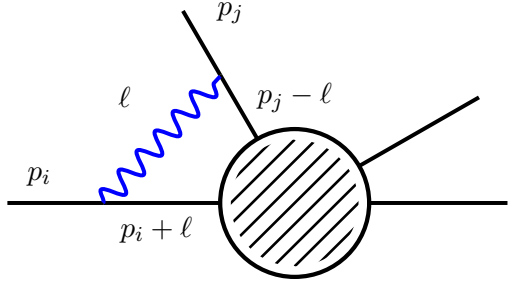
\begin{figure}[h]
\begin{center}
\begin{tikzpicture}[line width=1.5 pt, scale=1.4]

\begin{scope}[shift={(0,0)}]

\draw[realscalar] (.7,0)--(2,0);

\draw[realscalar][rotate=30] (.7,0)--(2,0);

\draw[realscalar][rotate=180] (1.8,0)--(2.7,0);
\draw[realscalar][rotate=180] (.7,0)--(1.8,0);

\draw[realscalar][rotate=120] (.7,0)--(1.4,0);
\draw[realscalar][rotate=120] (1.4,0)--(2.1,0);

\draw[graviton] (-1.8,0)--(-0.7,1.2);


 \node at (-1.6,1) {$\ell $}; 
  \node at (-2.4,.25) {$p_i $}; 
 \node at (-1.3,-.25) {$p_i +\ell $};
\node at (-.6,1.8) {$p_j $}; 
\node at (0,1) {$p_j-\ell  $};

\begin{scope}[shift={(0,0)}, scale=2]
    \draw [ultra thick] (0,0) circle (.35);
    \clip (0,0) circle (.3cm);
    \foreach \x in {-.9,-.8,...,.3}
    \draw[line width=1 pt] (\x,-.3) -- (\x+.6,.3);
\end{scope}
 
\end{scope}

\end{tikzpicture}
\end{center}
\caption{Infrared divergence in Supergravity III} 
\label{fig:gravitinoirdivergenceIII}
\end{figure}
 
  The contribution from diagram $(c)$ in fig \ref{fig:gravitinoirdivergenceII} is given by \cite{Weinberg:1965nx}
\begin{eqnarray}
 \Gamma^{(i;j \neq i,k \neq i)}_{N+1}(k,\{p_i\})&=&    A(p_j,p_k)\left[\frac{\kappa p_i \cdot \epsilon^\alpha\scharge_\alpha}{p_i\cdot k}\right] \, 
\Gamma_N(\{p_i\})
\label{softgravitinoirdiv7.11}
\end{eqnarray}
Putting \eqref{softgravitinoirdiv7} and \eqref{softgravitinoirdiv7.11} in  \eqref{totgamma}, we obtain
\begin{eqnarray}
\Gamma_{N+1}(k,\{p_i\})&=&  \Bigg[\kappa\sum_{i=1}^N\frac{ p_i \cdot \epsilon^\alpha\scharge_\alpha}{p_i\cdot k}\Bigg]   \left[ \sum_{j=1}^N\sum_{k=1,\ne j }^{N}  A(p_j,p_k)
    \right]\Gamma_N(\{p_i\})
\end{eqnarray}
Here we found that the soft gravitino factor just factors out from the IR divergent integral. 

\begin{figure}[h]
\begin{center}
\begin{tikzpicture}[line width=1.5 pt, scale=1.4]

\begin{scope}[shift={(0,0)}]

\draw[realscalar] (.7,0)--(2,0);

\draw[realscalar][rotate=30] (.7,0)--(2,0);

\draw[majorana][rotate=180] (2.6,0)--(3.0,0);
\draw[realscalar][rotate=180] (.7,0)--(2.6,0);

\draw[realscalar][rotate=120] (.7,0)--(1.4,0);
\draw[realscalar][rotate=120] (1.4,0)--(2.1,0);

 \draw[gravitino][rotate=180] (2.6,0)--(2.6,1);
\draw[graviton] (-1.8,0)--(-0.7,1.2); 

\draw[graviton] (-2.2,0)--(-.9,1.6);


 \node at (-1,.5) {$\ell_1 $}; 
 \node at (-1.9,1) {$\ell_2 $}; 

 \node at (-2.8,-1.2) {$k $}; 
  \node at (-2.4,.25) {$p_i $}; 
\node at (-.6,1.8) {$p_j $}; 

\begin{scope}[shift={(0,0)}, scale=2]
    \draw [ultra thick] (0,0) circle (.35);
    \clip (0,0) circle (.3cm);
    \foreach \x in {-.9,-.8,...,.3}
    \draw[line width=1 pt] (\x,-.3) -- (\x+.6,.3);
\end{scope}
 
\end{scope}

\begin{scope}[shift={(6,0)}]

\draw[realscalar] (.7,0)--(2,0);

\draw[realscalar][rotate=30] (.7,0)--(2,0);

\draw[majorana][rotate=180] (1.8,0)--(3.0,0);
\draw[realscalar][rotate=180] (.7,0)--(2,0);

\draw[realscalar][rotate=120] (.7,0)--(1.4,0);
\draw[realscalar][rotate=120] (1.4,0)--(2.1,0);

 \draw[gravitino][rotate=180] (2.0,0)--(2.0,1);
\draw[graviton] (-1.8,0)--(-0.7,1.2); 

\draw[graviton] (-2.2,0)--(-.9,1.6);


 \node at (-1,.5) {$\ell_1 $}; 
 \node at (-1.9,1) {$\ell_2 $}; 

 \node at (-2.0,-1.2) {$k $}; 
  \node at (-2.4,.25) {$p_i $}; 
\node at (-.6,1.8) {$p_j $}; 

\begin{scope}[shift={(0,0)}, scale=2]
    \draw [ultra thick] (0,0) circle (.35);
    \clip (0,0) circle (.3cm);
    \foreach \x in {-.9,-.8,...,.3}
    \draw[line width=1 pt] (\x,-.3) -- (\x+.6,.3);
\end{scope}
 
\end{scope}
 
\begin{scope}[shift={(3,-3)}]

 \draw[realscalar] (.7,0)--(2,0);

\draw[realscalar][rotate=30] (.7,0)--(2,0);

 \draw[majorana][rotate=180] (1.5,0)--(3.0,0);
\draw[realscalar][rotate=180] (.7,0)--(1.5,0);

\draw[realscalar][rotate=120] (.7,0)--(1.4,0);
\draw[realscalar][rotate=120] (1.4,0)--(2.1,0);

 \draw[gravitino][rotate=180] (1.5,0)--(1.5,1);
\draw[graviton] (-1.8,0)--(-0.7,1.2); 

\draw[graviton] (-2.2,0)--(-.9,1.6);


 \node at (-1,.5) {$\ell_1 $}; 
 \node at (-1.9,1) {$\ell_2 $}; 

 \node at (-1.5,-1.2) {$k $}; 
  \node at (-2.4,.25) {$p_i $}; 
\node at (-.6,1.8) {$p_j $}; 

\begin{scope}[shift={(0,0)}, scale=2]
    \draw [ultra thick] (0,0) circle (.35);
    \clip (0,0) circle (.3cm);
    \foreach \x in {-.9,-.8,...,.3}
    \draw[line width=1 pt] (\x,-.3) -- (\x+.6,.3);
\end{scope}
 
\end{scope}

\end{tikzpicture}
\end{center}
\caption{Infrared divergence in Supergravity IV} 
\label{fig:gravitinoirdivergenceIV}
\end{figure}
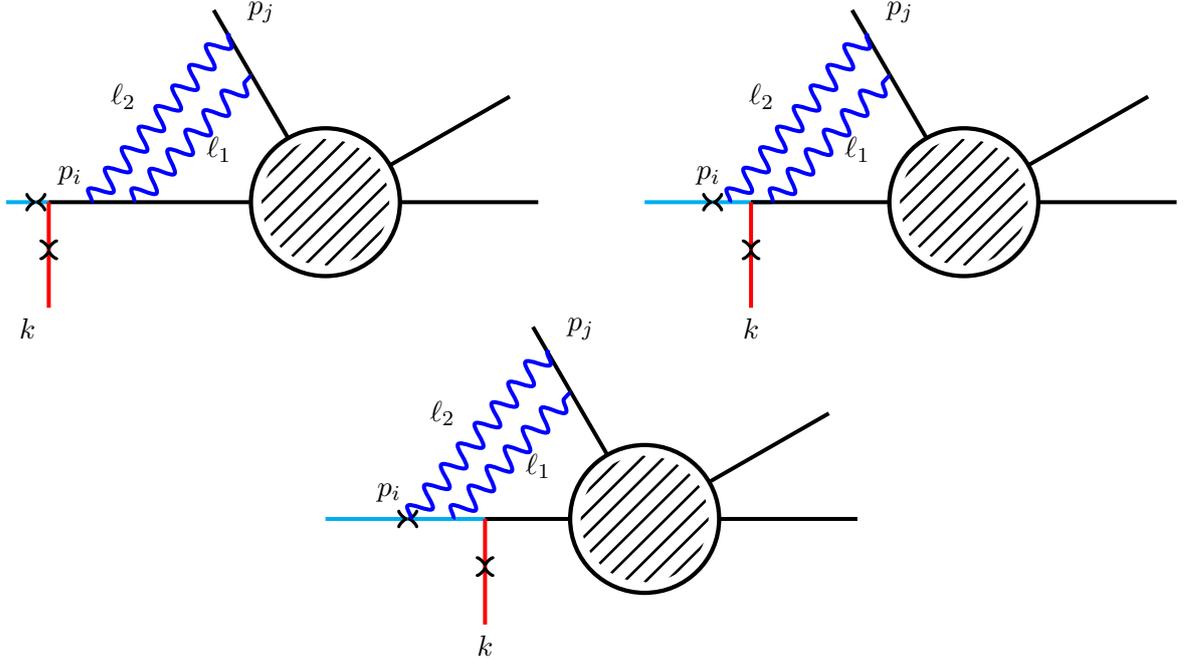

Now we will compute the two loop contributions to IR divergence. We depicted the corresponding Feynman diagrams in figure \ref{fig:gravitinoirdivergenceIV}. The contribution from these diagrams are given by 
\begin{eqnarray}
\Gamma^{(i;i,j)}_{N+1}(k,\{p_i\}) = \Bigg[\sum_{a=1}^6 \int d^4\ell_1 d^4\ell_2 \, I^{(a)} \Bigg]\, 
[\kappa\,  p_i \cdot \epsilon^\alpha\scharge_\alpha] \, 
\Gamma_N(\{p_i\})
\end{eqnarray}
where $I^{(1,2,3)}$ are obtained as the integrands from the three diagrams shown below. The other three integrands are obtained from the non-planar diagram. The explicit expressions for these integrands are given by 
\begin{eqnarray}
I^{(1)} &=&   \kappa^2   \frac{1}{p_i\cdot k}\, \frac{\beta_{ij}}{ \ell_2^2}\frac{1}{p_i\cdot (k + \ell_2)} \frac{\beta_{ij}}{ \ell_1^2}\frac{1}{p_i\cdot (k+ \ell_1+ \ell_2)}\frac{1}{p_j\cdot \ell_2 }\frac{1}{p_j\cdot (\ell_1+\ell_2)}\nonumber\\
I^{(2)}
&=& \kappa^2 \frac{\beta_{ij}}{ \ell_2^2}\frac{1}{p_i\cdot l_2}\frac{1}{p_i\cdot (k+ \ell_2)}\frac{\beta_{ij}}{p_i\cdot (k+ \ell_1+ \ell_2)}\frac{1}{p_j\cdot \ell_2 }\frac{1}{ \ell_1^2}\frac{1}{p_j\cdot (\ell_1+\ell_2)} \nonumber\\
I^{(3)} &=&  \kappa^2 \frac{\beta_{ij}}{p_i\cdot \ell_2}\frac{\beta_{ij}}{p_i\cdot (\ell_1+ \ell_2)}\frac{1}{p_i\cdot (k + \ell_1+ \ell_2)}\frac{1}{p_j\cdot \ell_2 }\frac{1}{ \ell_1^2}\frac{1}{ \ell_2^2}\frac{1}{p_j\cdot (\ell_1+\ell_2)}
\label{softgravitinoirdiv24}
\end{eqnarray}

Adding the three contributions above and the contributions from the non-planar diagrams  we obtain,
\begin{eqnarray}
I (p_i,p_j;k)&=& \frac{1}{p_i\cdot k} \int d^4\ell_1 d^4\ell_2 \, I(p_i,p_j;\ell_1,\ell_2)
\label{softgravitinoirdiv25}
\end{eqnarray}
where $I(p_i,p_j;\ell_1,\ell_2)$ is given by 
\begin{eqnarray}
I(p_i,p_j;\ell_1,\ell_2)&=&     \kappa^2 \frac{1}{p_i\cdot (\ell_1+ \ell_2)}\frac{1}{p_i\cdot \ell_2}\frac{1}{p_j\cdot \ell_2 }\frac{\beta_{ij}}{ \ell_1^2}\frac{\beta_{ij}}{ \ell_2^2}\frac{1}{p_j\cdot \ell_1}
\label{softgravitinoirdiv26}
\end{eqnarray}
 which is the same two loop integrand we get when there is no soft gravitino. 
There are other two loop diagrams that we have not depicted here, for example, the diagrams in which two virtual gravitons attach to different legs etc. Adding contribution from all two loop diagrams we obtain
\begin{eqnarray} 
\Gamma_{N+1}(k,\{p_i\})
&=&    \frac{1}{2}    \left[\sum_{j=1}^N\sum_{k=1;\ne j}^N A(p_j,p_k)\right]^2\left[\kappa\, \sum_{i=1}^N  \frac{ p_i \cdot \epsilon^\alpha\scharge_\alpha}{p_i\cdot k}\right]
\Gamma_N(\{p_i\})
\label{softgravitinoirdiv28}
\end{eqnarray}
Note that the soft factor appears just as a multiplicative factor with the infrared divergent piece. One can show that the contribution due to $N$- virtual soft-gravitons and an external soft gravitino comes out to be
\begin{eqnarray}
\Gamma_{N+1}(k,\{p_i\})
&=& \left[\kappa\, \sum_{i=1}^N  \frac{p_i \cdot \epsilon^\alpha\scharge_\alpha}{p_i\cdot k} \right] \left[\sum_{N=0}^\infty \frac{1}{N!}     \left[\sum_{j=1}^N\sum_{k=1;k\ne j}^N A(p_j,p_k)\right]^N\right] \Gamma_N(\{p_i\})
\label{softgravitinoirdiv29}
\end{eqnarray}
This implies the soft theorem is not affected by the IR divergence.
 
\subsubsection{Single real soft gravitino in presence of virtual graviphoton } 
\label{subsec:irdivergencevirtualgraviphoton}

In the presence of graviphoton, there are new IR divergent diagrams due to graviphoton running in the loops. These diagrams can be obtained by replacing graviton with graviphoton in the fig  \ref{fig:gravitinoirdivergenceII} and in the fig \ref{fig:gravitinoirdivergenceIV}. The computation is very similar to the one presented in subsection \ref{subsec:irdivergencevirtualgravitino}. The infrared divergence due to graviphoton is given by 
\begin{eqnarray}
   \left[\sum_{N=0}^\infty \frac{1}{N!}     \left[ B(p_i,p_j)\right]^N\right] 
\end{eqnarray}
where $B(p_i,p_j)$ is given by 
\begin{eqnarray}
    B(p_i,p_j)= \kappa^2 e_ie_j \int \frac{d^4\ell}{(2\pi)^4}  \frac{1}{\ell^2} \frac{1}{p_i\cdot \ell}\frac{1}{p_j\cdot \ell }\frac{1}{p_i\cdot (k+\ell) }
\end{eqnarray}
In presence of graviphoton, equation \eqref{softgravitinoirdiv29} will be replaced by the following equation 
\begin{eqnarray}
\Gamma_{N+1}(k,\{p_i\})
&=& \left[\kappa\,\sum_{i=1}^N  \frac{p_i \cdot \epsilon^\alpha\scharge_\alpha}{p_i\cdot k} \right] \left[\sum_{N=0}^\infty \frac{1}{N!}     \left[ \sum_{j=1}^N\sum_{k=1;k\ne j}^N (A(p_j,p_k) + B(p_j,p_k))\right]^N\right] \Gamma_N(\{p_i\})
\label{softgravitinoirdiv29}
\nonumber\\
\end{eqnarray}
Again we can see that the soft factor is not affected by the IR divergence.

\subsubsection{Massless matter}

Now we concentrate on the special case when some (or all) of the matter fields are massless \footnote{We are thankful to the unknown referee for pointing out this issue}. Weinberg in \cite{Weinberg:1965nx} showed that in the presence of massless matter the IR divergence due to virtual graviton cancels. However, there are irremovable IR divergences in QED with the massless matter.

In this case, the IR divergence comes from the presence of virtual graviton and virtual graviphoton.  The ones due to virtual graviton cancel due to Weinberg's argument. However, in the presence of graviphoton, there might be some non-removable IR divergences.
Graviphoton gauges the symmetries generated by the central charge. The central charge puts a lower bound on the mass of the particle (the BPS bound). The graviphoton only couples to matter with non-zero central charge and hence with non-zero mass. So there is no irremovable IR divergence in this case.

In presence of both the vector multiplet(s) and the massless matter multiplet(s) charged under the vector multiplet(s), there are irremovable IR divergences in $D=4$ due to the photon/gluon (of the vector multiplet) running in the loop. Since there is no vector multiplet in $\mathcal{N}=8$ supergravity, our analysis implies that there are no irremovable IR divergences in $\mathcal{N}=8$ supergravity (and in type II string theory).

\section{Conclusion}
\label{sec:conclusion}
In this paper, we have computed the multiple soft gravitini theorem at leading order in soft momenta for an arbitrary theory of supergravity. One natural question to ask is that what is the structure of the sub-leading soft gravitino theorem and how the structure of the subleading soft gravitino theorem is related to that of sub-leading and sub-subleading soft graviton theorem. One can use this approach to compute soft photino theorem and correction to soft photino theorem in the presence of gravitino, photon, and graviton. Another interesting question is to derive the result for multiple soft gravitini from the analysis of asymptotic symmetries and from CFT living on $\scri^\pm $ following \cite{Pasterski:2016qvg, Pasterski:2017kqt, Cheung:2016iub, Donnay:2018neh}. Following \cite{Higuchi:2018vyu} one could also try to verify this result from world-sheet methods. We leave these questions for future work.

\paragraph{Acknowledgement} 

We are thankful to Subramanya Hegde, Chandan Jana and Arnab P  Saha for collaborating at the early stage of the project. We are thankful to Matteo Bertolini, Atish Dabholkar, Kyriakos Papadodimas, Giovanni Tambalo, Cumrun Vafa and especially Ashoke Sen for discussions at various stages.  We are grateful to Atish Dabholkar, Nima Doroud, Subramanya Hegde, Chandan Jana, R Loganayagam, Usman Naseer and Mritunjay K Verma for reading the draft.  We are grateful to Ashoke Sen going through the draft meticulously and for suggesting many improvements to an earlier version of this draft. Finally, we are grateful to the anonymous referee for many suggestions to improve this paper.

\appendix
\section{Notation and convention}
\label{sec:gravitinonotation}
Our notation is as follows 
\begin{subequations}    
\begin{eqnarray} 
\textrm{Curved space indices}\qquad\qquad &&     \mu, \nu, \rho ,\sigma  
\\
\textrm{Tangent space indices}\qquad\qquad &&     a, b 
\\
\textrm{$SO(d,1)$ spinor indices}\qquad\qquad &&     \alpha, \beta 
\\
\textrm{Soft-particle indices}
\qquad \qquad && u,v
\\
\textrm{Hard-particle indices} 
\qquad \qquad && i,j
\\
\textrm{Number of Soft-particles}
\qquad \qquad &&  M
\\
\textrm{Number of Hard-particles} 
\qquad \qquad &&  N 
\\
\textrm{Polarization of the graviton}
\qquad \qquad &&  \zeta_{\mu \nu}
\\
\textrm{Polarization of the gravitino} 
\qquad \qquad &&  \epsilon_{\mu \alpha}
\end{eqnarray}
\end{subequations}

\subsection{Gamma matrix and spinor convention}

We use the following the gamma matrix convention  
\begin{eqnarray}
    \{\gamma^a,\gamma^b\}=-2\, \eta^{ab }
\label{gammaconvention1}
\end{eqnarray}    
and we get 
\begin{eqnarray}
    [\gamma^a,\gamma^{bc}]=-2\eta^{ab}\gamma^c+2\eta^{ac}\gamma^b
\label{gammaconvention2}
\end{eqnarray}
The spinors have the following index structure 
\begin{eqnarray}
    \psi_\alpha 
\label{gammaconvention3}
\end{eqnarray}
and the gamma matrix index structure is 
\begin{eqnarray}
    {{(\gamma^\mu)}_\alpha}^\beta 
\label{gammaconvention4}
\end{eqnarray}
We raise and lower the indices as follows (NW-SE convention)
\begin{eqnarray}
    \psi^\alpha = \mathcal{C}^{\alpha \beta }\psi_\beta 
    \qquad,\qquad  
    \psi_\alpha = \psi^\beta     \mathcal{C}_{\beta \alpha  }
\label{gammaconvention5}
\end{eqnarray}
Here $\mathcal{C}^{\alpha \beta }$ satisfies 
\begin{eqnarray}
    \mathcal{C}^{\alpha \beta }\mathcal{C}_{\gamma \beta }=\delta^\alpha_\gamma \qquad  
        \mathcal{C}_{\beta  \alpha }\mathcal{C}^{ \beta\gamma}=\delta^\gamma_\alpha
\label{gammaconvention6}
\end{eqnarray}
$(\gamma^\mu)_{\alpha \beta}$ is given by 
\begin{eqnarray}
\gamma^\mu_{\alpha \beta}=         {{(\gamma^\mu)}_\alpha}^\gamma\,  \mathcal{C}_{\gamma \beta  }
\label{gammaconvention7}
\end{eqnarray}

\subsection{ Majorana spinor }
For two Majorana spinors $\psi_1$ and $\psi_2$
\begin{eqnarray}
(\psi_1)^\alpha(\psi_2)_\alpha= (\psi_2)^\alpha(\psi_1)_\alpha
\end{eqnarray}
 
\section{Three soft gravitini} 
\label{sec:threesoft}

In this appendix, we present the explicit computation for three soft gravitini. This computation is instructive to understand the soft factor for multiple gravitini, described in section \ref{sec:arbitrarygravitino}. In this section, we write only $\Gamma_{N+3}$ instead of $\Gamma_{N+3} (\{p_i\}, k_1,k_2,k_3)$ to denote the amplitude with the soft gravitini and similarly we write $\Gamma_{N }$ instead of $\Gamma_{N}(\{p_i\})$ to denote the amplitudes involving only the hard-particles.   For three soft gravitini, the different contributions are as follows:
\begin{itemize}
\item First consider Feynman diagrams where all three gravitini attach to separate external legs (figure \ref{fig:triplegravitino1}). In this case the contribution will be just the multiplication of individual soft factors. So we get
\begin{equation} 
\Gamma_{N+3}^{(1)}=  \kappa^3 \sum_{i=1}^N  \frac{ p_i^{\mu} \epsilon^{(1); \alpha_1}_{\mu}}{p_i\cdot k_1} \scharge_{\alpha_1} \sum_{j=1, j\ne i }^N  \frac{ p_j^{\mu} \epsilon^{(2); \alpha_2}_{\mu} }{p_j\cdot k_2} \scharge_{\alpha_2} \sum_{k=1, k\ne i, j}^N  \frac{p_k^{\mu} \epsilon^{(3)\alpha_3}_{\mu}}{p_k\cdot k_3} \scharge_{\alpha_3}\,  \Gamma_{N}(\{p_i\})
\label{threegravitino1}
\end{equation}
    \item Now we consider the case when two gravitini attach to the same leg and the third one on different leg as shown in figure \ref{fig:triplegravitino1a}. The contribution from such configurations is given by 
\begin{equation} 
\Gamma_{N+3}^{uv|w;1} =  \kappa^3 \sum_{i=1}^N  \frac{ p_i^{\mu}\epsilon^{(u)\alpha_u}_{\mu}}{p_i\cdot k_u} \scharge_{\alpha_u}  \frac{ p_i^{\mu} \epsilon^{(v)\alpha_v}_{\mu}}{p_i\cdot (k_u + k_v)} \scharge_{\alpha_v} \sum_{j=1, j\ne i}^N  \frac{ p^{\mu}_j \epsilon^{(w)\alpha_w}_{\mu }}{p_j\cdot k_w} \scharge_{\alpha_w} \Gamma_N(\{p_i\})
\label{threegravitino1a}
\end{equation}
where $u,v,w$ can take values $1,2,3$. We can have different contributions depending on the order in which gravitini attach.
    \item The third possibility consists of the diagrams when all gravitini are being attached to the same external leg (figure \ref{fig:triplegravitinoII}). The contribution is given by

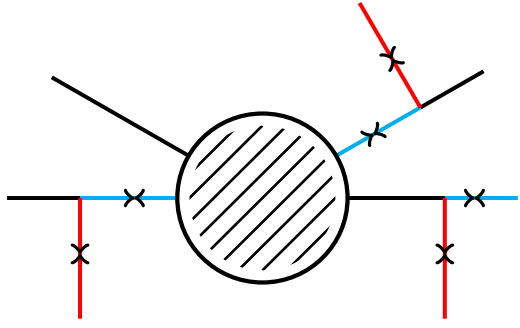
\begin{figure}[h]
\begin{center}
\begin{tikzpicture}[line width=1.5 pt, scale=.8]

\begin{scope}[shift={(0,0)}, scale=2]

\draw[realscalar] (.7,0)--(2,0);

\draw[realscalar][rotate=30] (1.5,0)--(2.1,0);
\draw[majorana][rotate=30] (.7,0)--(1.5,0);
\draw[realscalar][rotate=180] (1.5,0)--(2.1,0);

\draw[majorana][rotate=180] (.7,0)--(1.5,0);

\draw[realscalar][rotate=0] (.7,0)--(1.5,0);

\draw[majorana][rotate=0] (1.5,0)--(2.1,0);

\draw[realscalar][rotate=150] (.7,0)--(2,0);

\draw[gravitino][rotate=180] (1.5,0)--(1.5,1);

\draw[gravitinobar][rotate=0] (1.5,0)--(1.5,-1);
\draw[gravitino][rotate=30] (1.5,0)--(1.5,1);


\begin{scope}[shift={(0,0)}, scale=2]
    \draw [ultra thick] (0,0) circle (.35);
    \clip (0,0) circle (.3cm);
    \foreach \x in {-.9,-.8,...,.3}
    \draw[line width=1 pt] (\x,-.3) -- (\x+.6,.3);
\end{scope}
 
\end{scope}

\end{tikzpicture}
\end{center}
\caption{Feynman diagram for three soft gravitini - I}
\label{fig:triplegravitino1}
\end{figure}
\begin{figure}[h]
\begin{center}
\begin{tikzpicture}[line width=1.5 pt, scale=.8]

\begin{scope}[shift={(0,0)}, scale=2]

\draw[realscalar] (.7,0)--(2,0);

\draw[realscalar][rotate=30] (.7,0)--(2.1,0);

\draw[realscalar][rotate=180] (1.5,0)--(2.1,0);

\draw[majorana][rotate=180] (.9,0)--(1.5,0);
\draw[realscalar][rotate=180] (.7,0)--(.9,0);
\draw[realscalar][rotate=0] (.7,0)--(1.5,0);

\draw[majorana][rotate=0] (1.5,0)--(2.1,0);

\draw[realscalar][rotate=150] (.7,0)--(2,0);

\draw[gravitino][rotate= 180] (1.5,0)--(1.5,1);

\draw[gravitinobar][rotate= 180] (0.9,0)--(0.9,1);
\draw[gravitino][rotate= 0] (1.5,0)--(1.5,-1);


\begin{scope}[shift={(0,0)}, scale=2]
    \draw [ultra thick] (0,0) circle (.35);
    \clip (0,0) circle (.3cm);
    \foreach \x in {-.9,-.8,...,.3}
    \draw[line width=1 pt] (\x,-.3) -- (\x+.6,.3);
\end{scope}
 
\end{scope}

\end{tikzpicture}
\end{center}
\caption{Feynman diagram for three soft gravitini - II}
\label{fig:triplegravitino1a}
\end{figure}
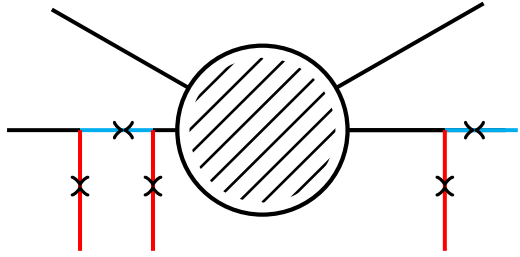
\begin{figure}[h]
\begin{center}
\begin{tikzpicture}[line width=1.5 pt, scale=.8]

\begin{scope}[shift={(0,0)}, scale=2]

\draw[realscalar] (.7,0)--(2,0);

\draw[realscalar][rotate=30] (.7,0)--(2,0);

\draw[realscalar][rotate=180] (.7,0)--(1.1,0);

\draw[majorana][rotate=180] (1.6,0)--(1.1,0);

\draw[realscalar][rotate=180] (1.6,0)--(2.1,0);
\draw[majorana][rotate=180] (2.1,0)--(2.6,0);

\draw[realscalar][rotate=150] (.7,0)--(2,0);

\draw[gravitino][rotate=180] (1.1,0)--(1.1,1);

\draw[gravitinobar][rotate=180] (1.6,0)--(1.6,1);
\draw[gravitino][rotate=180] (2.1,0)--(2.1,1);


\begin{scope}[shift={(0,0)}, scale=2]
    \draw [ultra thick] (0,0) circle (.35);
    \clip (0,0) circle (.3cm);
    \foreach \x in {-.9,-.8,...,.3}
    \draw[line width=1 pt] (\x,-.3) -- (\x+.6,.3);
\end{scope}
 
\end{scope}

\end{tikzpicture} 
\end{center}
\caption{Feynman diagram for three soft gravitini  - III}
\label{fig:triplegravitinoII}
\end{figure}
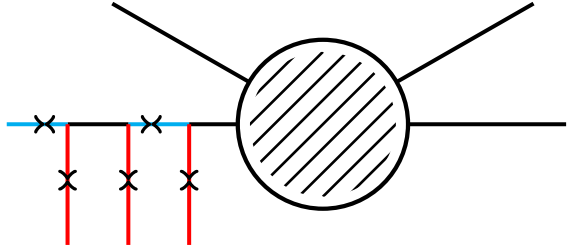
\begin{eqnarray} 
\Gamma_{N+3}^{uvw} &=&   \kappa^3 \sum_{i=1}^N  \frac{ \epsilon^{(u)\alpha_u}_{\mu}p_i^{\mu}}{p_i\cdot k_u} \frac{ \epsilon^{(v)\alpha_v}_{\nu}p_i^{\nu}}{p_i\cdot (k_u+k_v)}\frac{ \epsilon^{(w)\alpha_w}_{\rho}p_i^{\rho}}{p_i\cdot (k_u+k_v+k_w)} \scharge_{\alpha_u} \scharge_{\alpha_v} \scharge_{\alpha_w} \Gamma_N(\{p_i\})
\label{threegravitino11}
\end{eqnarray}
We have 6 diagrams which can be obtained by interchanging the external soft gravitini.

\item Now we consider the diagrams in which any two soft gravitini combine to give a soft graviton and then the soft graviton attaches to the external leg; the left-over (lonely !) third one directly attaches to the external leg. This can also give rise to two scenarios, i.e. the internal soft graviton and the leftover lonely gravitino can attach to same hard particles or different hard particles. 
\end{itemize}
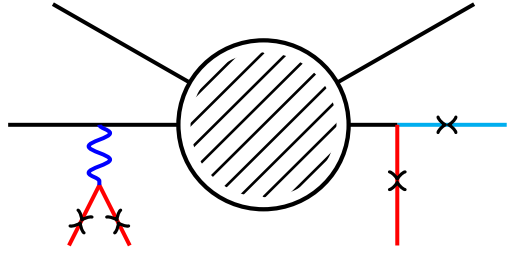
\begin{figure}[h]
\begin{center}
\begin{tikzpicture}[line width=1.5 pt, scale=.8]

\begin{scope}[shift={(0,-12)}, scale=2]

\draw[majorana] (1.1,0)--(2,0);
\draw[realscalar] (.7,0)--(1.1,0);
\draw[realscalar][rotate=30] (.7,0)--(2,0);

\draw[realscalar][rotate=180] (.7,0)--(2.1,0);

\draw[realscalar][rotate=150] (.7,0)--(2,0);

\draw[gravitinobar][rotate=180] (1.6,1)--(1.35,.5);


\draw[gravitino][rotate=180] (1.1,1)--(1.35,.5);

\draw[graviton][rotate=180] (1.35,0)--(1.35,.5);
\draw[gravitino] (1.1,0)--(1.1, - 1);


\begin{scope}[shift={(0,0)}, scale=2]
    \draw [ultra thick] (0,0) circle (.35);
    \clip (0,0) circle (.3cm);
    \foreach \x in {-.9,-.8,...,.3}
    \draw[line width=1 pt] (\x,-.3) -- (\x+.6,.3);
\end{scope}
 
\end{scope}

\end{tikzpicture}
\end{center}
\caption{Feynman diagram for three soft gravitini - IV}
\label{fig:triplegravitinoIII}
\end{figure}
In the case when they attach on separate legs as shown in figure \ref{fig:triplegravitinoIII}, we just have the multiplication of two factors:
\begin{eqnarray}
    \Gamma_{N+3}^{uv|w;2}&=&  \kappa^3  \sum_{i=1}^N   
\left[\frac{\cfactor_{uv}(p_i) }{ p_i\cdot (k_u + k_v) }\right]\sum_{j=1,j\ne i}^N   \frac{ \epsilon^{(w)\alpha_w}_\mu p_j^{\mu}}{p_j\cdot k_w} \scharge_{\alpha_w} \Gamma_N(\{p_i\})
\label{threegravitino21}
\end{eqnarray}
Since any two gravitini can combine to give the internal soft graviton  (and the third one will attach to the separate leg), there are three possibilities. 
 
Now we can have the case when both the internal soft graviton and the left-over soft gravitino attach to same external leg as shown in figure \ref{fig:triplegravitinoIV} 
\begin{figure}[h]
\begin{center}
\begin{tikzpicture}[line width=1.5 pt, scale=.8]

\begin{scope}[shift={(0,-12)}, scale=2]

\draw[realscalar] (0.7,0)--(2,0);
\draw[realscalar][rotate=30] (.7,0)--(2,0);

\draw[majorana][rotate=180] (1.85,0)--(2.5,0);
\draw[realscalar][rotate=180] (.7,0)--(1.35,0);
\draw[realscalar][rotate=180] (1.35,0)--(1.85,0);

\draw[realscalar][rotate=150] (.7,0)--(2,0);

\draw[gravitinobar][rotate=180] (1.6,1)--(1.35,.5);


\draw[gravitino][rotate=180] (1.1,1)--(1.35,.5);

\draw[graviton][rotate=180] (1.35,0)--(1.35,.5);
\draw[gravitino][rotate=180]  (1.85,0)--(1.85, 1);


\begin{scope}[shift={(0,0)}, scale=2]
    \draw [ultra thick] (0,0) circle (.35);
    \clip (0,0) circle (.3cm);
    \foreach \x in {-.9,-.8,...,.3}
    \draw[line width=1 pt] (\x,-.3) -- (\x+.6,.3);
\end{scope}
 
\end{scope}

\end{tikzpicture}
\end{center}
\caption{Feynman diagram for three soft gravitini - V}
\label{fig:triplegravitinoIV}
\end{figure}
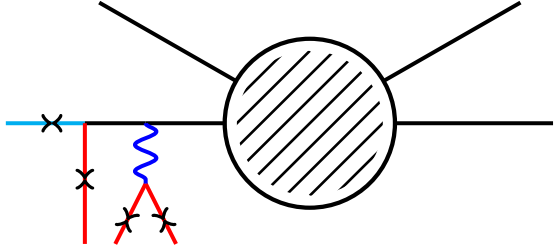
\begin{eqnarray} 
    \Gamma_{N+3}^{uv|w;3} &=&  \kappa^3  \sum_{i=1}^N   \left[\frac{ \epsilon^{(w)\alpha_w}_{\mu}p_j^{\mu}}{p_j\cdot k_w} \scharge_{\alpha_w} \, 
 \frac{\cfactor_{uv} (p_i)}{( p_i\cdot (k_1+k_2+k_3)) }\right] \Gamma_N(\{p_i\})
\label{threegravitino22}
\end{eqnarray}
We will have another diagram in which the graviton attaches to the external leg first and then the gravitino attaches to the external leg i.e.
\begin{figure}[h]
\begin{center}
\begin{tikzpicture}[line width=1.5 pt, scale=.8]

\begin{scope}[shift={(0,-12)}, scale=2]

\draw[realscalar] (0.7,0)--(2,0);
\draw[realscalar][rotate=30] (.7,0)--(2,0);

\draw[majorana][rotate=180] (1.85,0)--(2.5,0);
\draw[majorana][rotate=180] (1.35,0)--(1.85,0);

\draw[realscalar][rotate=180] (.7,0)--(1.35,0);

\draw[realscalar][rotate=150] (.7,0)--(2,0);

\draw[gravitinobar][rotate=180] (2.0,1)--(1.85,.5);


\draw[gravitino][rotate=180] (1.6,1)--(1.85,.5);

\draw[graviton][rotate=180] (1.85,0)--(1.85,.5);
\draw[gravitino][rotate=180]  (1.35,0)--(1.35, 1);


\begin{scope}[shift={(0,0)}, scale=2]
    \draw [ultra thick] (0,0) circle (.35);
    \clip (0,0) circle (.3cm);
    \foreach \x in {-.9,-.8,...,.3}
    \draw[line width=1 pt] (\x,-.3) -- (\x+.6,.3);
\end{scope}
 
\end{scope}

\end{tikzpicture}
\end{center}
\caption{Feynman diagram for three soft gravitini - VI}
\label{fig:triplegravitinoV}
\end{figure}
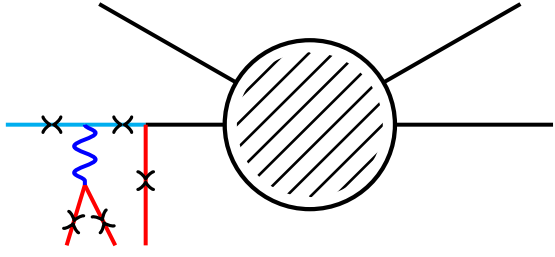
\begin{eqnarray} 
   \Gamma_{N+3}^{uv|w;4}&=& \kappa^3  \sum_{i=1}^N  
\left[\frac{ \cfactor_{uv} (p_i)}{p_i\cdot (k_u+k_v)} \frac{ \epsilon^{(w)\alpha_w}_{\mu}p_j^{\mu}}{ p_i\cdot (k_1+k_2+k_3) } \scharge_{\alpha_w} \, \right] \Gamma_N(\{p_i\})
\label{threegravitino23}
\end{eqnarray}
Adding the contributions from  \eqref{threegravitino21}, \eqref{threegravitino22} and \eqref{threegravitino23}, we get
\begin{eqnarray}
    \Gamma_{N+3}^{uv|w;4}&=&\Gamma_{N+3}^{uv|w;2}+\Gamma_{N+3}^{uv|w;3}+\Gamma_{N+3}^{uv|w;4}
 \nonumber\\   
  &=&    \kappa^2  \left[ \left[ \sum_{i=1}^N   
 \frac{\cfactor_{uv} (p_i)}{( p_i\cdot (k_u+k_v)) } \right]  \sfactor_w \right] \Gamma_N(\{p_i\})
\label{threegravitino33}
\end{eqnarray}
Now we write the contributions due to the six diagrams shown in figure \ref{fig:triplegravitinoII}. We choose a particular ordering. We choose $\scharge_\gamma$ to be the right-most. The $\Gamma_{N+3}^{123}$ remains the same 
\begin{eqnarray}
\Gamma_{N+3}^{123} &=&  \kappa^3 \sum_{i=1}^N   \frac{ \epsilon^{(1)\alpha}_{\mu}p_i^{\mu}}{p_i\cdot k_1} \frac{ \epsilon^{(2)\beta}_{\nu}p_i^{\nu}}{p_i\cdot (k_1+k_2)}\frac{ \epsilon^{(3)\gamma}_{\rho}p_i^{\rho}}{p_i\cdot (k_1+k_2+k_3)} \scharge_\alpha \scharge_\beta \scharge_\gamma\,  \Gamma_N(\{p_i\})
\label{threegravitino41}
\end{eqnarray}
Now we bring any other expression into this particular ordering by using \eqref{drsusyalgebra1}. For example, 
\begin{eqnarray}
\Gamma_{N+3}^{132} &=&   \kappa^3 \sum_{i=1}^N   \frac{ \epsilon^{(1)\alpha}_{\mu}p_i^{\mu}}{p_i\cdot k_1} \frac{ \epsilon^{(3)\gamma}_{\nu}p_i^{\nu}}{p_i\cdot (k_1+k_3)}\frac{ \epsilon^{(2)\beta}_{\rho}p_i^{\rho}}{p_i\cdot (k_1+k_2+k_3)} \scharge_\alpha \scharge_\gamma \scharge_\beta\,  \Gamma_N(\{p_i\})
\nonumber\\
&=& \kappa^3 \sum_{i=1}^N \frac{ \epsilon^{(1)\alpha}_{\mu}p_i^{\mu}}{p_i\cdot k_1} \frac{ \epsilon^{(2)\beta}_{\nu}p_i^{\nu}}{p_i\cdot (k_1+k_3)}\frac{ \epsilon^{(3)\gamma}_{\rho}p_i^{\rho}}{p_i\cdot (k_1+k_2+k_3)}\Big[\scharge_{\alpha} \scharge_\beta \scharge_{\gamma} + \frac{1}{2} (\slashed{p}_i)_{\beta\gamma}\scharge_\alpha\Big]\Gamma_N(\{p_i\})
\nonumber\\
\label{threegravitino42}
\end{eqnarray}
Following the same philosophy, we obtain 
\begin{subequations}
\begin{eqnarray}
\Gamma_{N+3}^{213} 
&=& \kappa^3 \sum_{i=1}^N  \frac{ \epsilon^{(1)\alpha}_{\mu}p_i^{\mu}}{p_i\cdot k_2} \frac{ \epsilon^{(2)\beta}_{\nu}p_i^{\nu}}{p_i\cdot (k_1+k_2)}\frac{ \epsilon^{(3)\gamma}_{\rho}p_i^{\rho}}{p_i\cdot (k_1+k_2+k_3)}\Big[\scharge_{\alpha} \scharge_\beta \scharge_{\gamma} + \frac{1}{2} (\slashed{p}_i)_{\alpha\beta} \scharge_\gamma\Big]\Gamma_N(\{p_i\})
\label{threegravitino43a}
\\
\Gamma_{N+3}^{231}  
&=& \kappa^3  \sum_{i=1}^N \frac{ \epsilon^{(1)\alpha}_{\mu}p_i^{\mu}}{p_i\cdot k_2} \frac{ \epsilon^{(2)\beta}_{\nu}p_i^{\nu}}{p_i\cdot (k_3+k_2)}\frac{ \epsilon^{(3)\gamma}_{\rho}p_i^{\rho}}{p_i\cdot (k_1+k_2+k_3)}\Big[ \scharge_{\alpha} \scharge_\beta \scharge_{\gamma} - \frac{1}{2} (\slashed{p}_i)_{\alpha\gamma}\scharge_\beta +\frac{1}{2} (\slashed{p}_i)_{\alpha\beta}\scharge_\gamma\Big]\Gamma_N(\{p_i\})
\nonumber\\
\label{threegravitino43b}
\\
\Gamma_{N+3}^{321}  
&=& \kappa^3  \sum_{i=1}^N \frac{ \epsilon^{(1)\alpha}_{\mu}p_i^{\mu}}{p_i\cdot k_3} \frac{ \epsilon^{(2)\beta}_{\nu}p_i^{\nu}}{p_i\cdot (k_3+k_2)}\frac{ \epsilon^{(3)\gamma}_{\rho}p_i^{\rho}}{p_i\cdot (k_1+k_2+k_3)}\Big[\scharge_{\alpha} \scharge_\beta \scharge_{\gamma}- \frac{1}{2} (\slashed{p}_i)_{\alpha\gamma} \scharge_\beta 
\nonumber\\
&&+\frac{1}{2} (\slashed{p}_i)_{\alpha\beta}\scharge_\gamma + \frac{1}{2} (\slashed{p}_i)_{\beta\gamma}\scharge_\alpha \Big]\Gamma_N(\{p_i\})
\label{threegravitino43c}
\\
\Gamma_{N+3}^{312}  
&=& \kappa^3  \sum_{i=1}^N \frac{ \epsilon^{(1)\alpha}_{\mu}p_i^{\mu}}{p_i\cdot k_3} \frac{ \epsilon^{(2)\beta}_{\nu}p_i^{\nu}}{p_i\cdot (k_3+k_1)}\frac{ \epsilon^{(3)\gamma}_{\rho}p_i^{\rho}}{p_i\cdot (k_1+k_2+k_3)}\Big[ \scharge_{\alpha} \scharge_\beta \scharge_{\gamma}- \frac{1}{2} (\slashed{p}_i)_{\alpha\gamma} \scharge_\beta + \frac{1}{2} (\slashed{p}_i)_{\beta\gamma} \scharge_\alpha\Big]\Gamma_N(\{p_i\})
\nonumber\\
\label{threegravitino43d}
\end{eqnarray}
\end{subequations}
Adding all the contributions from equation \eqref{threegravitino41}, \eqref{threegravitino42} \eqref{threegravitino43a}-\eqref{threegravitino43d} we  get:
\begin{eqnarray}\label{eqn_three_exchange}
\Gamma_{N+3}^{ex} &=&  \kappa^3  \sum_{i=1}^N  \frac{ \epsilon^{(1)\alpha}_{\mu}p_i^{\mu}}{p_i\cdot k_1} \frac{ \epsilon^{(2)\beta}_{\nu}p_i^{\nu}}{p_i\cdot k_2}\frac{ \epsilon^{(3)\gamma}_{\rho}p_i^{\rho}}{p_i\cdot k_3}\scharge_{\alpha} \scharge_\beta \scharge_{\gamma} \Gamma_N(p_i)  + \frac{1}{2}\kappa^3  \sum_{i=1}^N  \Bigg[\frac{ \epsilon^{(1)\alpha}_{\mu}p_i^{\mu}}{p_i\cdot k_1}(\slashed{p_i})_{\beta\gamma}\scharge_\alpha \frac{ \epsilon^{(2)\beta}_{\nu}p_i^{\nu}}{p_i\cdot (k_2+k_3)}
\nonumber\\
&&- \frac{ \epsilon^{(1)\alpha}_{\mu}p_i^{\mu}}{p_i\cdot (k_1+k_3)}\frac{ \epsilon^{(2)\beta}_{\nu}p_i^{\nu}}{p_i\cdot k_2}(\slashed{p_i})_{\alpha\gamma}\scharge_\beta - \frac{ \epsilon^{(1)\alpha}_{\mu}p_i^{\mu}}{p_i\cdot (k_3+k_2)} \frac{ \epsilon^{(2)\beta}_{\nu}p_i^{\nu}}{p_i\cdot k_2} (\slashed{p_i})_{\alpha\beta}\scharge_\gamma \Bigg]\frac{ \epsilon^{(3)\gamma}_{\rho}p_i^{\rho}}{p_i\cdot k_3}\Gamma_N(\{p_i\})
 \nonumber\\
\label{threegravitino44}
\end{eqnarray} 
We add the contributions from \eqref{threegravitino1}, \eqref{threegravitino1a} and \eqref{threegravitino33} to get the full answer. The full result can be written as
\begin{eqnarray}
    \Gamma_{N+3}(\{p_i\},\{k_u\}) =   \Big[\sfactor_1\, \sfactor_2\,  \sfactor_3 + \mfactor_{12}\, \sfactor_3 + \mfactor_{23}\, \sfactor_1 + \mfactor_{13}\, \sfactor_2 \Big]\Gamma_N(\{p_i\})
\label{threegravitino45}
\end{eqnarray}
where $\sfactor_u$ and $\mfactor_{uv} $ is defined in \eqref{singlesoftgrav11} and \eqref{doublesoftgravi62} respectively. The above answer matches with the proposed answer \eqref{multiplegravitinosoft} with $M=3$. 

\subsection*{Rearrangement} We have written the answer for a particular ordering ($1$-$2$-$3$). In this case, we explicitly demonstrate the rearrangement. Let's say we want to write in the order $1$-$3$-$2$.  We apply the identity \eqref{doublesoftgravi65} for $u=2, v=3$  
\begin{eqnarray}
    \Gamma_{N+3}(\{p_i\},\{k_u\}) &=&   \Big[\sfactor_1 (\sfactor_3 \sfactor_2-\mfactor_{23}+\mfactor_{32}) + \mfactor_{12}\, \sfactor_3 + \mfactor_{23}\, \sfactor_1 + \mfactor_{13}\, \sfactor_2 \Big]\Gamma_{N}(\{p_i\})
\nonumber\\
&=&   \Big[\sfactor_1\,  \sfactor_3\,  \sfactor_2 + \mfactor_{12}\, \sfactor_3\,  + \mfactor_{32}\, \sfactor_1 + \mfactor_{13}\, \sfactor_2 \Big]\Gamma_{N}(\{p_i\})
\label{threegravitino46}
\end{eqnarray}

\subsection{Gauge invariance} The gauge invariance of \eqref{threegravitino45} is the easiest to show if we put pure gauge polarization for the last one, for example, 3rd gravitino in \eqref{threegravitino45}, 2nd gravitino in \eqref{threegravitino46}. Because the answer can always be rearranged to any particular ordering, we can always bring any particular gravitino to be the last entry.  So, it's sufficient to show that the gauge invariance for the pure gauge polarization of the last one. 

Let us consider \eqref{threegravitino45} and pure gauge polarization for the 3rd gravitino. The first and the second term vanish as in equation \eqref{singlesoftgrav41} and the third \& the fourth term vanish because of \eqref{doublesoftgravi173a}. 

\paragraph{Symmetric form }
Now we write the answer \eqref{threegravitino46} in the form which is manifestly symmetric in all the gravitini 
\begin{eqnarray}
    \Gamma_{N+3}(\{p_i\},\{k_u\}) &=&   \left[\frac{1}{3!}\sfactor_{(1}\, \sfactor_2\, \sfactor_{3)}
    \right.+  
\nonumber\\
&&\left. 
  \kappa^2 \sum_{r\neq u \neq s}\sum_{i=1}^N  \frac{1 }{ p_i\cdot (k_r + k_u)}\left(\frac{(\epsilon^{(r)}\cdot p_i)\slashed{p_i}(\epsilon^{(u)}\cdot p_i)p_{i}\cdot (k_r-k_u)}{4(p_i\cdot k_r)( p_i \cdot k_u)} + \cfactor_{ur}(p_i)\right)\sfactor_s\right]\Gamma_N(\{p_i\})
 \nonumber\\
\label{threegravitino147}
\end{eqnarray}

%



\addcontentsline{toc}{section}{References}
\bibliographystyle{utphys}
\bibliography{jrsoftgravitinodraft}  

\providecommand{\href}[2]{#2}\begingroup\raggedright\begin{thebibliography}{10}

\bibitem{Sen:2017xjn}
A.~Sen, ``{Soft Theorems in Superstring Theory},''
  \href{http://dx.doi.org/10.1007/JHEP06(2017)113}{{\em JHEP} {\bfseries 06}
  (2017) 113},
\href{http://arxiv.org/abs/1702.03934}{{\ttfamily arXiv:1702.03934 [hep-th]}}.

\bibitem{GellMann:1954kc}
M.~Gell-Mann and M.~L. Goldberger, ``{Scattering of low-energy photons by
  particles of spin 1/2},''
\href{http://dx.doi.org/10.1103/PhysRev.96.1433}{{\em Phys. Rev.} {\bfseries
  96} (1954) 1433--1438}.

\bibitem{Low:1954kd}
F.~E. Low, ``{Scattering of light of very low frequency by systems of spin
  1/2},''
\href{http://dx.doi.org/10.1103/PhysRev.96.1428}{{\em Phys. Rev.} {\bfseries
  96} (1954) 1428--1432}.

\bibitem{Low:1958sn}
F.~E. Low, ``{Bremsstrahlung of very low-energy quanta in elementary particle
  collisions},''
\href{http://dx.doi.org/10.1103/PhysRev.110.974}{{\em Phys. Rev.} {\bfseries
  110} (1958) 974--977}.

\bibitem{Weinberg:1964ew}
S.~Weinberg, ``{Photons and Gravitons in s Matrix Theory: Derivation of Charge
  Conservation and Equality of Gravitational and Inertial Mass},''
\href{http://dx.doi.org/10.1103/PhysRev.135.B1049}{{\em Phys. Rev.} {\bfseries
  135} (1964) B1049--B1056}.

\bibitem{Weinberg:1965nx}
S.~Weinberg, ``{Infrared photons and gravitons},''
\href{http://dx.doi.org/10.1103/PhysRev.140.B516}{{\em Phys. Rev.} {\bfseries
  140} (1965) B516--B524}.

\bibitem{Gross:1968in}
D.~J. Gross and R.~Jackiw, ``{Low-Energy Theorem for Graviton Scattering},''
\href{http://dx.doi.org/10.1103/PhysRev.166.1287}{{\em Phys. Rev.} {\bfseries
  166} (1968) 1287--1292}.

\bibitem{Jackiw:1968zza}
R.~Jackiw, ``{Low-Energy Theorems for Massless Bosons: Photons and
  Gravitons},''
\href{http://dx.doi.org/10.1103/PhysRev.168.1623}{{\em Phys. Rev.} {\bfseries
  168} (1968) 1623--1633}.

\bibitem{Burnett:1967km}
T.~H. Burnett and N.~M. Kroll, ``{Extension of the low soft photon theorem},''
\href{http://dx.doi.org/10.1103/PhysRevLett.20.86}{{\em Phys. Rev. Lett.}
  {\bfseries 20} (1968) 86}.

\bibitem{Bell:1969yw}
J.~S. Bell and R.~Van~Royen, ``{On the low-burnett-kroll theorem for
  soft-photon emission},''
\href{http://dx.doi.org/10.1007/BF02823297}{{\em Nuovo Cim.} {\bfseries A60}
  (1969) 62--68}.

\bibitem{Saito:1970yq}
S.~Saito, ``{Low-energy theorem for Compton scattering},''
\href{http://dx.doi.org/10.1103/PhysRev.184.1894}{{\em Phys. Rev.} {\bfseries
  184} (1969) 1894--1902}.

\bibitem{DelDuca:1990gz}
V.~Del~Duca, ``{High-energy Bremsstrahlung Theorems for Soft Photons},''
\href{http://dx.doi.org/10.1016/0550-3213(90)90392-Q}{{\em Nucl. Phys.}
  {\bfseries B345} (1990) 369--388}.

\bibitem{Strominger:2013lka}
A.~Strominger, ``{Asymptotic Symmetries of Yang-Mills Theory},''
  \href{http://dx.doi.org/10.1007/JHEP07(2014)151}{{\em JHEP} {\bfseries 07}
  (2014) 151},
\href{http://arxiv.org/abs/1308.0589}{{\ttfamily arXiv:1308.0589 [hep-th]}}.

\bibitem{Strominger:2013jfa}
A.~Strominger, ``{On BMS Invariance of Gravitational Scattering},''
  \href{http://dx.doi.org/10.1007/JHEP07(2014)152}{{\em JHEP} {\bfseries 07}
  (2014) 152},
\href{http://arxiv.org/abs/1312.2229}{{\ttfamily arXiv:1312.2229 [hep-th]}}.

\bibitem{He:2014laa}
T.~He, V.~Lysov, P.~Mitra, and A.~Strominger, ``{BMS supertranslations and
  Weinberg?s soft graviton theorem},''
  \href{http://dx.doi.org/10.1007/JHEP05(2015)151}{{\em JHEP} {\bfseries 05}
  (2015) 151},
\href{http://arxiv.org/abs/1401.7026}{{\ttfamily arXiv:1401.7026 [hep-th]}}.

\bibitem{Strominger:2014pwa}
A.~Strominger and A.~Zhiboedov, ``{Gravitational Memory, BMS Supertranslations
  and Soft Theorems},'' \href{http://dx.doi.org/10.1007/JHEP01(2016)086}{{\em
  JHEP} {\bfseries 01} (2016) 086},
\href{http://arxiv.org/abs/1411.5745}{{\ttfamily arXiv:1411.5745 [hep-th]}}.

\bibitem{Kapec:2015vwa}
D.~Kapec, V.~Lysov, S.~Pasterski, and A.~Strominger, ``{Higher-Dimensional
  Supertranslations and Weinberg's Soft Graviton Theorem},''
\href{http://arxiv.org/abs/1502.07644}{{\ttfamily arXiv:1502.07644 [gr-qc]}}.

\bibitem{Strominger:2017zoo}
A.~Strominger, ``{Lectures on the Infrared Structure of Gravity and Gauge
  Theory},''
\href{http://arxiv.org/abs/1703.05448}{{\ttfamily arXiv:1703.05448 [hep-th]}}.

\bibitem{Kapec:2014zla}
D.~Kapec, V.~Lysov, and A.~Strominger, ``{Asymptotic Symmetries of Massless QED
  in Even Dimensions},''
  \href{http://dx.doi.org/10.4310/ATMP.2017.v21.n7.a6}{{\em Adv. Theor. Math.
  Phys.} {\bfseries 21} (2017) 1747--1767},
\href{http://arxiv.org/abs/1412.2763}{{\ttfamily arXiv:1412.2763 [hep-th]}}.

\bibitem{Pate:2017fgt}
M.~Pate, A.-M. Raclariu, and A.~Strominger, ``{Gravitational Memory in Higher
  Dimensions},'' \href{http://dx.doi.org/10.1007/JHEP06(2018)138}{{\em JHEP}
  {\bfseries 06} (2018) 138},
\href{http://arxiv.org/abs/1712.01204}{{\ttfamily arXiv:1712.01204 [hep-th]}}.

\bibitem{White:2011yy}
C.~D. White, ``{Factorization Properties of Soft Graviton Amplitudes},''
  \href{http://dx.doi.org/10.1007/JHEP05(2011)060}{{\em JHEP} {\bfseries 05}
  (2011) 060},
\href{http://arxiv.org/abs/1103.2981}{{\ttfamily arXiv:1103.2981 [hep-th]}}.

\bibitem{Casali:2014xpa}
E.~Casali, ``{Soft sub-leading divergences in Yang-Mills amplitudes},''
  \href{http://dx.doi.org/10.1007/JHEP08(2014)077}{{\em JHEP} {\bfseries 08}
  (2014) 077},
\href{http://arxiv.org/abs/1404.5551}{{\ttfamily arXiv:1404.5551 [hep-th]}}.

\bibitem{Schwab:2014xua}
B.~U.~W. Schwab and A.~Volovich, ``{Subleading Soft Theorem in Arbitrary
  Dimensions from Scattering Equations},''
  \href{http://dx.doi.org/10.1103/PhysRevLett.113.101601}{{\em Phys. Rev.
  Lett.} {\bfseries 113} no.~10, (2014) 101601},
\href{http://arxiv.org/abs/1404.7749}{{\ttfamily arXiv:1404.7749 [hep-th]}}.

\bibitem{Bern:2014oka}
Z.~Bern, S.~Davies, and J.~Nohle, ``{On Loop Corrections to Subleading Soft
  Behavior of Gluons and Gravitons},''
  \href{http://dx.doi.org/10.1103/PhysRevD.90.085015}{{\em Phys. Rev.}
  {\bfseries D90} no.~8, (2014) 085015},
\href{http://arxiv.org/abs/1405.1015}{{\ttfamily arXiv:1405.1015 [hep-th]}}.

\bibitem{He:2014bga}
S.~He, Y.-t. Huang, and C.~Wen, ``{Loop Corrections to Soft Theorems in Gauge
  Theories and Gravity},''
  \href{http://dx.doi.org/10.1007/JHEP12(2014)115}{{\em JHEP} {\bfseries 12}
  (2014) 115},
\href{http://arxiv.org/abs/1405.1410}{{\ttfamily arXiv:1405.1410 [hep-th]}}.

\bibitem{Larkoski:2014hta}
A.~J. Larkoski, ``{Conformal Invariance of the Subleading Soft Theorem in Gauge
  Theory},'' \href{http://dx.doi.org/10.1103/PhysRevD.90.087701}{{\em Phys.
  Rev.} {\bfseries D90} no.~8, (2014) 087701},
\href{http://arxiv.org/abs/1405.2346}{{\ttfamily arXiv:1405.2346 [hep-th]}}.

\bibitem{White:2014qia}
C.~D. White, ``{Diagrammatic insights into next-to-soft corrections},''
  \href{http://dx.doi.org/10.1016/j.physletb.2014.08.041}{{\em Phys. Lett.}
  {\bfseries B737} (2014) 216--222},
\href{http://arxiv.org/abs/1406.7184}{{\ttfamily arXiv:1406.7184 [hep-th]}}.

\bibitem{Cachazo:2014dia}
F.~Cachazo and E.~Y. Yuan, ``{Are Soft Theorems Renormalized?},''
\href{http://arxiv.org/abs/1405.3413}{{\ttfamily arXiv:1405.3413 [hep-th]}}.

\bibitem{Afkhami-Jeddi:2014fia}
N.~Afkhami-Jeddi, ``{Soft Graviton Theorem in Arbitrary Dimensions},''
\href{http://arxiv.org/abs/1405.3533}{{\ttfamily arXiv:1405.3533 [hep-th]}}.

\bibitem{Schwab:2014fia}
B.~U.~W. Schwab, ``{Subleading Soft Factor for String Disk Amplitudes},''
  \href{http://dx.doi.org/10.1007/JHEP08(2014)062}{{\em JHEP} {\bfseries 08}
  (2014) 062},
\href{http://arxiv.org/abs/1406.4172}{{\ttfamily arXiv:1406.4172 [hep-th]}}.

\bibitem{DiVecchia:2015oba}
P.~Di~Vecchia, R.~Marotta, and M.~Mojaza, ``{Soft theorem for the graviton,
  dilaton and the Kalb-Ramond field in the bosonic string},''
  \href{http://dx.doi.org/10.1007/JHEP05(2015)137}{{\em JHEP} {\bfseries 05}
  (2015) 137},
\href{http://arxiv.org/abs/1502.05258}{{\ttfamily arXiv:1502.05258 [hep-th]}}.

\bibitem{Bianchi:2014gla}
M.~Bianchi, S.~He, Y.-t. Huang, and C.~Wen, ``{More on Soft Theorems: Trees,
  Loops and Strings},''
  \href{http://dx.doi.org/10.1103/PhysRevD.92.065022}{{\em Phys. Rev.}
  {\bfseries D92} no.~6, (2015) 065022},
\href{http://arxiv.org/abs/1406.5155}{{\ttfamily arXiv:1406.5155 [hep-th]}}.

\bibitem{Broedel:2014fsa}
J.~Broedel, M.~de~Leeuw, J.~Plefka, and M.~Rosso, ``{Constraining subleading
  soft gluon and graviton theorems},''
  \href{http://dx.doi.org/10.1103/PhysRevD.90.065024}{{\em Phys. Rev.}
  {\bfseries D90} no.~6, (2014) 065024},
\href{http://arxiv.org/abs/1406.6574}{{\ttfamily arXiv:1406.6574 [hep-th]}}.

\bibitem{Vera:2014tda}
A.~Sabio~Vera and M.~A. Vazquez-Mozo, ``{The Double Copy Structure of Soft
  Gravitons},'' \href{http://dx.doi.org/10.1007/JHEP03(2015)070}{{\em JHEP}
  {\bfseries 03} (2015) 070},
\href{http://arxiv.org/abs/1412.3699}{{\ttfamily arXiv:1412.3699 [hep-th]}}.

\bibitem{Zlotnikov:2014sva}
M.~Zlotnikov, ``{Sub-sub-leading soft-graviton theorem in arbitrary
  dimension},'' \href{http://dx.doi.org/10.1007/JHEP10(2014)148}{{\em JHEP}
  {\bfseries 10} (2014) 148},
\href{http://arxiv.org/abs/1407.5936}{{\ttfamily arXiv:1407.5936 [hep-th]}}.

\bibitem{Du:2014eca}
Y.-J. Du, B.~Feng, C.-H. Fu, and Y.~Wang, ``{Note on Soft Graviton theorem by
  KLT Relation},'' \href{http://dx.doi.org/10.1007/JHEP11(2014)090}{{\em JHEP}
  {\bfseries 11} (2014) 090},
\href{http://arxiv.org/abs/1408.4179}{{\ttfamily arXiv:1408.4179 [hep-th]}}.

\bibitem{Cachazo:2015ksa}
F.~Cachazo, S.~He, and E.~Y. Yuan, ``{New Double Soft Emission Theorems},''
  \href{http://dx.doi.org/10.1103/PhysRevD.92.065030}{{\em Phys. Rev.}
  {\bfseries D92} no.~6, (2015) 065030},
\href{http://arxiv.org/abs/1503.04816}{{\ttfamily arXiv:1503.04816 [hep-th]}}.

\bibitem{Kalousios:2014uva}
C.~Kalousios and F.~Rojas, ``{Next to subleading soft-graviton theorem in
  arbitrary dimensions},''
  \href{http://dx.doi.org/10.1007/JHEP01(2015)107}{{\em JHEP} {\bfseries 01}
  (2015) 107},
\href{http://arxiv.org/abs/1407.5982}{{\ttfamily arXiv:1407.5982 [hep-th]}}.

\bibitem{Bern:2014vva}
Z.~Bern, S.~Davies, P.~Di~Vecchia, and J.~Nohle, ``{Low-Energy Behavior of
  Gluons and Gravitons from Gauge Invariance},''
  \href{http://dx.doi.org/10.1103/PhysRevD.90.084035}{{\em Phys. Rev.}
  {\bfseries D90} no.~8, (2014) 084035},
\href{http://arxiv.org/abs/1406.6987}{{\ttfamily arXiv:1406.6987 [hep-th]}}.

\bibitem{Bonocore:2014wua}
D.~Bonocore, E.~Laenen, L.~Magnea, L.~Vernazza, and C.~D. White, ``{The method
  of regions and next-to-soft corrections in Drell–Yan production},''
  \href{http://dx.doi.org/10.1016/j.physletb.2015.02.008}{{\em Phys. Lett.}
  {\bfseries B742} (2015) 375--382},
\href{http://arxiv.org/abs/1410.6406}{{\ttfamily arXiv:1410.6406 [hep-ph]}}.

\bibitem{Schwab:2014sla}
B.~U.~W. Schwab, ``{A Note on Soft Factors for Closed String Scattering},''
  \href{http://dx.doi.org/10.1007/JHEP03(2015)140}{{\em JHEP} {\bfseries 03}
  (2015) 140},
\href{http://arxiv.org/abs/1411.6661}{{\ttfamily arXiv:1411.6661 [hep-th]}}.

\bibitem{Klose:2015xoa}
T.~Klose, T.~McLoughlin, D.~Nandan, J.~Plefka, and G.~Travaglini,
  ``{Double-Soft Limits of Gluons and Gravitons},''
  \href{http://dx.doi.org/10.1007/JHEP07(2015)135}{{\em JHEP} {\bfseries 07}
  (2015) 135},
\href{http://arxiv.org/abs/1504.05558}{{\ttfamily arXiv:1504.05558 [hep-th]}}.

\bibitem{Lipstein:2015rxa}
A.~E. Lipstein, ``{Soft Theorems from Conformal Field Theory},''
  \href{http://dx.doi.org/10.1007/JHEP06(2015)166}{{\em JHEP} {\bfseries 06}
  (2015) 166},
\href{http://arxiv.org/abs/1504.01364}{{\ttfamily arXiv:1504.01364 [hep-th]}}.

\bibitem{Volovich:2015yoa}
A.~Volovich, C.~Wen, and M.~Zlotnikov, ``{Double Soft Theorems in Gauge and
  String Theories},'' \href{http://dx.doi.org/10.1007/JHEP07(2015)095}{{\em
  JHEP} {\bfseries 07} (2015) 095},
\href{http://arxiv.org/abs/1504.05559}{{\ttfamily arXiv:1504.05559 [hep-th]}}.

\bibitem{Rao:2016tgx}
J.~Rao and B.~Feng, ``{Note on Identities Inspired by New Soft Theorems},''
  \href{http://dx.doi.org/10.1007/JHEP04(2016)173}{{\em JHEP} {\bfseries 04}
  (2016) 173},
\href{http://arxiv.org/abs/1604.00650}{{\ttfamily arXiv:1604.00650 [hep-th]}}.

\bibitem{DiVecchia:2015bfa}
P.~Di~Vecchia, R.~Marotta, and M.~Mojaza, ``{Double-soft behavior for scalars
  and gluons from string theory},''
  \href{http://dx.doi.org/10.1007/JHEP12(2015)150}{{\em JHEP} {\bfseries 12}
  (2015) 150},
\href{http://arxiv.org/abs/1507.00938}{{\ttfamily arXiv:1507.00938 [hep-th]}}.

\bibitem{Bianchi:2015yta}
M.~Bianchi and A.~L. Guerrieri, ``{On the soft limit of open string disk
  amplitudes with massive states},''
  \href{http://dx.doi.org/10.1007/JHEP09(2015)164}{{\em JHEP} {\bfseries 09}
  (2015) 164},
\href{http://arxiv.org/abs/1505.05854}{{\ttfamily arXiv:1505.05854 [hep-th]}}.

\bibitem{Guerrieri:2015eea}
A.~L. Guerrieri, ``{Soft behavior of string amplitudes with external massive
  states},'' \href{http://dx.doi.org/10.1393/ncc/i2016-16221-2}{{\em Nuovo
  Cim.} {\bfseries C39} no.~1, (2016) 221},
\href{http://arxiv.org/abs/1507.08829}{{\ttfamily arXiv:1507.08829 [hep-th]}}.

\bibitem{Huang:2015sla}
Y.-t. Huang and C.~Wen, ``{Soft theorems from anomalous symmetries},''
  \href{http://dx.doi.org/10.1007/JHEP12(2015)143}{{\em JHEP} {\bfseries 12}
  (2015) 143},
\href{http://arxiv.org/abs/1509.07840}{{\ttfamily arXiv:1509.07840 [hep-th]}}.

\bibitem{Alston:2015gea}
S.~D. Alston, D.~C. Dunbar, and W.~B. Perkins, ``{$n$-point amplitudes with a
  single negative-helicity graviton},''
  \href{http://dx.doi.org/10.1103/PhysRevD.92.065024}{{\em Phys. Rev.}
  {\bfseries D92} no.~6, (2015) 065024},
\href{http://arxiv.org/abs/1507.08882}{{\ttfamily arXiv:1507.08882 [hep-th]}}.

\bibitem{Bianchi:2015lnw}
M.~Bianchi and A.~L. Guerrieri, ``{On the soft limit of closed string
  amplitudes with massive states},''
  \href{http://dx.doi.org/10.1016/j.nuclphysb.2016.02.005}{{\em Nucl. Phys.}
  {\bfseries B905} (2016) 188--216},
\href{http://arxiv.org/abs/1512.00803}{{\ttfamily arXiv:1512.00803 [hep-th]}}.

\bibitem{DiVecchia:2015srk}
P.~Di~Vecchia, R.~Marotta, and M.~Mojaza, ``{Soft Theorems from String
  Theory},'' \href{http://dx.doi.org/10.1002/prop.201500068}{{\em Fortsch.
  Phys.} {\bfseries 64} (2016) 389--393},
\href{http://arxiv.org/abs/1511.04921}{{\ttfamily arXiv:1511.04921 [hep-th]}}.

\bibitem{Bianchi:2016tju}
M.~Bianchi and A.~L. Guerrieri,
  \href{http://dx.doi.org/10.1142/9789813226609_0555}{``{On the soft limit of
  tree-level string amplitudes},''} in {\em {Proceedings, 14th Marcel Grossmann
  Meeting on Recent Developments in Theoretical and Experimental General
  Relativity, Astrophysics, and Relativistic Field Theories (MG14) (In 4
  Volumes): Rome, Italy, July 12-18, 2015}}, vol.~4, pp.~4157--4163.
\newblock 2017.
\newblock
\href{http://arxiv.org/abs/1601.03457}{{\ttfamily arXiv:1601.03457 [hep-th]}}.
\newblock

\bibitem{DiVecchia:2016amo}
P.~Di~Vecchia, R.~Marotta, and M.~Mojaza, ``{Subsubleading soft theorems of
  gravitons and dilatons in the bosonic string},''
  \href{http://dx.doi.org/10.1007/JHEP06(2016)054}{{\em JHEP} {\bfseries 06}
  (2016) 054},
\href{http://arxiv.org/abs/1604.03355}{{\ttfamily arXiv:1604.03355 [hep-th]}}.

\bibitem{He:2016vfi}
S.~He, Z.~Liu, and J.-B. Wu, ``{Scattering Equations, Twistor-string Formulas
  and Double-soft Limits in Four Dimensions},''
  \href{http://dx.doi.org/10.1007/JHEP07(2016)060}{{\em JHEP} {\bfseries 07}
  (2016) 060},
\href{http://arxiv.org/abs/1604.02834}{{\ttfamily arXiv:1604.02834 [hep-th]}}.

\bibitem{Cachazo:2016njl}
F.~Cachazo, P.~Cha, and S.~Mizera, ``{Extensions of Theories from Soft
  Limits},'' \href{http://dx.doi.org/10.1007/JHEP06(2016)170}{{\em JHEP}
  {\bfseries 06} (2016) 170},
\href{http://arxiv.org/abs/1604.03893}{{\ttfamily arXiv:1604.03893 [hep-th]}}.

\bibitem{Saha:2016kjr}
A.~P. Saha, ``{Double Soft Theorem for Perturbative Gravity},''
  \href{http://dx.doi.org/10.1007/JHEP09(2016)165}{{\em JHEP} {\bfseries 09}
  (2016) 165},
\href{http://arxiv.org/abs/1607.02700}{{\ttfamily arXiv:1607.02700 [hep-th]}}.

\bibitem{DiVecchia:2016szw}
P.~Di~Vecchia, R.~Marotta, and M.~Mojaza, ``{Soft behavior of a closed massless
  state in superstring and universality in the soft behavior of the dilaton},''
  \href{http://dx.doi.org/10.1007/JHEP12(2016)020}{{\em JHEP} {\bfseries 12}
  (2016) 020},
\href{http://arxiv.org/abs/1610.03481}{{\ttfamily arXiv:1610.03481 [hep-th]}}.

\bibitem{DiVecchia:2017uqn}
P.~Di~Vecchia, R.~Marotta, and M.~Mojaza, ``{Double-soft behavior of the
  dilaton of spontaneously broken conformal invariance},''
  \href{http://dx.doi.org/10.1007/JHEP09(2017)001}{{\em JHEP} {\bfseries 09}
  (2017) 001},
\href{http://arxiv.org/abs/1705.06175}{{\ttfamily arXiv:1705.06175 [hep-th]}}.

\bibitem{Cheung:2016drk}
C.~Cheung, K.~Kampf, J.~Novotny, C.-H. Shen, and J.~Trnka, ``{A Periodic Table
  of Effective Field Theories},''
  \href{http://dx.doi.org/10.1007/JHEP02(2017)020}{{\em JHEP} {\bfseries 02}
  (2017) 020},
\href{http://arxiv.org/abs/1611.03137}{{\ttfamily arXiv:1611.03137 [hep-th]}}.

\bibitem{Luna:2016idw}
A.~Luna, S.~Melville, S.~G. Naculich, and C.~D. White, ``{Next-to-soft
  corrections to high energy scattering in QCD and gravity},''
  \href{http://dx.doi.org/10.1007/JHEP01(2017)052}{{\em JHEP} {\bfseries 01}
  (2017) 052},
\href{http://arxiv.org/abs/1611.02172}{{\ttfamily arXiv:1611.02172 [hep-th]}}.

\bibitem{Elvang:2016qvq}
H.~Elvang, C.~R.~T. Jones, and S.~G. Naculich, ``{Soft Photon and Graviton
  Theorems in Effective Field Theory},''
  \href{http://dx.doi.org/10.1103/PhysRevLett.118.231601}{{\em Phys. Rev.
  Lett.} {\bfseries 118} no.~23, (2017) 231601},
\href{http://arxiv.org/abs/1611.07534}{{\ttfamily arXiv:1611.07534 [hep-th]}}.

\bibitem{Saha:2017yqi}
A.~P. Saha, ``{Double soft limit of the graviton amplitude from the
  Cachazo-He-Yuan formalism},''
  \href{http://dx.doi.org/10.1103/PhysRevD.96.045002}{{\em Phys. Rev.}
  {\bfseries D96} no.~4, (2017) 045002},
\href{http://arxiv.org/abs/1702.02350}{{\ttfamily arXiv:1702.02350 [hep-th]}}.

\bibitem{Chakrabarti:2017ltl}
S.~Chakrabarti, S.~P. Kashyap, B.~Sahoo, A.~Sen, and M.~Verma, ``{Subleading
  Soft Theorem for Multiple Soft Gravitons},''
\href{http://arxiv.org/abs/1707.06803}{{\ttfamily arXiv:1707.06803 [hep-th]}}.

\bibitem{Hamada:2018vrw}
Y.~Hamada and G.~Shiu, ``{Infinite Set of Soft Theorems in Gauge-Gravity
  Theories as Ward-Takahashi Identities},''
  \href{http://dx.doi.org/10.1103/PhysRevLett.120.201601}{{\em Phys. Rev.
  Lett.} {\bfseries 120} no.~20, (2018) 201601},
\href{http://arxiv.org/abs/1801.05528}{{\ttfamily arXiv:1801.05528 [hep-th]}}.

\bibitem{Li:2018gnc}
Z.-Z. Li, H.-H. Lin, and S.-Q. Zhang, ``{Infinite Soft Theorems from Gauge
  Symmetry},'' \href{http://dx.doi.org/10.1103/PhysRevD.98.045004}{{\em Phys.
  Rev.} {\bfseries D98} no.~4, (2018) 045004},
\href{http://arxiv.org/abs/1802.03148}{{\ttfamily arXiv:1802.03148 [hep-th]}}.

\bibitem{Sen:2017nim}
A.~Sen, ``{Subleading Soft Graviton Theorem for Loop Amplitudes},''
\href{http://arxiv.org/abs/1703.00024}{{\ttfamily arXiv:1703.00024 [hep-th]}}.

\bibitem{Laddha:2017ygw}
A.~Laddha and A.~Sen, ``{Sub-subleading Soft Graviton Theorem in Generic
  Theories of Quantum Gravity},''
\href{http://arxiv.org/abs/1706.00759}{{\ttfamily arXiv:1706.00759 [hep-th]}}.

\bibitem{Cachazo:2014fwa}
F.~Cachazo and A.~Strominger, ``{Evidence for a New Soft Graviton Theorem},''
\href{http://arxiv.org/abs/1404.4091}{{\ttfamily arXiv:1404.4091 [hep-th]}}.

\bibitem{AtulBhatkar:2018kfi}
S.~Atul~Bhatkar and B.~Sahoo, ``{Sub-leading Soft Theorem for arbitrary number
  of external soft photons and gravitons},''
\href{http://arxiv.org/abs/1809.01675}{{\ttfamily arXiv:1809.01675 [hep-th]}}.

\bibitem{Coleman:1967ad}
S.~R. Coleman and J.~Mandula, ``{All Possible Symmetries of the S Matrix},''
\href{http://dx.doi.org/10.1103/PhysRev.159.1251}{{\em Phys. Rev.} {\bfseries
  159} (1967) 1251--1256}.

\bibitem{Cachazo:2013hca}
F.~Cachazo, S.~He, and E.~Y. Yuan, ``{Scattering of Massless Particles in
  Arbitrary Dimensions},''
  \href{http://dx.doi.org/10.1103/PhysRevLett.113.171601}{{\em Phys. Rev.
  Lett.} {\bfseries 113} no.~17, (2014) 171601},
\href{http://arxiv.org/abs/1307.2199}{{\ttfamily arXiv:1307.2199 [hep-th]}}.

\bibitem{Cachazo:2013iea}
F.~Cachazo, S.~He, and E.~Y. Yuan, ``{Scattering of Massless Particles:
  Scalars, Gluons and Gravitons},''
  \href{http://dx.doi.org/10.1007/JHEP07(2014)033}{{\em JHEP} {\bfseries 07}
  (2014) 033},
\href{http://arxiv.org/abs/1309.0885}{{\ttfamily arXiv:1309.0885 [hep-th]}}.

\bibitem{Cachazo:2014xea}
F.~Cachazo, S.~He, and E.~Y. Yuan, ``{Scattering Equations and Matrices: From
  Einstein To Yang-Mills, DBI and NLSM},''
  \href{http://dx.doi.org/10.1007/JHEP07(2015)149}{{\em JHEP} {\bfseries 07}
  (2015) 149},
\href{http://arxiv.org/abs/1412.3479}{{\ttfamily arXiv:1412.3479 [hep-th]}}.

\bibitem{Cachazo:2013gna}
F.~Cachazo, S.~He, and E.~Y. Yuan, ``{Scattering equations and
  Kawai-Lewellen-Tye orthogonality},''
  \href{http://dx.doi.org/10.1103/PhysRevD.90.065001}{{\em Phys. Rev.}
  {\bfseries D90} no.~6, (2014) 065001},
\href{http://arxiv.org/abs/1306.6575}{{\ttfamily arXiv:1306.6575 [hep-th]}}.

\bibitem{Dumitrescu:2015fej}
T.~T. Dumitrescu, T.~He, P.~Mitra, and A.~Strominger, ``{Infinite-Dimensional
  Fermionic Symmetry in Supersymmetric Gauge Theories},''
\href{http://arxiv.org/abs/1511.07429}{{\ttfamily arXiv:1511.07429 [hep-th]}}.

\bibitem{Grisaru:1976vm}
M.~T. Grisaru, H.~N. Pendleton, and P.~van Nieuwenhuizen, ``{Supergravity and
  the S Matrix},''
\href{http://dx.doi.org/10.1103/PhysRevD.15.996}{{\em Phys. Rev.} {\bfseries
  D15} (1977) 996}.

\bibitem{Grisaru:1977kk}
M.~T. Grisaru and H.~N. Pendleton, ``{Soft Spin 3/2 Fermions Require Gravity
  and Supersymmetry},''
\href{http://dx.doi.org/10.1016/0370-2693(77)90383-5}{{\em Phys. Lett.}
  {\bfseries 67B} (1977) 323--326}.

\bibitem{Grisaru:1977px}
M.~T. Grisaru and H.~N. Pendleton, ``{Some Properties of Scattering Amplitudes
  in Supersymmetric Theories},''
\href{http://dx.doi.org/10.1016/0550-3213(77)90277-2}{{\em Nucl. Phys.}
  {\bfseries B124} (1977) 81--92}.

\bibitem{Lysov:2015jrs}
V.~Lysov, ``{Asymptotic Fermionic Symmetry From Soft Gravitino Theorem},''
\href{http://arxiv.org/abs/1512.03015}{{\ttfamily arXiv:1512.03015 [hep-th]}}.

\bibitem{Avery:2015iix}
S.~G. Avery and B.~U.~W. Schwab, ``{Residual Local Supersymmetry and the Soft
  Gravitino},'' \href{http://dx.doi.org/10.1103/PhysRevLett.116.171601}{{\em
  Phys. Rev. Lett.} {\bfseries 116} no.~17, (2016) 171601},
\href{http://arxiv.org/abs/1512.02657}{{\ttfamily arXiv:1512.02657 [hep-th]}}.

\bibitem{Kulish:1970ut}
P.~P. Kulish and L.~D. Faddeev, ``{Asymptotic conditions and infrared
  divergences in quantum electrodynamics},''
  \href{http://dx.doi.org/10.1007/BF01066485}{{\em Theor. Math. Phys.}
  {\bfseries 4} (1970) 745}.
[Teor. Mat. Fiz.4,153(1970)].

\bibitem{Kapec:2017tkm}
D.~Kapec, M.~Perry, A.-M. Raclariu, and A.~Strominger, ``{Infrared Divergences
  in QED, Revisited},''
  \href{http://dx.doi.org/10.1103/PhysRevD.96.085002}{{\em Phys. Rev.}
  {\bfseries D96} no.~8, (2017) 085002},
\href{http://arxiv.org/abs/1705.04311}{{\ttfamily arXiv:1705.04311 [hep-th]}}.

\bibitem{Gabai:2016kuf}
B.~Gabai and A.~Sever, ``{Large gauge symmetries and asymptotic states in
  QED},'' \href{http://dx.doi.org/10.1007/JHEP12(2016)095}{{\em JHEP}
  {\bfseries 12} (2016) 095},
\href{http://arxiv.org/abs/1607.08599}{{\ttfamily arXiv:1607.08599 [hep-th]}}.

\bibitem{Choi:2017ylo}
S.~Choi and R.~Akhoury, ``{BMS Supertranslation Symmetry Implies Faddeev-Kulish
  Amplitudes},'' \href{http://dx.doi.org/10.1007/JHEP02(2018)171}{{\em JHEP}
  {\bfseries 02} (2018) 171},
\href{http://arxiv.org/abs/1712.04551}{{\ttfamily arXiv:1712.04551 [hep-th]}}.

\bibitem{Ware:2013zja}
J.~Ware, R.~Saotome, and R.~Akhoury, ``{Construction of an asymptotic S matrix
  for perturbative quantum gravity},''
  \href{http://dx.doi.org/10.1007/JHEP10(2013)159}{{\em JHEP} {\bfseries 10}
  (2013) 159},
\href{http://arxiv.org/abs/1308.6285}{{\ttfamily arXiv:1308.6285 [hep-th]}}.

\bibitem{Choi:2017bna}
S.~Choi, U.~Kol, and R.~Akhoury, ``{Asymptotic Dynamics in Perturbative Quantum
  Gravity and BMS Supertranslations},''
  \href{http://dx.doi.org/10.1007/JHEP01(2018)142}{{\em JHEP} {\bfseries 01}
  (2018) 142},
\href{http://arxiv.org/abs/1708.05717}{{\ttfamily arXiv:1708.05717 [hep-th]}}.

\bibitem{Sen:2017szq}
A.~Sen, ``{Background Independence of Closed Superstring Field Theory},''
  \href{http://dx.doi.org/10.1007/JHEP02(2018)155}{{\em JHEP} {\bfseries 02}
  (2018) 155},
\href{http://arxiv.org/abs/1711.08468}{{\ttfamily arXiv:1711.08468 [hep-th]}}.

\bibitem{Ferrara:1976kg}
S.~Ferrara, D.~Z. Freedman, P.~van Nieuwenhuizen, P.~Breitenlohner, F.~Gliozzi,
  and J.~Scherk, ``{Scalar Multiplet Coupled to Supergravity},''
\href{http://dx.doi.org/10.1103/PhysRevD.15.1013}{{\em Phys. Rev.} {\bfseries
  D15} (1977) 1013}.

\bibitem{Zachos:1978iw}
C.~K. Zachos, ``{$N=2$ Supergravity Theory With a Gauged Central Charge},''
\href{http://dx.doi.org/10.1016/0370-2693(78)90799-2}{{\em Phys. Lett.}
  {\bfseries 76B} (1978) 329--332}.

\bibitem{Pasterski:2016qvg}
S.~Pasterski, S.-H. Shao, and A.~Strominger, ``{Flat Space Amplitudes and
  Conformal Symmetry of the Celestial Sphere},''
  \href{http://dx.doi.org/10.1103/PhysRevD.96.065026}{{\em Phys. Rev.}
  {\bfseries D96} no.~6, (2017) 065026},
\href{http://arxiv.org/abs/1701.00049}{{\ttfamily arXiv:1701.00049 [hep-th]}}.

\bibitem{Pasterski:2017kqt}
S.~Pasterski and S.-H. Shao, ``{Conformal basis for flat space amplitudes},''
  \href{http://dx.doi.org/10.1103/PhysRevD.96.065022}{{\em Phys. Rev.}
  {\bfseries D96} no.~6, (2017) 065022},
\href{http://arxiv.org/abs/1705.01027}{{\ttfamily arXiv:1705.01027 [hep-th]}}.

\bibitem{Cheung:2016iub}
C.~Cheung, A.~de~la Fuente, and R.~Sundrum, ``{4D scattering amplitudes and
  asymptotic symmetries from 2D CFT},''
  \href{http://dx.doi.org/10.1007/JHEP01(2017)112}{{\em JHEP} {\bfseries 01}
  (2017) 112},
\href{http://arxiv.org/abs/1609.00732}{{\ttfamily arXiv:1609.00732 [hep-th]}}.

\bibitem{Donnay:2018neh}
L.~Donnay, A.~Puhm, and A.~Strominger, ``{Conformally Soft Photons and
  Gravitons},''
\href{http://arxiv.org/abs/1810.05219}{{\ttfamily arXiv:1810.05219 [hep-th]}}.

\bibitem{Higuchi:2018vyu}
S.~Higuchi and H.~Kawai, ``{Universality of soft theorem from locality of soft
  vertex operators},''
\href{http://arxiv.org/abs/1805.11079}{{\ttfamily arXiv:1805.11079 [hep-th]}}.

\end{thebibliography}\endgroup

\end{document}